\documentclass[twocolumn]{myaastex631} 
\usepackage{graphicx,url}
\usepackage{amsmath,amssymb}
\usepackage{hyperref}
\hypersetup{colorlinks,
	linkcolor=blue, 
	urlcolor=magenta, 
	anchorcolor=blue,
	citecolor=blue
}
\usepackage{bm}
\usepackage{xcolor}
\usepackage{mathrsfs}
\usepackage{natbib}
\usepackage{soul} 
\newcommand{\norm}[1]{\left\lVert#1\right\rVert}

\newcommand\blfootnotetext[1]{%
	\begingroup
	\renewcommand\thefootnote{}\footnotetext{#1}%
	\addtocounter{footnote}{-1}%
	\endgroup}

\definecolor{greenW}{rgb}{0.0, 0.55, 0.1}
\definecolor{orangeW}{rgb}{1.0, 0.5, 0.05}

\defcitealias{Wanggj:2020}{W20}
\defcitealias{Wanggj:2022}{W22}

\shorttitle{CoLFI: Neural Density Estimator Inference}
\shortauthors{Wang et al.}

\begin{document}
\title{CoLFI: Cosmological Likelihood-free Inference with Neural Density Estimators}

\author{Guo-Jian Wang}
\affil{School of Chemistry and Physics, University of KwaZulu-Natal, Westville Campus, Private Bag X54001, Durban, 4000, South Africa}
\affil{NAOC-UKZN Computational Astrophysics Centre (NUCAC), University of KwaZulu-Natal, Durban, 4000, South Africa}
\affil{National Institute for Theoretical and Computational Sciences (NITheCS), South Africa}

\author{Cheng Cheng}
\affil{School of Chemistry and Physics, University of KwaZulu-Natal, Westville Campus, Private Bag X54001, Durban, 4000, South Africa}
\affil{NAOC-UKZN Computational Astrophysics Centre (NUCAC), University of KwaZulu-Natal, Durban, 4000, South Africa}
\affil{Xinjiang Astronomical Observatories, Chinese Academy of Sciences, Urumqi 830011, People's Republic of China}

\author{Yin-Zhe Ma$^{\dagger}$}
\blfootnotetext{$^{\dagger}$ Corresponding author: ma@ukzn.ac.za}
\affil{School of Chemistry and Physics, University of KwaZulu-Natal, Westville Campus, Private Bag X54001, Durban, 4000, South Africa}
\affil{NAOC-UKZN Computational Astrophysics Centre (NUCAC), University of KwaZulu-Natal, Durban, 4000, South Africa}
\affil{National Institute for Theoretical and Computational Sciences (NITheCS), South Africa}
\affil{Department of Physics, Stellenbosch University, Matieland 7602, South Africa}

\author{Jun-Qing Xia}
\affil{Department of Astronomy, Beijing Normal University, Beijing 100875, People's Republic of China}

\author{Amare Abebe}
\affil{National Institute for Theoretical and Computational Sciences (NITheCS), South Africa}
\affil{Centre for Space Research, North-West University, Potchefstroom 2520, South Africa}

\author{Aroonkumar Beesham}
\affil{National Institute for Theoretical and Computational Sciences (NITheCS), South Africa}
\affil{Department of Mathematical Sciences, University of Zululand, Private Bag X1001, Kwa-Dlangezwa 3886, South Africa}
\affil{Faculty of Natural Sciences, Mangosuthu University of Technology, P O Box 12363, Jacobs 4052, South Africa}

\received{2022 September 5}
\revised{2023 May 16}
\accepted{2023 June 18}
\published{2023 August 22}

\begin{abstract}
In previous works, we proposed to estimate cosmological parameters with the artificial neural network (ANN) and the mixture density network (MDN). In this work, we propose an improved method called the mixture neural network (MNN) to achieve parameter estimation by combining ANN and MDN, which can overcome shortcomings of the ANN and MDN methods. Besides, we propose sampling parameters in a hyper-ellipsoid for the generation of the training set, which makes the parameter estimation more efficient. A high-fidelity posterior distribution can be obtained using $\mathcal{O}(10^2)$ forward simulation samples. In addition, we develop a code-named {\it CoLFI} for parameter estimation, which incorporates the advantages of MNN, ANN, and MDN, and is suitable for any parameter estimation of complicated models in a wide range of scientific fields. CoLFI provides a more efficient way for parameter estimation, especially for cases where the likelihood function is intractable or cosmological models are complex and resource-consuming. It can learn the conditional probability density $p(\boldsymbol\theta|\boldsymbol{d})$ using samples generated by models, and the posterior distribution $p(\boldsymbol\theta|\boldsymbol{d}_0)$ can be obtained for a given observational data $\boldsymbol{d}_0$. We tested the MNN using power spectra of the cosmic microwave background and  Type Ia supernovae and obtained almost the same result as the Markov Chain Monte Carlo method. The numerical difference only exists at the level of $\mathcal{O}(10^{-2}\sigma)$. The method can be extended to higher-dimensional data.
\end{abstract}

\section{Introduction}\label{sec:introduction}

Parameter estimation is one of the most important steps in cosmology to understand the physical processes in the universe. The main method for parameter estimation is Bayesian inference based on Bayes' theorem:
\begin{equation}\label{equ:bayes_theorem}
p(\bm\theta|\bm{d}_0, \mathcal{M}) = \frac{p(\bm{d}_0|\bm\theta, \mathcal{M}) p(\bm\theta|\mathcal{M})}{p(\bm{d}_0|\mathcal{M})},
\end{equation}
where $\bm{d}_0$ is the observational data, $\mathcal{M}$ is the corresponding model, $p(\bm\theta|\bm{d}_0, \mathcal{M})$ is the posterior distribution, $p(\bm{d}_0|\bm\theta, \mathcal{M})$ is the likelihood function, $p(\bm\theta|\mathcal{M})$ is the prior distribution, and $p(\bm{d}_0|\mathcal{M})$ is the normalization constant, which is also called Bayesian evidence. For standard Bayesian inference, the posterior distribution is usually explored using the Markov Chain Monte Carlo (MCMC) sampling method, variational inference, or other Bayesian computation methods (see \citealt{Gelman:2013} for a review). These {\it traditional} methods generally require the computation of the likelihood function $p(\bm{d}_0|\bm\theta, \mathcal{M})$ for the given model $\mathcal{M}$ and parameters $\bm\theta$. However, for some complex and resource-consuming models, the likelihood function may be intractable in practice, and simulations may also consume a lot of time and computational resources, making parameter inference unpragmatic. Therefore, any parameter estimation that can avoid or solve these problems will be of great help in the study of cosmology.

Likelihood-free inference (LFI) is emerging as a new paradigm for performing Bayesian inference under very complex generative models, using only forward simulations. Traditional approaches to LFI are based on approximate Bayesian computation (ABC, \citealt{Lintusaari:2017}), which uses simulated data sets to bypass the computation of the likelihood function. ABC explores the prior model parameter space and compares simulated and observational data (or summary statistics $\bm{t}$ of the data) using a distance metric. An approximate Bayesian posterior distribution is then obtained by accepting samples whose distance metric is less than a given threshold \citep{Blum:2013,Akeret:2015,Hahn:2017}. This method is practical for cosmological data analysis and is widely used for parameter estimation in cosmology and astrophysics \citep{Cameron:2012,Weyant:2013,Ishidaa:2015,Jennings:2017,Aufort:2020,Tortorelli:2020}.

However, the ABC method typically requires a large number of simulations, which grow exponentially with the number of model parameters. Therefore, it is unfeasible to use the ABC method if the simulation is moderately expensive \citep{Alsing:2019}. To solve this problem, methods based on density-estimation likelihood-free inference (DELFI, \citealt{Fan:2013,Papamakarios:2016,Lueckmann:2017,Alsing:2018,Lueckmann:2019,Papamakarios:2019}; \citealt{Wanggj:2020}, hereafter \citetalias{Wanggj:2020}; \citealt{Wangyc:2021,Zhang:2021,Zhao:2022a,Zhao:2022b}; \citealt{Wanggj:2022}, hereafter \citetalias{Wanggj:2022}) have been proposed to train a flexible density estimator for the posterior distribution (or synthetic likelihood) using a series of simulated data (summary)-parameter pairs $\{\bm{t}, \bm\theta\}$. This approach enables high-fidelity posterior inferences, requiring numbers of simulation samples several orders of magnitude smaller than those from the traditional ABC-based methods. There are, in principle, three ways for parameter inference in DELFI \citep{Alsing:2019}:
\begin{itemize}
	\item[(1)] Fit a model to the joint probability density $p(\bm\theta, \bm{t})$. The posterior distribution is then obtained by evaluating the joint probability density at the observational summary $\bm{t}_0$, $p(\bm\theta|\bm{t}_0)\propto p(\bm\theta, \bm{t}{\rm =}\bm{t}_0)$ \citep{Alsing:2018}.
	\item[(2)] Fit a model to the conditional probability density $p(\bm\theta|\bm{t})$. Then, obtain the posterior distribution at the observational summary $\bm{t}_0$ \citep{Papamakarios:2016,Lueckmann:2017}.
	\item[(3)] Fit a model to the conditional probability density $p(\bm{t}|\bm\theta)$. Then, obtain the likelihood $p(\bm{t}_0|\bm\theta)$ by evaluating at the observational summary $\bm{t}_0$. Finally, the posterior distribution is obtained by multiplying the likelihood and the prior as $p(\bm\theta|\bm{t}_0)\propto p(\bm{t}_0|\bm\theta)\times p(\bm\theta)$ \citep{Alsing:2019,Lueckmann:2019,Papamakarios:2019}.
\end{itemize}

In the literature, the mixture density network (MDN; \citealt{Bishop:1994}) is used to model the conditional probability density $p(\bm\theta|\bm{d})$ with a mixture model (\citealt{Papamakarios:2016,Lueckmann:2017}; \citetalias{Wanggj:2022}). It is found that MDN can improve LFI in several ways, such as representing the posterior distribution parametrically, as opposed to as a set of samples; targeting an exact posterior distribution rather than an approximation of it; effectively utilizing simulation by interpolating between samples and gradually focusing on reasonable parameter regions, instead of rejecting samples. In \citetalias{Wanggj:2020}, we proposed estimating parameters using the artificial neural network (ANN) by learning the conditional probability density $p(\bm\theta|\bm{d})$. Then, the posterior distribution can be obtained at the observational data $\bm{d}_0$. However, for data with covariance, the conditional probability density $p(\bm\theta|\bm{d})$ is not well learned. In \citetalias{Wanggj:2022}, we then proposed to estimate parameters using the MDN to model the conditional probability density $p(\bm\theta|\bm{d})$ with a mixture model. The MDN method can solve the problem in \citetalias{Wanggj:2020}. However, we found that multiple components should be used, especially for parameters that deviate from Gaussian distribution. Using multiple components will increase the training time, and the MDN is sometimes unstable in the  training process, resulting in failure to obtain a posterior distribution.

In this work, we propose an improved method called the mixture neural network (MNN) to achieve parameter estimation by combining the ANN and MDN, which can overcome limitations in the ANN and MDN methods. The MNN method is designed to learn  the conditional probability density $p(\bm\theta|\bm{d})$ using a series of simulated data-parameter pairs $\{\bm{d}, \bm\theta\}$. Then, the posterior distribution can be obtained with high accuracy at the observational data point $\bm{d}_0$. Furthermore, we propose an efficient parameter space sampling method by considering the covariance between parameters, which makes it possible to train  a network with $\mathcal{O}(10^2)$ forward simulation samples. In addition, a code called Cosmological Likelihood-free Inference (CoLFI\footnote{\url{https://github.com/Guo-Jian-Wang/colfi}}) is developed to achieve parameter inference. We test the MNN method by estimating parameters of the $\Lambda$ cold dark matter (CDM) and $w$CDM cosmological models using Type Ia supernovae (SN Ia) and angular power spectra of the cosmic microwave background (CMB).

This paper is organized as follows: In Section \ref{sec:methodology}, we illustrate the method of estimating parameters using the MNN, which includes a subsequent introduction to the ANN, MDN, MNN, generation of the training  set, and training  and parameter estimation method. Section \ref{sec:application_to_CMB} shows the application of the MNN method to the {\it Planck} CMB data. Section \ref{sec:application_to_SN} shows the application of the MNN method to the Pantheon SN Ia data. Section \ref{sec:joint_constraint} presents a joint constraint on parameters using the power spectra of the {\it Planck} CMB and the Pantheon SN Ia data. Section \ref{sec:effect_of_hyperparameters} shows the effect of hyperparameters of the network on parameter estimations. Discussions about the MNN method and CoLFI are present in Section \ref{sec:discussions}. We conclude in Section \ref{sec:conclusions}.

\section{Methodology}\label{sec:methodology}

In this section, we will first review the general principle of estimating parameters using the ANN and MDN in Sections \ref{sec:ann} and \ref{sec:mdn}. Then, we introduce the method of estimating parameters with the  MNN in Section \ref{sec:mnn}. In Section \ref{sec:training_data}, the generation and preprocessing methods of training set are introduced. Finally, we illustrate the detailed training and parameter estimation process in Section \ref{sec:training_and_parameter_estimation}.

\subsection{Artificial Neural Networks}\label{sec:ann}

An ANN, also called a neural network (NN), is a mathematical model that contains a collection of input, hidden, and output layers. Each layer contains many neurons, which are the basic elements of an NN. Each neuron transforms the input from other neurons and gives an output
\begin{equation}\label{equ:neuron_function}
y = f\left(\sum_i w_i x_i + b\right)~,
\end{equation}
where $x$ is the input, $f(\cdot)$ is a nonlinear function, which is usually called an activation function, $w$ and $b$ are parameters to be learned  by the network. In general, the batch normalization technique \citep{Ioffe:2015} is applied before the activation function to facilitate optimization and speed up convergence.

In supervised learning  tasks, the network should be trained  using a training  set before conducting estimations. For the parameter estimation task, the training  set contains a collection of measurements that are labeled corresponding to ground-truth  parameters, where the measurements are generated by a specific model (e.g., a cosmological model). Therefore, the measurements $\bm{d}$ are fed to the input layer, then the information of the measurements passes through each hidden layer, and finally, the estimated parameters $\bm\theta$ are computed from the output layer. In the training  process, the network will be trained  by minimizing a loss function $\mathcal{L}$, which quantifies the difference between the predicted result and the ground truth. For the ANN method proposed by \citetalias{Wanggj:2020}, the least absolute deviation is used as the loss function:
\begin{equation}\label{equ:loss_L1}
\mathcal{L} = \mathbb{E}\left( \frac{1}{N}\sum_{i=1}^{N}|\bm\theta_i - \hat{\bm\theta}_i| \right) ,
\end{equation}
where $N$ is the number of cosmological parameters, $\hat{\bm\theta}$ is the ground truth (i.e. the target) in the training set, and $\bm\theta$ here is the estimated cosmological parameters, which should be considered as a point of the parameter space. Therefore, the $\bm\theta$ here can be interpreted as point estimates of the cosmological parameters. The losses here are averaged over cosmological parameters and also averaged over the minibatch samples fed to the network. For more details, we refer interested readers to \citetalias{Wanggj:2020}.

\subsection{Mixture Density Network}\label{sec:mdn}

An MDN is a combination of an ANN and a mixture model. The mixture model here is a probabilistic model that assumes that all data points are generated from a mixture of a finite number of distributions with unknown parameters, where the distribution can be any kind of probability distribution (e.g. Gaussian distribution or Beta distribution). Therefore, for measurement $\bm{d}$ and cosmological parameters $\bm\theta$, the probability density of $\bm\theta$ with $K$ components has the form \citepalias{Wanggj:2022}
\begin{equation}\label{equ:pdf_of_mixture_model}
p(\bm\theta|\bm{d}) = \sum_{i=1}^K \omega_i p_i(\bm\theta|\bm{d})~,
\end{equation} 
where the nonnegative, normalized $\omega_i$ is a mixture weight representing the probability that $\bm\theta$ belongs to the $i$th component ($\sum_{i=1}^{K}\omega_i = 1$).

For the MDN with Gaussian components, Equation~(\ref{equ:pdf_of_mixture_model}) becomes (see also, e.g.~\citetalias{Wanggj:2022})
\begin{align}\label{equ:pdf_of_gaussian_1}
\nonumber p(\theta|\bm{d}) &= \sum_{i=1}^K \omega_i\mathcal{N}(\theta; \mu_i, \sigma_i) \\
&= \sum_{i=1}^K \omega_i\cdot\frac{1}{\sqrt{2\pi\sigma^2_i}}e^{-\frac{(\theta-\mu_i)^2}{2\sigma^2_i}}~,
\end{align}
for the case of only one parameter, and
\begin{align}\label{equ:pdf_of_gaussian_multi}
\nonumber p(\bm\theta|\bm{d}) &= \sum_{i=1}^K \omega_i\mathcal{N}(\bm\theta; \bm\mu_i, \bm\Sigma_i) \\
&= \sum_{i=1}^K \omega_i\cdot\frac{\exp{\left( -\tfrac{1}{2} (\bm\theta - \bm\mu_i)^\top \bm\Sigma_i^{-1} (\bm\theta - \bm\mu_i) \right)}}{\sqrt{\left( 2\pi \right)^N |\bm\Sigma_i|}}~,
\end{align}
for the case of $N$ parameters. The purpose of parameter estimation using the MDN is to obtain the parameters of the mixture model. Therefore, an MDN with Gaussian components actually learns a mapping between the measurement $\bm{d}$ and the parameters of the Gaussian mixture model ($\omega$, $\bm\mu$, and $\bm\Sigma$; or $\sigma$ for the case of only one cosmological parameter). Therefore, the parameters ($w$ and $b$ in Equation~(\ref{equ:neuron_function})) of an MDN with Gaussian components can be optimized by minimizing the loss function
\begin{align}\label{equ:loss_mdn_gaussian_1}
\mathcal{L} &= \mathbb{E}\left[ -\ln\left(\sum_{i=1}^K \omega_i\cdot\frac{1}{\sqrt{2\pi\sigma^2_i}}e^{-\frac{(\hat{\theta}-\mu_i)^2}{2\sigma^2_i}}\right) \right] ~,
\end{align}
\begin{align}\label{equ:loss_mdn_gaussian_multi}
\nonumber\mathcal{L} &= \mathbb{E}\left[ -\ln\left( \sum_{i=1}^K \omega_i \right.\right. \\
&\left.\left.\times\frac{\exp{\left( -\tfrac{1}{2} (\hat{\bm\theta} - \bm\mu_i)^\top \bm\Sigma_i^{-1} (\hat{\bm\theta} - \bm\mu_i) \right)}}{\sqrt{\left( 2\pi \right)^N |\bm\Sigma_i|}} \right) \right] ~,
\end{align}
for one and multiple parameters respectively, where $\hat{\bm\theta}$ (or $\hat{\theta}$) is the ground truth (i.e. the target) in the training set. ($\omega$, $\bm\mu$, or $\mu$; and $\bm\Sigma$, or $\sigma$) is the parameter set of the Gaussian mixture model. The losses here are averaged over the minibatch samples fed to the network.

After the training process, we can obtain the parameters of the Gaussian components; thus, we can finally obtain the posterior distribution by generating samples via Equations~(\ref{equ:pdf_of_gaussian_1}) and (\ref{equ:pdf_of_gaussian_multi}). However, it should be noted that, to make the convergence of the network more stable, the MDN actually learns  the upper Cholesky factor $\bm U$ of the precision matrix $\bm\Sigma^{-1}$ \citepalias{Wanggj:2022}. Even so, when using a very large number of components for multiple-parameter cases, the learned upper Cholesky factor $\bm U$ may cause the covariance matrix $\bm\Sigma$ to be nonpositive definite, which makes the MDN unstable and unable to constrain these components. Therefore, it is difficult to use the MDN method for some multiple-parameter cases, especially for parameters with a non-Gaussian distribution that requires more components.

\begin{figure*}
	\centering
	\includegraphics[width=0.45\textwidth]{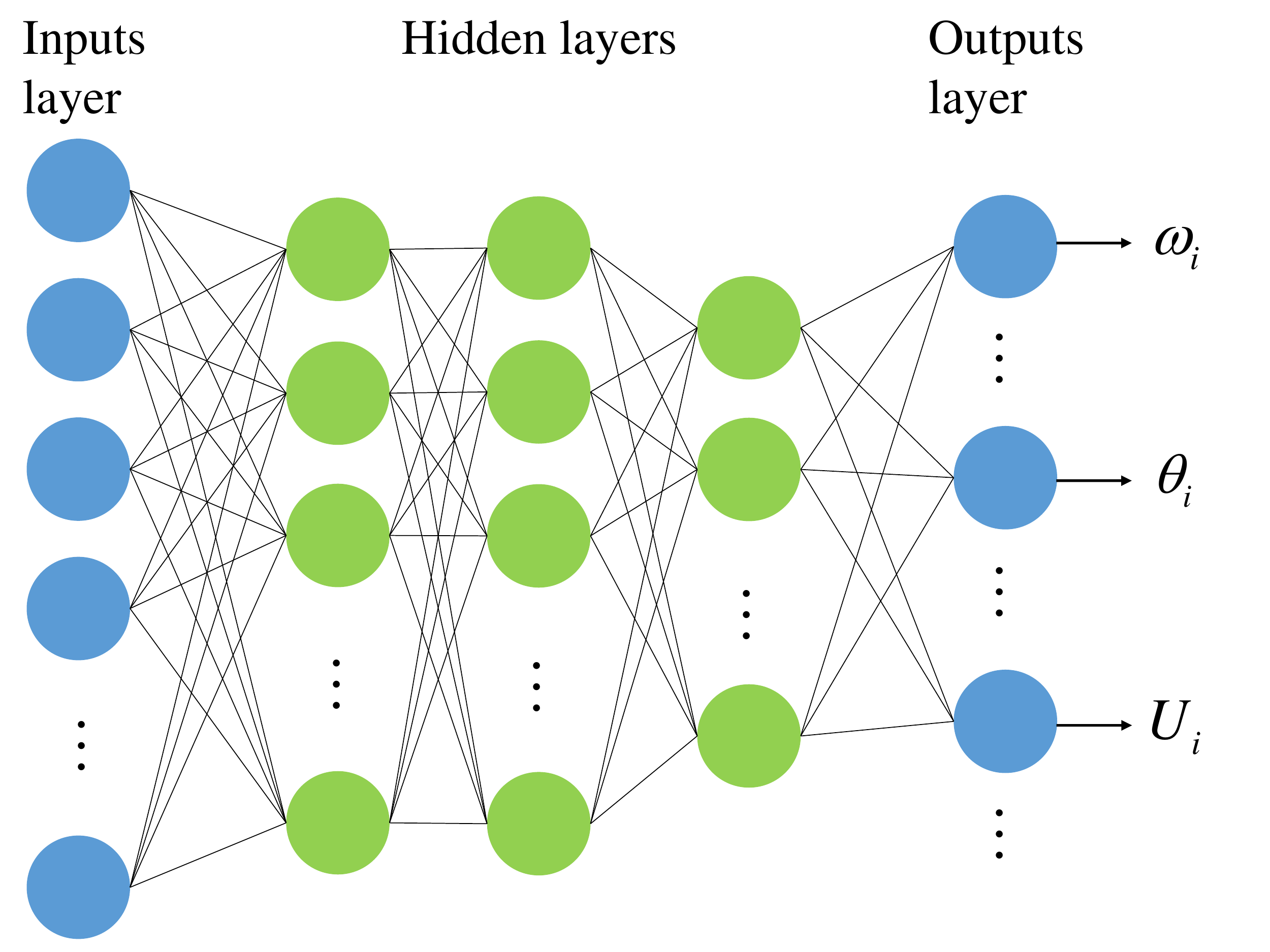}\quad
	\includegraphics[width=0.45\textwidth]{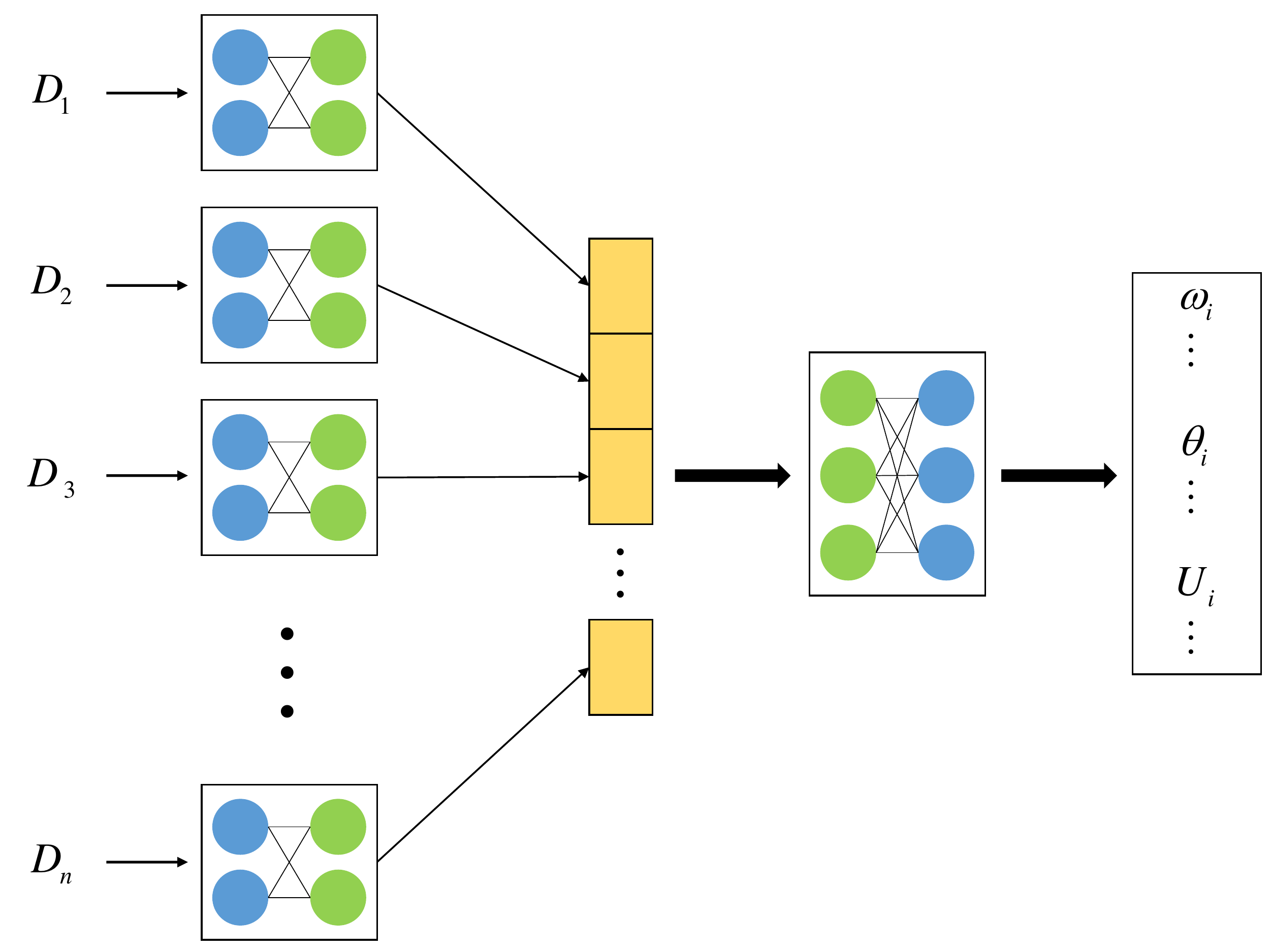}
	\caption{The network structure used for parameter estimation. The structure of the left panel is for one data set, while the multibranch network of the right panel is for multiple data sets $\{D_1, D_2, D_3, ..., D_n\}$ that are from different experiments.}\label{fig:mnn}
\end{figure*}

\subsection{Mixture Neural Network}\label{sec:mnn}

The basic principle of the MDN method is the assumption that the posterior distribution is a mixture of some unknown distributions, which then learns the mixture model using an ANN. Therefore, we should first get the parameters of the mixture model and then obtain the posterior distribution by generating samples based on the mixture model (Equations~(\ref{equ:pdf_of_gaussian_1}) and (\ref{equ:pdf_of_gaussian_multi})). However, for cosmological parameters that may deviate from Gaussian distribution, multiple components should be used to obtain the correct posterior distribution \citepalias{Wanggj:2022}. This will take more time to train the network and also increase the instability of the network, which makes it difficult to learn the parameters of the mixture model.

Inspired by \citetalias{Wanggj:2020}, we propose that the cosmological parameters can be obtained directly by the NN instead of sampling via a mixture model. In this case, we slightly modify the loss function in  Equations~(\ref{equ:loss_mdn_gaussian_1}) and (\ref{equ:loss_mdn_gaussian_multi}) to the following:
\begin{eqnarray}
\mathcal{L} &=& \mathbb{E}\left[ -\ln\left(\sum_{i=1}^K \omega_i\cdot\frac{1}{\sqrt{2\pi\sigma^2_i}}e^{-\frac{(\theta_i-\hat{\theta})^2}{2\sigma^2_i}}\right) \right] \label{equ:loss_mnn_1} \\
\nonumber\mathcal{L} &=& \mathbb{E}\left[ -\ln\left( \sum_{i=1}^K \omega_i \right.\right. \nonumber \\
& \times &\left.\left.\frac{\exp{\left( -\tfrac{1}{2} (\bm\theta_i - \hat{\bm\theta})^\top \bm\Sigma_i^{-1} (\bm\theta_i - \hat{\bm\theta}) \right)}}{\sqrt{\left( 2\pi \right)^N |\bm\Sigma_i|}} \right) \right] \label{equ:loss_mnn_multi}, 
\end{eqnarray}
where $\bm\theta$ (or $\theta$) are the estimated cosmological parameters, which should be considered as a point of the parameter space, $\hat{\bm\theta}$ (or $\hat{\theta}$) is the ground truth (i.e., the target) in the training set, $\omega$ is the mixture weight, and $\bm\Sigma$ (or $\sigma$) is the covariance matrix (or standard deviation) of the cosmological parameters. Similar to the ANN method, the $\bm\theta_i$ here should be interpreted as the point estimates of the cosmological parameters. This method has the same formula of posterior probability density as the MDN method (i.e., Equation (\ref{equ:pdf_of_mixture_model})). Therefore, the posterior distribution can be finally obtained by Equation (\ref{equ:pdf_of_mixture_model}). It looks like the posterior distribution is a mixture of the output of the ANN. Therefore, we call this method the MNN.

It should be noted that, although the loss function formula of MNN is similar to that of MDN, their main ideas are different. For the MDN method, we should assume a specific mixture model (e.g., Gaussian mixture model or Beta mixture model), and the posterior distribution should be obtained in two steps: (a) obtain parameters of the mixture model using an NN; (b) generate samples using the mixture model to obtain the posterior distribution. But for the MNN method, there is no explicit form of the mixture model while the cosmological parameters can be obtained directly from an NN. In this case, we can obtain the posterior distribution by feeding the network a series of data-like samples. Besides, the inputs (mainly the noise type and inference input) of MNN are different from that of the MDN method, which will be shown in Section \ref{sec:add_noise} and Table \ref{tab:compare_with_MDN_ANN}. Therefore, the interpretation of the prediction or how to obtain the posterior distribution differs for MNN and MDN. More details on obtaining the posterior distribution will be shown in Section~\ref{sec:training_and_parameter_estimation}, and a systematic comparison of the methods will be shown in Section~\ref{sec:comparing_with_MDN_and_ANN}. Notice that, if we consider an NN as an implicit mixture model, MNN can be considered as a special kind of MDN, with different training and prediction procedures.

Considering numerical stability, following \citetalias{Wanggj:2022}, we use the log-sum-exp trick and carry out the calculations in the logarithmic domain. Therefore, Equation~(\ref{equ:loss_mnn_1}) can be rewritten as
\begin{align}
\mathcal{L} &= \mathbb{E}\left[ \ln\left(\sum_{i=1}^K e^{\left[\ln(\omega_i) + \ln(p_i(\theta_i|\bm{d}))\right]}\right) \right] ~,
\end{align}
where
\begin{equation}
\ln\left[p_i(\theta_i|\bm{d})\right] = -\frac{(\theta_i-\hat{\theta})^2}{2\sigma_i^2} - \ln(\sigma_i) - \frac{\ln(2\pi)}{2}~.
\end{equation}
Equation~(\ref{equ:loss_mnn_multi}) can be rewritten as
\begin{align}
\mathcal{L} &= \mathbb{E}\left[ -\ln\left(\sum_{i=1}^K e^{\left[\ln(\omega_i) + \ln(p_i(\bm\theta_i|\bm{d}))\right]}\right) \right]~,
\end{align}
where
\begin{align}\label{equ:pdf_of_gaussian_multi_log}
\nonumber \ln[p_i(\bm\theta_i|\bm{d})] &= -\frac{1}{2} (\bm\theta_i - \hat{\bm\theta})^\top \bm\Sigma_i^{-1} (\bm\theta_i - \hat{\bm\theta}) \\
&+ \ln\left(|\bm\Sigma_i^{-1}|^{\frac{1}{2}}\right) - \ln\left(\sqrt{(2\pi)^N}\right)~.
\end{align}

\begin{figure*}
	\centering
	\includegraphics[width=0.45\textwidth]{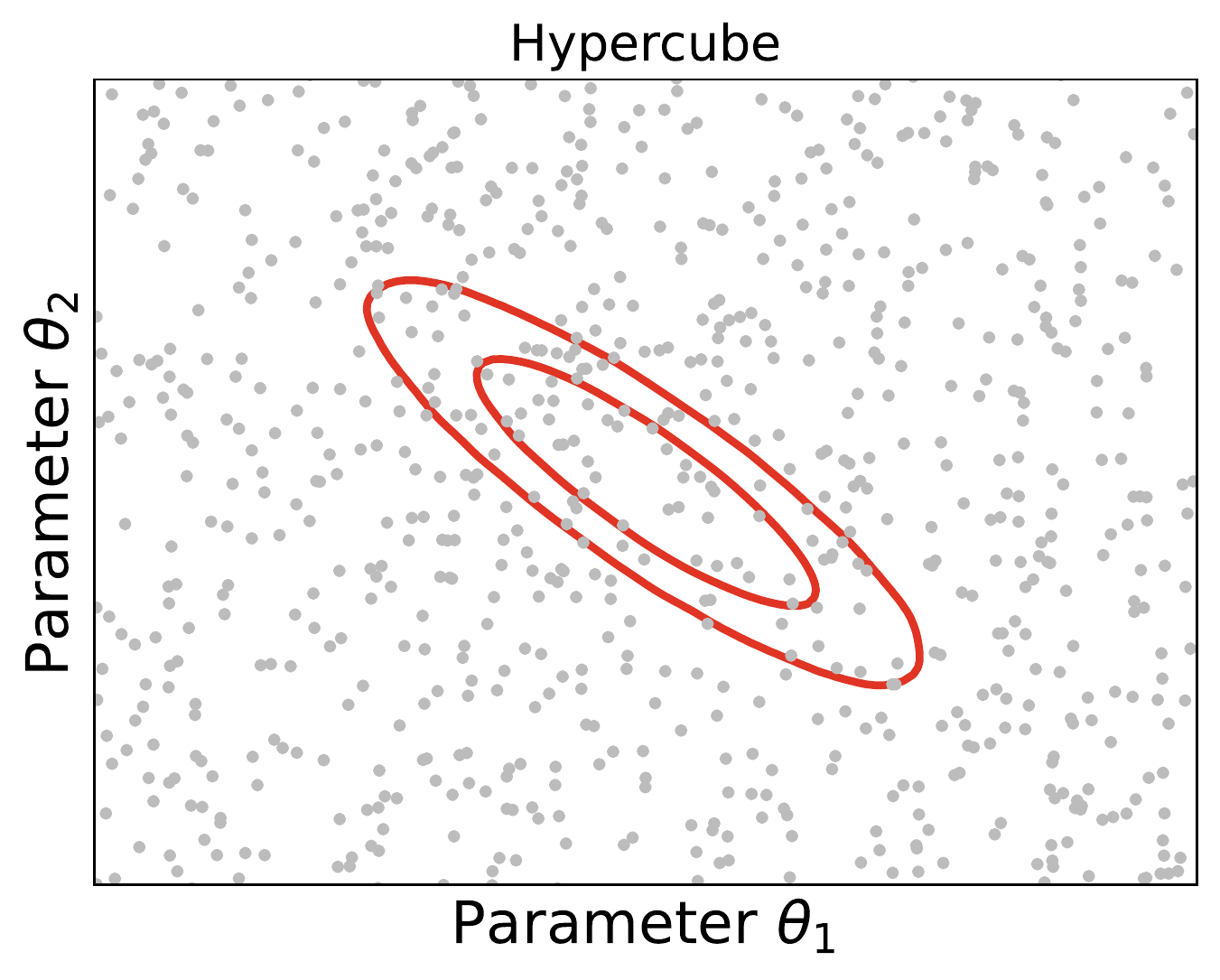}\quad
	\includegraphics[width=0.45\textwidth]{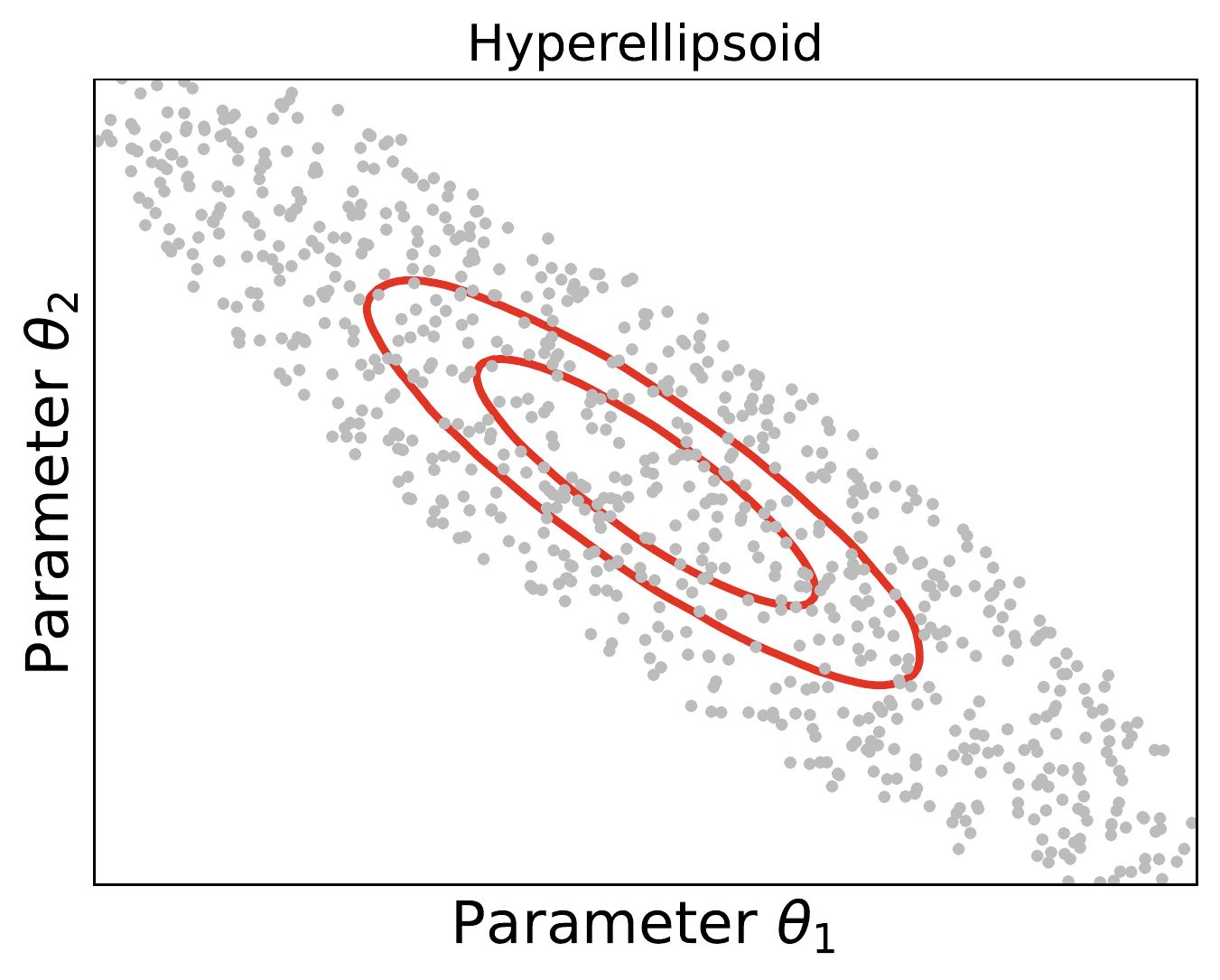}
	\caption{Two sampling methods of parameter space. The left panel stands for sampling in a hypercube uniformly, while the right panel refers to sampling in a hyper-ellipsoid uniformly by considering covariance between parameters. The red contours stand for the $1\sigma$ and $2\sigma$ contours of the posterior distribution, and the gray points refer to samples generated in the corresponding parameter space.}\label{fig:sampling_methods}
\end{figure*}

We note that the precision matrix $\bm\Sigma^{-1}_i$ in Equation~(\ref{equ:pdf_of_gaussian_multi_log}) can be characterized by its Cholesky factor $\bm{U}_i$ \citepalias{Wanggj:2022}:
\begin{align}\label{equ:cov_inv_cholesky}
\bm\Sigma^{-1}_i = \bm{U}^\top_i\bm{U}_i~,
\end{align}
where $\bm{U}_i$ is an upper triangular matrix with strictly positive diagonal entries. There are $N(N+1)/2$ nonzero entries for $\bm{U}_i$, which is much less than that of $\bm\Sigma^{-1}_i$ ($N^2$ entries). Thus, if the MNN learns $\bm\Sigma^{-1}_i$ instead of $\bm{U}_i$, there will be more neurons in the network, which will increase the training time. Besides, the output of the network may also make $\bm\Sigma^{-1}_i$ a non-positive-definite matrix. Therefore, to increase the stability of the network and to reduce the training time, the MNN is actually learning  $\bm{U}_i$. Since the output of the network can be either positive or negative, we use a Softplus function (\citealt{softplus}; see also \citetalias{Wanggj:2022}) to enforce the positiveness of the diagonal entries of $\bm{U}_i$ by taking
\begin{align}
(\bm{U}_i)_{jk} &=
\begin{cases}
\text{Softplus}(\widetilde{\bm{U}}_i)_{jk}~, & \text{if}~j=k \\
(\widetilde{\bm{U}}_i)_{jk}~, & \text{otherwise}
\end{cases}
\end{align}
where $\widetilde{\bm{U}}_i$ is output of the network.
\begin{equation}\label{equ:softplus}
\text{Softplus}(x) = \frac{1}{\beta}\ln(1 + e^{\beta x})~,
\end{equation}
where $\beta$ is a parameter, and we set it to unity throughout the paper. The Cholesky decomposition in Equation~(\ref{equ:cov_inv_cholesky}) offers an efficient way to calculate the matrix in Equation~(\ref{equ:pdf_of_gaussian_multi_log}). Putting Equation~(\ref{equ:cov_inv_cholesky}) into Equation~(\ref{equ:pdf_of_gaussian_multi_log}), we can obtain a new form:
\begin{align}
\nonumber \ln[p_i(\bm\theta_i|\bm{d})] &= -\frac{1}{2}\norm{\bm{U}_i (\bm\theta_i - \hat{\bm\theta})}^2_2 + \sum_{j=1}^N\ln\left(\text{diag}(\bm{U}_i)_j\right) \\
&\quad - \ln\left(\sqrt{(2\pi)^N}\right)~.
\end{align} 
The matrix $\bm{U}$ contains information of covariance between cosmological parameters. It will be optimized in the training process to ensure the estimated cosmological parameters have correct correlations. However, the matrix $\bm U$ is not used when estimating cosmological parameters via Equation (\ref{equ:pdf_of_mixture_model}) because the NN in MNN can output cosmological parameters directly, which is different from the MDN method (see Section \ref{sec:mdn}).

In Figure~\ref{fig:mnn}, we show the general structures of MNN used for parameter estimation. The left panel is for the case of one data set, while the multibranch network in the right panel is for the case of multiple data sets. The activation function used here is the Softplus function (Equation~(\ref{equ:softplus})). The number of neurons in each hidden layer is decreased proportionally based on the architecture proposed by \citetalias{Wanggj:2020}. Specifically, the number of neurons in the $i$th hidden layer is
\begin{equation}
N_i = \frac{N_{\text{in}}}{F^i}~,
\end{equation}
where $N_{\text{in}}$ is the number of neurons of the input layer. $F$ is a decreasing factor that is defined by
\begin{equation}
F = \left(\frac{N_{\text{in}}}{N_{\text{out}}} \right)^{\frac{1}{n+1}}~,
\end{equation}
where $n$ is the number of hidden layers.
\begin{equation}
N_{\text{out}} = K + \frac{KN(N+3)}{2}
\end{equation}
is the number of neurons in the output layer, where $K$ is the number of components (in Equations~(\ref{equ:loss_mnn_1}) and (\ref{equ:loss_mnn_multi})), and $N$ is the number of cosmological parameters. In our analysis, we consider a network with three hidden layers and will discuss the effect of the number of hidden layers in Section \ref{sec:effect_of_hiddenLayer}. Besides, the number of components $K$ is set to 1, and a discussion of the choice of $K$ will be shown in Section \ref{sec:effect_of_components}.

The MNN method has several advantages over the MDN method. One is that the MNN method is more stable than the MDN method because the covariance matrix will not be used when obtaining posterior distribution since the NN will output the cosmological parameters directly. For the the MDN method on the other hand, the learned upper Cholesky factor $\bm U$ may cause the covariance matrix $\bm\Sigma$ to be nonpositive definite, which makes the MDN method unstable (see Section \ref{sec:mdn}). Another advantage is that, for parameters with non-Gaussian distribution, MNN can obtain high-fidelity posterior distribution using only one component (see Section \ref{sec:effect_of_components}). For the MDN method, we need many more components to obtain a robust and reliable posterior distribution, which will take more time to train  the network and also increase the instability of the network \citepalias{Wanggj:2022}. In addition, the MNN method is more accurate than the MDN and ANN methods for parameters with truncated distributions due to the direct output of cosmological parameters in the loss function (see Section \ref{sec:parameters_with_physical_limits}). Therefore, the MNN method is more stable and much easier to train than the MDN method.

\subsection{Training Data}\label{sec:training_data}

Training data plays an important role in parameter estimation using NNs. In this section, we will illustrate the generation of the training set and its preprocessing.

\subsubsection{Training Set}\label{sec:training_set}

For the parameter estimation tasks, the network model should be trained using data generated by a model (e.g., a cosmological model) before conducting parameter estimation. Since the network model used here is to learn the conditional probability density $p(\bm\theta|\bm{d})$, the observational data $\bm{d}_0$ (or the posterior distribution) should be covered by the data space (or parameter space) of the training set. Learning from~\citetalias{Wanggj:2020}, we set the range of parameters in the training set to $[P-5\sigma_{p}, P+5\sigma_{p}]$, where $P$ is the best-fit value of the posterior distribution, and $\sigma_p$ is the corresponding $1\sigma$ error. Note that the best-fit value here refers to the mode of the posterior distribution. In this range, two sampling methods are considered to generate cosmological parameters: sampling uniformly in a hypercube (the left panel of Figure~\ref{fig:sampling_methods}) and sampling uniformly in a hyper-ellipsoid by considering the covariance between parameters (the right panel of Figure~\ref{fig:sampling_methods}).

For the sampling in a hypercube, we first generate samples using a uniform distribution in the $\pm 5\sigma_p$ range of the posterior distribution for each cosmological parameter and then combine them randomly. Note that, for the cosmological parameters with physical limits (e.g., the sum of the neutrino masses must be positive, $\sum m_\nu>0$), the $\pm 5\sigma_p$ range may cross the physical boundary. Therefore, for this case, the $\pm 5\sigma_p$ range will be cut according to the physical limit.

For the sampling in a hyper-ellipsoid, samples can be generated via \citep{Harman:2010,Gammell:2014}
\begin{equation}
\bm{X}_{\text{ellipsoid}} = 5\bm{L}\bm{X}_{\text{ball}} + \bm{P},
\end{equation}
where $\bm{P}$ is the best-fit value set from the posterior distribution, and the transformation $\bm{L}$ is given by the Cholesky decomposition of the covariance matrix of the posterior distribution:
\begin{equation}\label{equ:cholesk_posterior_cov}
\bm{L}\bm{L}^T\equiv\bm\Sigma.
\end{equation}
$\bm{X}_{\text{ball}}$ here is for samples distributed uniformly in a unit $N$-dimensional ball; which can be obtained via
\begin{equation}
\bm{X}_{\text{ball}} = \bm{X}_\text{U}^{1/N}\cdot\bm{Y},
\end{equation}
where $\bm{X}_\text{U}\sim U(0,1)$ is generated via uniform distribution in the interval (0,1), $N$ is the number of cosmological parameters, and
\begin{eqnarray}
\bm{Y} \equiv \frac{\bm{X}_N}{\norm{\bm{X}_N}_2},
\end{eqnarray}
where $\bm{X}_N\sim\mathcal{N}(\bm{0}_N,\bm{I}_N)$ is generated using the uncorrelated multivariate normal distribution.

The covariance matrix of the posterior distribution $\bm\Sigma$ in Equation~(\ref{equ:cholesk_posterior_cov}) is calculated using an ANN chain (see Section \ref{sec:training_and_parameter_estimation}) via
\begin{equation}
\bm\Sigma (\bm{X}, \bm{Y}) = \frac{1}{n}\sum_{i=1}^{n}(\bm{X}_i - \bar{\bm{X}}) (\bm{Y}_i - \bar{\bm{Y}}),
\end{equation}
where $n$ is the number of samples in the ANN chain, and $\bar{\bm{X}}$ and $\bar{\bm{Y}}$ are the means of the random variables $\bm{X}$ and $\bm{Y}$, respectively. We note that, for some cosmological parameters, the posterior distribution may deviate from the Gaussian distribution, in which case their mean may not be equal to their best-fit values. Therefore, we are using the best-fit values of $\bar{\bm{X}}$ and $\bar{\bm{Y}}$ instead of their mean values, and the diagonal entries of $\bm\Sigma$ are replaced by $\sigma^2_{\text{max}} = \max(\sigma^2_\text{left}, \sigma^2_\text{right})$, where $\sigma_\text{left}$ and $\sigma_\text{right}$ are the left-side and the right-side standard deviations, respectively.

For cosmological parameters that have physical limits, the samples for $\bm{X}_{\text{ellipsoid}}$ generated here may have nonphysical values. Therefore, for this case, all the nonphysical samples in $\bm{X}_{\text{ellipsoid}}$ will be removed. Obviously, the sampling in a hyper-ellipsoid is more efficient than the sampling in a hypercube, which makes it possible to train fewer samples. In our analysis, we mainly use the method of sampling in the hyper-ellipsoid to generate the training set, and discussions about the parameter sampling method will be shown in Section \ref{sec:effect_of_spaceSamplingMethod}.

\begin{figure*}
	\centering
	\includegraphics[width=0.7\textwidth]{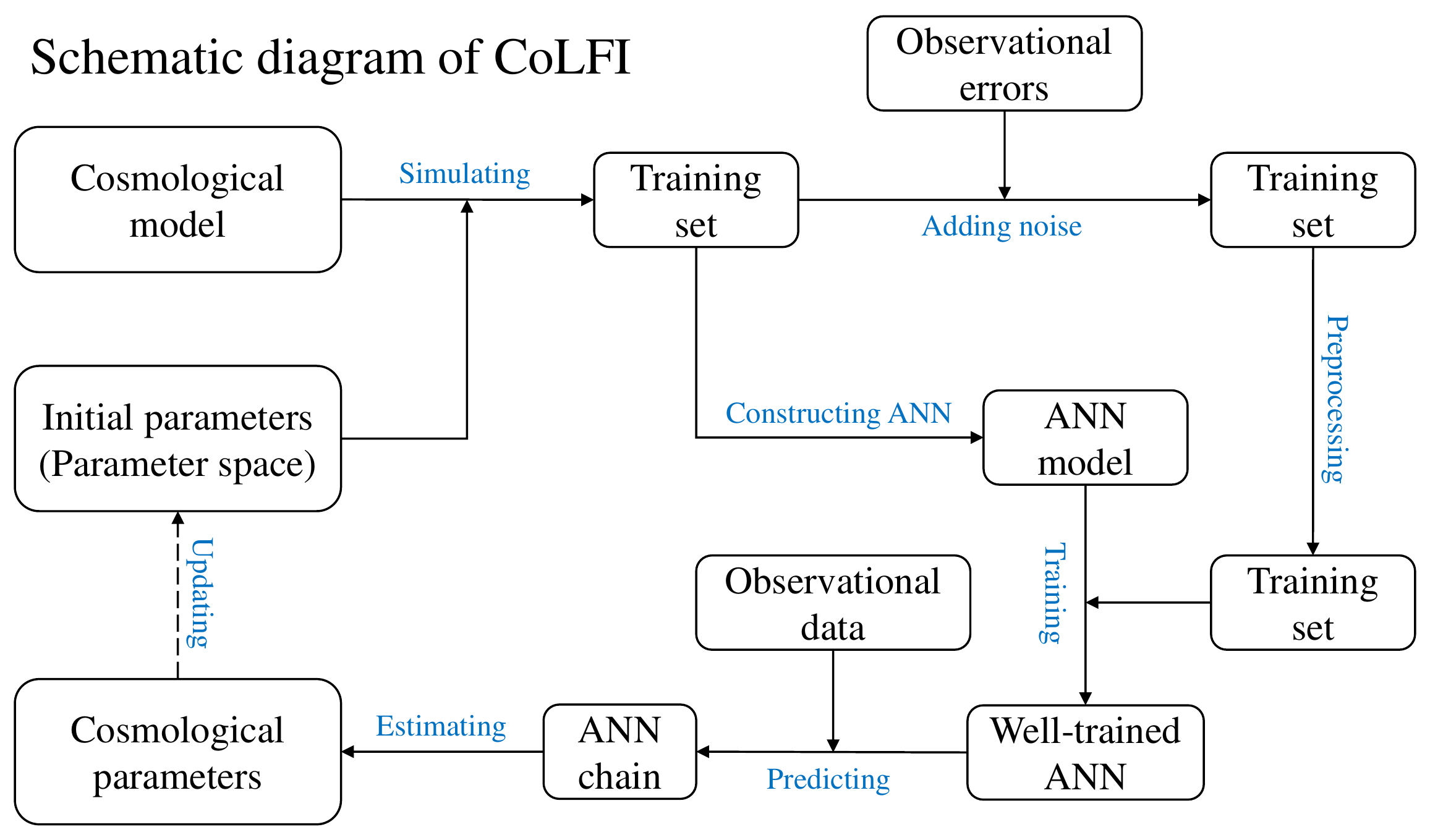}
	\caption{Schematic diagram of CoLFI. The initial parameters here can be set freely, and the parameter space to be learned will be updated after each estimation.}\label{fig:colfi_schematic}
\end{figure*}

\subsubsection{Add Noise}\label{sec:add_noise}

For supervised learning tasks, the training set should have the same distributions as the test set to get correct estimations. Therefore, the training set generated in Section \ref{sec:training_set} should have the same distributions as the observational data $\bm{d}_0$. Here, we assume that the $\bm{d}_0$ is subject to multivariate Gaussian distribution $\mathcal{N}(\bar{\bm{d}}_0, \bm\Sigma)$. Thus, the samples in the training set should be transformed to have the same distributions as the observational data by adding noise. Following \citetalias{Wanggj:2020}, we add Gaussian noise to the training set at each epoch of the training process. Specifically, the noise subject to $\mathcal{N}(0, A^2\bm\Sigma)$ (or $\mathcal{N}(0, A^2\sigma^2)$ for the measurements without covariance) are added to the training set, where $A$ is a coefficient subject to Gaussian distribution $\mathcal{N}(0, \sigma_A^2)$. The selection of $\sigma_A$ here should ensure that unity is covered by the range of $|A|$. In our analysis, $\sigma_A$ is set to 0.2 to ensure $|A|_\text{max}\sim 1$,\footnote{We do not recommend using very large values, especially for parameters with truncated distribution (see Figure~\ref{fig:effect_of_coefficient_planck}).} and discussions about the impact of $\sigma_A$ values will be shown in Section~\ref{sec:parameters_with_physical_limits}. Note that the covariance matrix here can be decomposed by Cholesky decomposition to speed up the generation of noise:
\begin{equation}\label{equ:cov_obs_cholesky}
A^2\bm\Sigma=(A\bm{L})(A\bm{L}^T),
\end{equation}
where $\bm{L}$ is the lower Cholesky factor, which has strictly positive diagonal entries. Since the samples generated by Gaussian distribution $\mathcal{N}(0, \sigma_A^2)$ may have zero values that make the diagonal of $A\bm{L}$ zero, we actually generate the coefficient $A$ by adding a small-positive number $\epsilon \sim \mathcal{O}(10^{-20})$. In order to improve the accuracy and performance of MNN, following \citetalias{Wanggj:2022}, we add multiple sets of noise to each sample, which means that multiple sets of noise will be generated and added to a sample to ensure that the MNN knows that the measurement may differ due to the presence of noise. Specifically, throughout the paper, we add five sets of noise to each sample in the training process.

\subsubsection{Data Preprocessing}\label{sec:data_preprocessing}

Data preprocessing is important for machine learning based on ANNs. Both the measurements (the input) and the cosmological parameters (the target) in the training set should be preprocessed before feeding to the network model. Specifically, for the cosmological parameters, we conduct two steps to normalize them \citepalias{Wanggj:2020,Wanggj:2022}: divide the cosmological parameters by their estimated values (here the mean of parameters in the training set is taken as the estimated value) to ensure that they all become the order of unity; then normalize them using the $z$-score normalization technique \citep{Kreyszig:2011}
\begin{eqnarray}
z = \frac{x-\mu}{\sigma}~,
\end{eqnarray}
where $\mu$ and $\sigma$ are the mean and standard deviation of the corresponding cosmological parameters. These steps make it possible for the network to deal with any kind of cosmological parameters with a different order of magnitude. Therefore, this makes the MNN a general method that can be applied to any cosmological parameters.

For the joint constraint on parameters using multiple data sets, the order of magnitude of measurements from different experiments may also be different. Therefore, the measurements should also be scaled by dividing the mean of the measurements in the training set to ensure that they all have the same order of magnitude. Besides, the measurements are also normalized by using the $z$-score normalization technique.

\subsection{Training and Parameter Estimation}\label{sec:training_and_parameter_estimation}

Here, we illustrate how to train an MNN model and conduct parameter estimation with the well-trained model. In Figure~\ref{fig:colfi_schematic}, we show the schematic diagram of CoLFI, which contains the main process of training and parameter estimation. First, we should build a class object for the cosmological model that contains the simulation method of the measurements, where the simulation method here is used to generate the training set. Then, the initial parameters (intervals of parameters) should be given. Note that the initial parameters here can be set freely, which means that the biased initial parameters that do not cover the best-fit value are acceptable. This is beneficial for parameters with poor knowledge. A discussion of this can be found in Section \ref{sec:update_parameter_space}. For some cosmological parameters, there are physical limits (e.g., the sum of the neutrino masses must be positive, $\sum m_\nu>0$, and the matter density parameter $\Omega_{\rm m}\in(0,1)$). Therefore, physical limits should be given for these parameters, which will be used in the simulation and estimation processes.

After passing the class object and the initial parameters to CoLFI, the training set will be simulated automatically using the method of Section \ref{sec:training_set}, where two sampling methods can be considered to generate cosmological parameters: sampling uniformly in a hypercube or a hyper-ellipsoid (see Figure \ref{fig:sampling_methods}). Generally speaking, the posterior distribution is unknown before the first estimation using the MNN model. Thus, the cosmological parameters here cannot be generated in a hyper-ellipsoid. Therefore, we are generating the cosmological parameters uniformly in a hypercube using the initial parameters. However, if prior knowledge of the distribution of the cosmological parameters is known (e.g., an ANN or MCMC chain), the chain will be taken as the initial parameter. In this case, the cosmological parameters will be generated uniformly in a hyper-ellipsoid. 

After that, an MNN model will be constructed automatically based on the training set. At the same time, the training  set will be preprocessed before training the network. Noise will then be automatically generated based on the observation errors and is added to the training set using the method of Section \ref{sec:add_noise}. Then, the training set  will be normalized using the method of Section \ref{sec:data_preprocessing}. Finally, the training set will be fed to the MNN model, and the model will be well-trained after thousands of epochs (e.g., 2000 epochs). 

\begin{figure}
	\centering
	\includegraphics[width=0.45\textwidth]{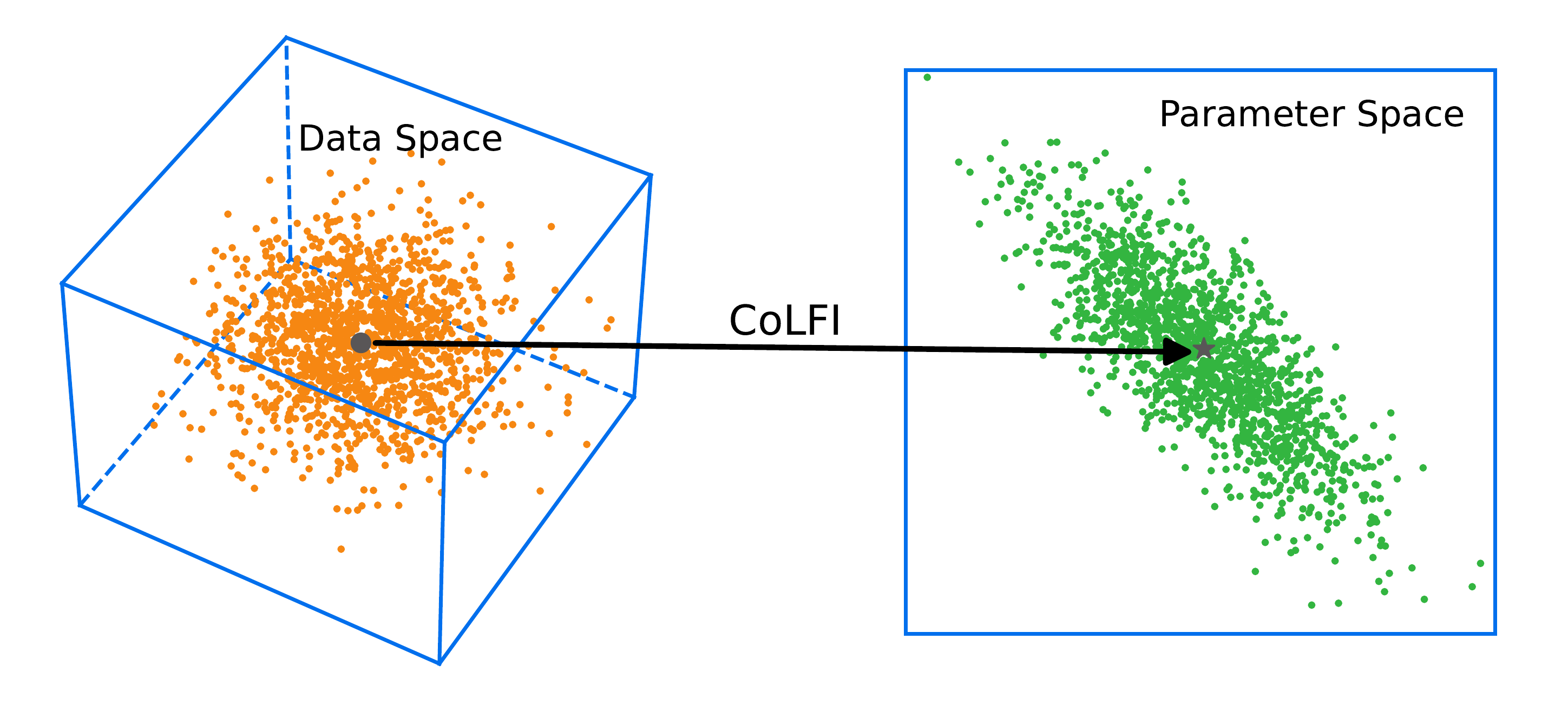}
	\caption{An example of CoLFI that learns a mapping between the data space of measurements and the cosmological parameter space.}\label{fig:colfi_mapping}
\end{figure}

We can then estimate cosmological parameters using the well-trained MNN model. The MNN model here actually learns a  mapping between the data space of the measurements and the parameter space of cosmological parameters (see Figure~\ref{fig:colfi_mapping}). Therefore, in order to obtain the posterior distribution, we should feed the distribution of the measurements to the MNN model. Specifically, we generate a large number of data-like samples (e.g., 10,000 samples) of the measurements using the observational data $\mathcal{N}(\bar{\bm{d}}_0, \bm\Sigma)$ and feed them to the MNN model to obtain an ANN chain. Note that Equation (\ref{equ:pdf_of_mixture_model}) was used when obtaining the ANN chain. Finally, the cosmological parameters can be estimated by using the ANN chain.

Note that the initial parameters passed to CoLFI are general ranges of parameters, which may not cover the true parameters. Therefore, the cosmological parameters obtained after the first estimation may be a biased estimation. Thus, the parameter space should be updated according to the estimation above, and the steps illustrated above should be repeated to ensure that the estimation tends to be stable. It should be noted that, when using the ANN chain to update the parameter space after the first estimation, the parameter space will be updated to $[P-5\sigma_{p}, P+5\sigma_{p}]$, and the cosmological parameters will be generated in a hyper-ellipsoid. A discussion about updating the parameter space will be shown in Section \ref{sec:update_parameter_space}. Finally, the ANN chains after burn-in can be used for parameter estimations. Note that cosmological parameters in the ANN chain may have nonphysical values, such as negative values of $\sum m_\nu$. Therefore, all the nonphysical values will be removed according to the boundaries that are physically sensible.

\begin{table}
	\centering
	\caption{Constraints on parameters of the $\Lambda$CDM model using the {\it Planck}-2015 CMB temperature power spectrum ($C^{\rm TT}_{\ell}$), quoted as the best-fit values with $1\sigma$ confidence level (C.L.).}\label{tab:params_planck_TT_6params}
	\begin{tabular}{c|c|c}
		\hline\hline
		& \multicolumn{2}{c}{Methods} \\
		\cline{2-3}
		Parameters & MCMC & MNN \\
		\hline
		$H_0$               & $67.987\pm1.233$    & $68.009\pm1.242$ \\
		$\Omega_{\rm b}h^2$ & $0.02238\pm0.00024$ & $0.02238\pm0.00024$ \\
		$\Omega_{\rm c}h^2$ & $0.11835\pm0.00276$ & $0.11831\pm0.00279$ \\
		$\tau$		        & $0.13641\pm0.03280$ & $0.13303\pm0.0323$ \\
		$10^9A_{\rm s}$     & $2.45765\pm0.15047$ & $2.43776\pm0.14814$ \\
		$n_{\rm s}$         & $0.96829\pm0.00696$ & $0.96880\pm0.00711$ \\
		\hline\hline
	\end{tabular}
\end{table}

\begin{figure*}
	\centering
	\includegraphics[width=0.9\textwidth]{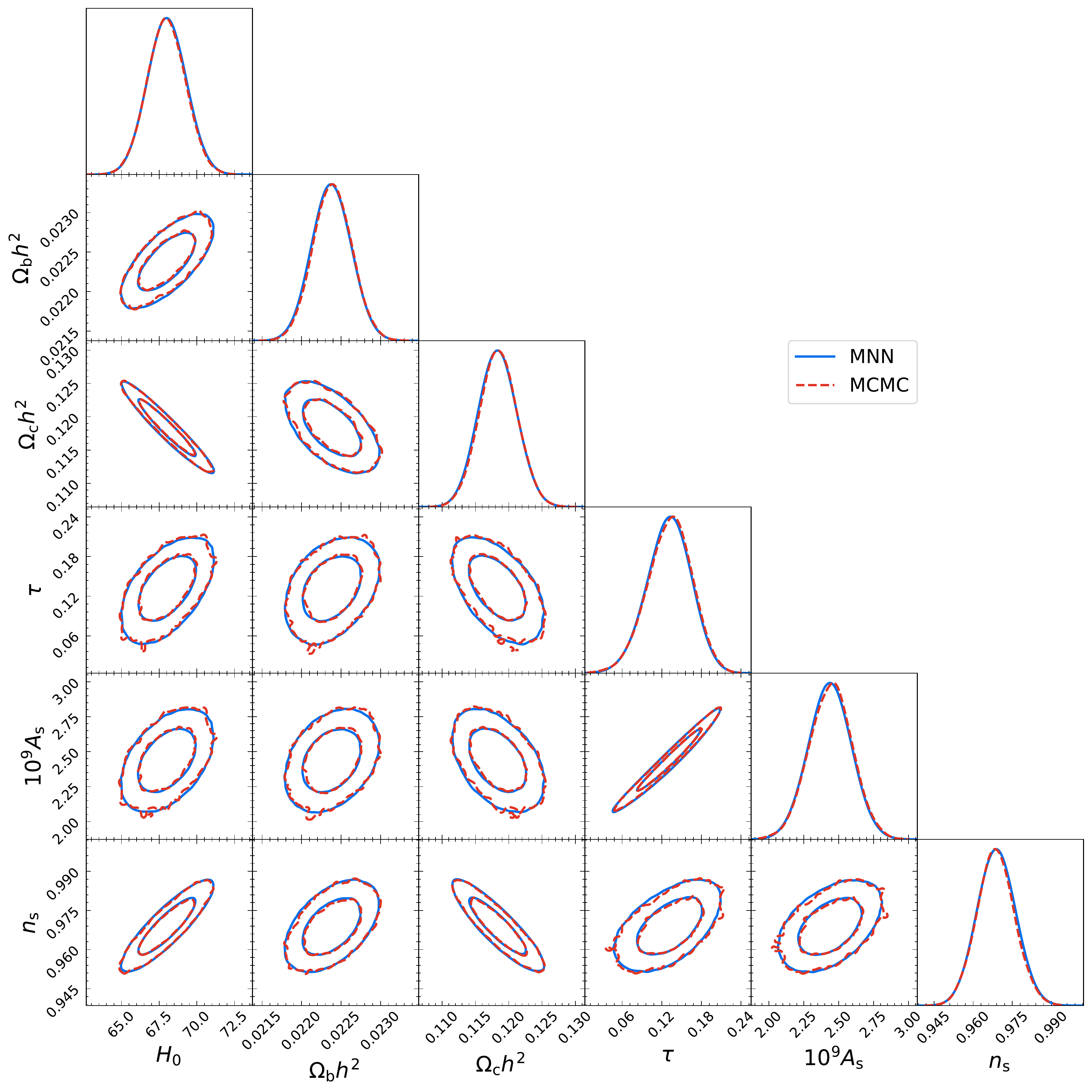}
	\caption{One-dimensional and two-dimensional marginalized distributions with 1$\sigma$ and 2$\sigma$ contours of $H_0$, $\Omega_{\rm b}h^2$, $\Omega_{\rm c}h^2$, $\tau$, $A_{\rm s}$, and $n_{\rm s}$ constrained from $C^{\rm TT}_{\ell}$ from {\it Planck}-2015.}\label{fig:contour_planck_TT_6params}
\end{figure*}

\begin{figure}
	\centering
	\includegraphics[width=0.45\textwidth]{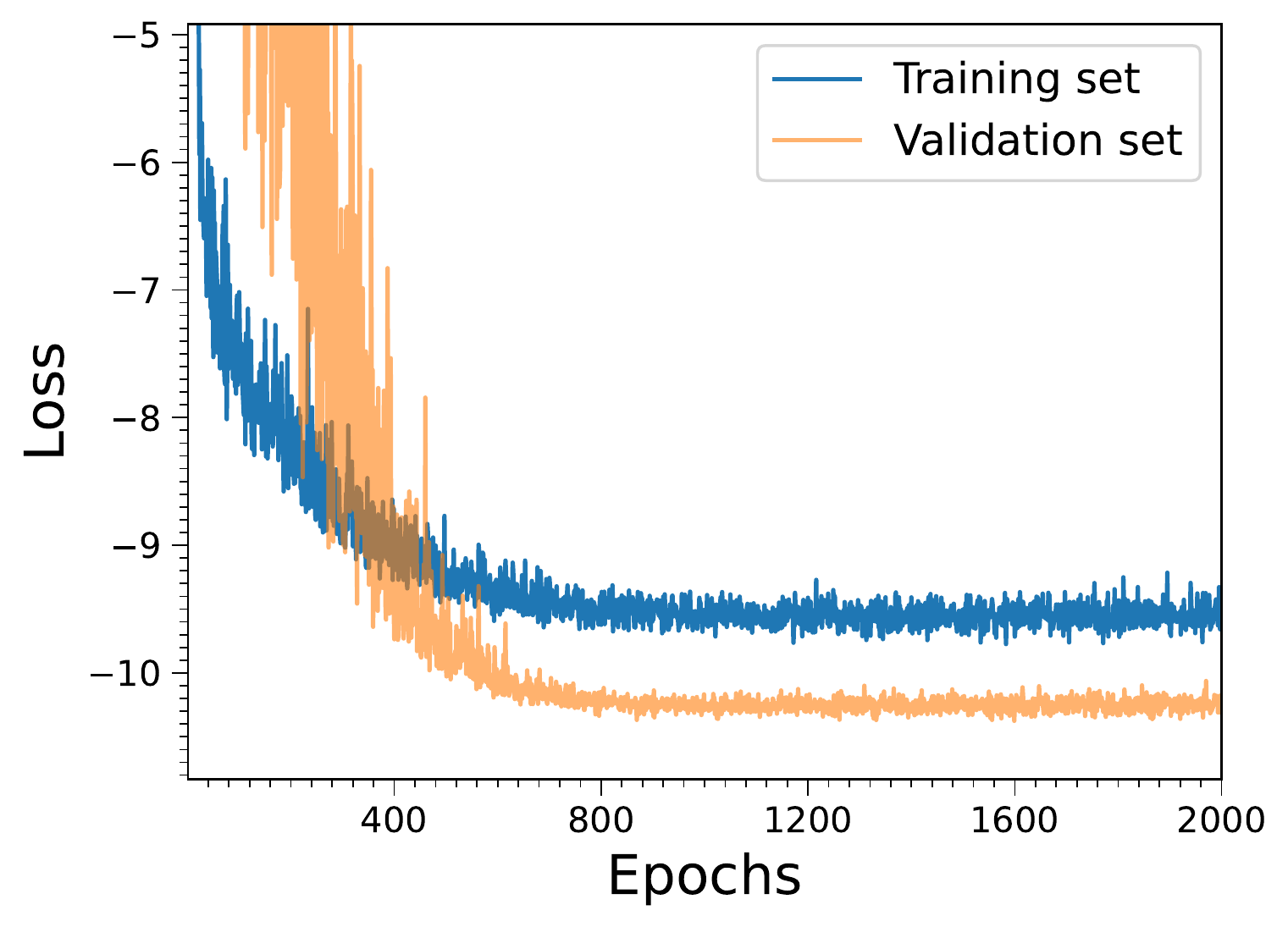}
	\caption{Losses of the training set and validation set. There are 3000 samples in the training set and 500 samples in the validation set.}\label{fig:loss_planck_tt}
\end{figure}

\section{Application to CMB}\label{sec:application_to_CMB}

In this section, we test the MNN method using the CMB observations. For simplicity, we only consider using the {\it Planck}-2015 $C^{\rm TT}_{\ell}$ {COM\_PowerSpect\_CMB\_R2.02.fits}\footnote{\url{http://pla.esac.esa.int/pla/\#cosmology}} to constrain parameters of the standard $\Lambda$CDM model: the Hubble constant $H_0$, the baryon density $\Omega_{\rm b}h^2$, the CDM density $\Omega_{\rm c}h^2$, the optical depth $\tau$, and the amplitude and spectral index of primordial curvature perturbations ($A_{\rm s}$, $n_{\rm s}$). The multipole of the power spectrum used here is in the range $\ell\in[30,2000]$, and the covariance between each multipole is not considered. First, we estimate these parameters using the MCMC method. Here, the public package {\tt emcee} \citep{Foreman-Mackey:2013}, a Python module that achieves the MCMC method, is used for parameter estimation. We generate an MCMC chain with 100,000 steps after burn-in. Then, the best-fit values with 1$\sigma$ errors of the parameters are calculated from the MCMC chain by using {\it coplot}\footnote{\url{https://github.com/Guo-Jian-Wang/coplot}}, as shown in Table \ref{tab:params_planck_TT_6params}. The corresponding one-dimensional and two-dimensional contours are shown in Figure~\ref{fig:contour_planck_TT_6params}, with the red dashed lines.

Then, we constrain these parameters using the MNN method, and the settings illustrated in Sections \ref{sec:mnn}, \ref{sec:training_data}, and \ref{sec:training_and_parameter_estimation} are used. There are 3000 samples of power spectra in the training set and 500 samples in the validation set. Three hidden layers are considered in the MNN, the Softplus (Equation (\ref{equ:softplus})) is taken as the activation function, and the number of epochs is 2000. Using the method illustrated in Section \ref{sec:training_and_parameter_estimation}, we obtain three ANN chains of the cosmological parameters after burn-in; each ANN chain contains 10,000 samples. Then, we calculate the best-fit values and the $1\sigma$ errors using these three ANN chains, as shown in Table \ref{tab:params_planck_TT_6params}. The corresponding one-dimensional and two-dimensional contours are shown in Figure~\ref{fig:contour_planck_TT_6params}, with the blue solid lines. We can see that the results of the MNN method are almost the same as those of the MCMC method. The deviation between the MNN results and the MCMC results can then be calculated using
\begin{equation}
\Delta P = \frac{|P_{\rm MCMC} - P_{\rm MNN}|}{\sigma} ,
\end{equation}
where $\sigma=\sqrt{\sigma^2_{\rm MCMC} + \sigma^2_{\rm MNN}}$, $P_{\rm MCMC}$ and $P_{\rm MNN}$ are the best-fit parameters of the MCMC and MNN methods, respectively. Specifically, the deviations for the parameters are $0.013\sigma$, $0.020\sigma$, $0.010\sigma$, $0.073\sigma$, $0.094\sigma$, and $0.051\sigma$, respectively. Obviously, these deviations are quite small.

We note that the initial parameters are a dominant factor in the burn-in phase, which determines how many times the network needs to adjust the parameter space before finding the true posterior. Besides, the settings of hyperparameters in MNN (e.g., the number of hidden layers, the number of training samples, the number of epochs, and the activation function) also have a slight influence on the burn-in phase. We do not show these details here, and we recommend readers to look at Figure \ref{fig:biased_initial_H0} in Section \ref{sec:update_parameter_space} to see how the network gradually finds the true posterior.

The hyperparameters of the MNN should be set manually before estimating cosmological parameters, and the network will not optimize them in the training process. Therefore, the hyperparameters should be set reasonably to ensure the rationality of the posterior, which can be checked from the loss function. Figure \ref{fig:loss_planck_tt} shows an example of training and validation losses taken from an MNN model after burn-in. It shows that there is no overfitting for the MNN. Obviously, the loss function is reasonable for the training and validation sets. The reason why the loss of the validation set is smaller than that of the training set is that multiple levels of noise and multiple sets of noise are added to each sample (see Section \ref{sec:add_noise}).

\section{Application to SN Ia}\label{sec:application_to_SN}

For the analysis in Section \ref{sec:application_to_CMB}, we did not consider the covariance matrix of the power spectrum. In this section, we test the capability of the MNN method in dealing with observational data with covariance by constraining $w$ and $\Omega_{\rm m}$ of the $w$CDM model using the Pantheon SN Ia data \citep{Scolnic:2018}. The Pantheon SN Ia data used here contains 1048 data points in the redshift range [0.01, 2.26]. The distance modulus of the Pantheon SN Ia is
\begin{equation}
\mu=m_{B,{\rm corr}}^* - M_{B} ~,
\end{equation}
where $m_{B,{\rm corr}}^* = m_{B}^*+\alpha\times x_1-\beta\times c + \Delta_B$ is the corrected apparent magnitude reported in \citet{Scolnic:2018}, and $M_B$ is the $B$-band absolute magnitude. $\alpha$ is the coefficient of the relation between luminosity and stretch, $\beta$ is the coefﬁcient of the relation between luminosity and color, and $\Delta_B$ is a distance correction based on predicted biases from simulations. In our analysis, the systematic uncertainties are considered; therefore, the systematic covariance matrix $\bm{C}_{\rm sys}$ \citep{Scolnic:2018} is used to add noise to the training set.

Since the measurements of the Pantheon SN Ia data are the corrected apparent magnitudes, the input of the network is $m_{B,{\rm corr}}^*$ for the observational SN Ia data, and $\mu + M_B$ for the simulated data generated by the $w$CDM model. Because of the strong correlation between the absolute magnitude $M_B$ and the Hubble constant $H_0$, following \citetalias{Wanggj:2022}, we combine $M_B$ and $H_0$ to be a new parameter and constrain it with the cosmological parameters simultaneously. Then, we have
\begin{align}
\mu + M_B &= 5\log_{10} \widetilde{D}_{\rm L}(z) + \mu_c~,
\end{align}
where $\mu_c \equiv 5\log_{10}\left(c/H_0/ {\rm Mpc}\right) + M_B + 25$ is taken as a new nuisance parameter to be estimated with the cosmological parameters together, and
\begin{eqnarray}
\widetilde{D}_{\rm L}(z) & \equiv & \frac{D_{\rm L}(z)H_{0}}{c}  \nonumber \\
&=&  (1+z)\int_0^z\frac{{\rm d}z^\prime}{E(z^\prime)}~,
\end{eqnarray}
is a dimensionless luminosity distance. The $E(z)$ function is 
\begin{equation}
E(z) = \sqrt{\Omega_{\rm m}(1+z)^3 + (1-\Omega_{\rm m})(1+z)^{3(1+w)}}~, \label{eq:Ez}
\end{equation}
where $\Omega_{\rm m}$ is the matter density parameter, and $w$ is the equation of state of dark energy. In Equation~(\ref{eq:Ez}), we assume spatial flatness. For more details, we refer interested readers to \citetalias{Wanggj:2022}.

\begin{table}
	\centering
	\caption{Constraints on parameters of the $w$CDM using the Pantheon SN Ia data, quoted by its best-fit values and $1\sigma$ C.L.}\label{tab:params_pantheon}
	\begin{tabular}{c|c|c}
		\hline\hline
		& \multicolumn{2}{c}{Methods} \\
		\cline{2-3}
		Parameters & MCMC & MNN \\
		\hline
		$w$              & $-1.011\pm0.216$ & $-1.009\pm0.213$ \\
		$\Omega_{\rm m}$ & $0.327\pm0.074$  & $0.325\pm0.074$ \\
		$\mu_c$          & $23.807\pm0.015$ & $23.808\pm0.015$ \\
		\hline\hline
	\end{tabular}
\end{table}
\begin{figure}
	\centering
	\includegraphics[width=0.45\textwidth]{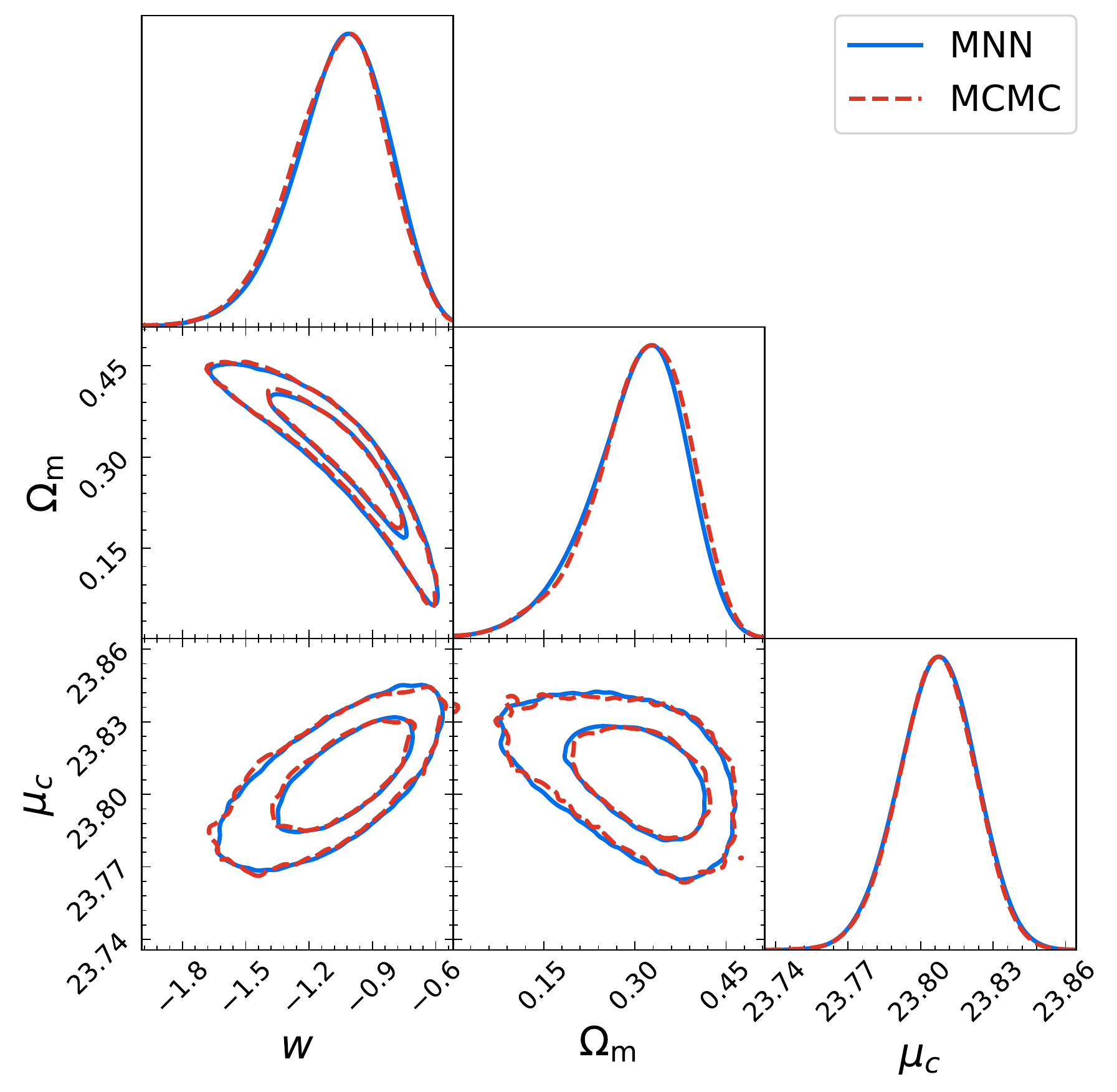}
	\caption{One-dimensional and two-dimensional marginalized distributions with 1$\sigma$ and 2$\sigma$ contours of $w$, $\Omega_{\rm m}$, and $\mu_c$ constrained from Pantheon SN Ia.}\label{fig:contour_pantheon}
\end{figure}

First, we constrain $w$, $\Omega_{\rm m}$, and $\mu_c$ using the MCMC method. We generate an MCMC chain with 100,000 steps after burn-in. Then, we calculate the best-fit values and the $1\sigma$ errors, as shown in Table \ref{tab:params_pantheon}. The corresponding one-dimensional and two-dimensional contours of these parameters are shown in Figure~\ref{fig:contour_pantheon}, with the red dashed line. With the same procedure and MNN settings as in Section \ref{sec:application_to_CMB}, we constrain these parameters using the MNN method. After burn-in, an ANN chain with 10,000 samples is obtained. Then, we calculate the best-fit values with $1\sigma$ errors, as shown in Table \ref{tab:params_pantheon}. The deviations of the MNN results and the MCMC results are $0.005\sigma$, $0.013\sigma$, and $0.044\sigma$, respectively. Obviously, these deviations are quite small. Furthermore, we plot the one-dimensional and two-dimensional contours in Figure~\ref{fig:contour_pantheon}, with the blue solid lines. We can see that the results of the constraints are very close to each other.

One can see that the MNN result is much better than that with one Gaussian component illustrated in \citetalias{Wanggj:2022}. The reason is that the ANN has strong nonlinearity, which enables it to learn complex distributions, even non-Gaussian distributions like $w$ and $\Omega_{\rm m}$. On the other hand, for the mixture model, it is impossible to fit $w$ and $\Omega_{\rm m}$ with only one Gaussian component. Besides, we can see that the MNN result is also much better than that illustrated in \citetalias{Wanggj:2020}. The reason is that $\mathcal{N}(0, \bm\Sigma)$ is used by \citetalias{Wanggj:2020} to generate noise instead of $\mathcal{N}(0, A^2\bm\Sigma)$, which makes it difficult to learn a good mapping between measurements and cosmological parameters.

\begin{table}
	\centering
	\caption{Same as Table~\ref{tab:params_planck_TT_6params} but including {\it Planck}-2015 CMB polarization power spectra ($C^{\rm TE}_{\ell}$ and $C^{\rm EE}_{\ell}$) and Pantheon SN Ia data.}\label{tab:params_planck_pantheon}
	\begin{tabular}{c|c|c}
		\hline\hline
		& \multicolumn{2}{c}{Methods} \\
		\cline{2-3}
		Parameters & MCMC & MNN \\
		\hline
		$H_0$               & $67.701\pm0.633$    & $67.690\pm0.614$    \\
		$\Omega_{\rm b}h^2$ & $0.02231\pm0.00015$ & $0.02229\pm0.00015$ \\
		$\Omega_{\rm c}h^2$ & $0.11866\pm0.00141$ & $0.11867\pm0.00138$ \\
		$\tau$		        & $0.06589\pm0.01346$ & $0.06587\pm0.01286$ \\
		$10^9A_{\rm s}$     & $2.13366\pm0.05636$ & $2.13862\pm0.05384$ \\
		$n_{\rm s}$         & $0.96817\pm0.00398$ & $0.96776\pm0.00391$ \\
		$M_B$		        & $-19.418\pm0.017$   & $-19.418\pm0.017$   \\
		\hline\hline
	\end{tabular}
\end{table}
\begin{figure*}
	\centering
	\includegraphics[width=0.9\textwidth]{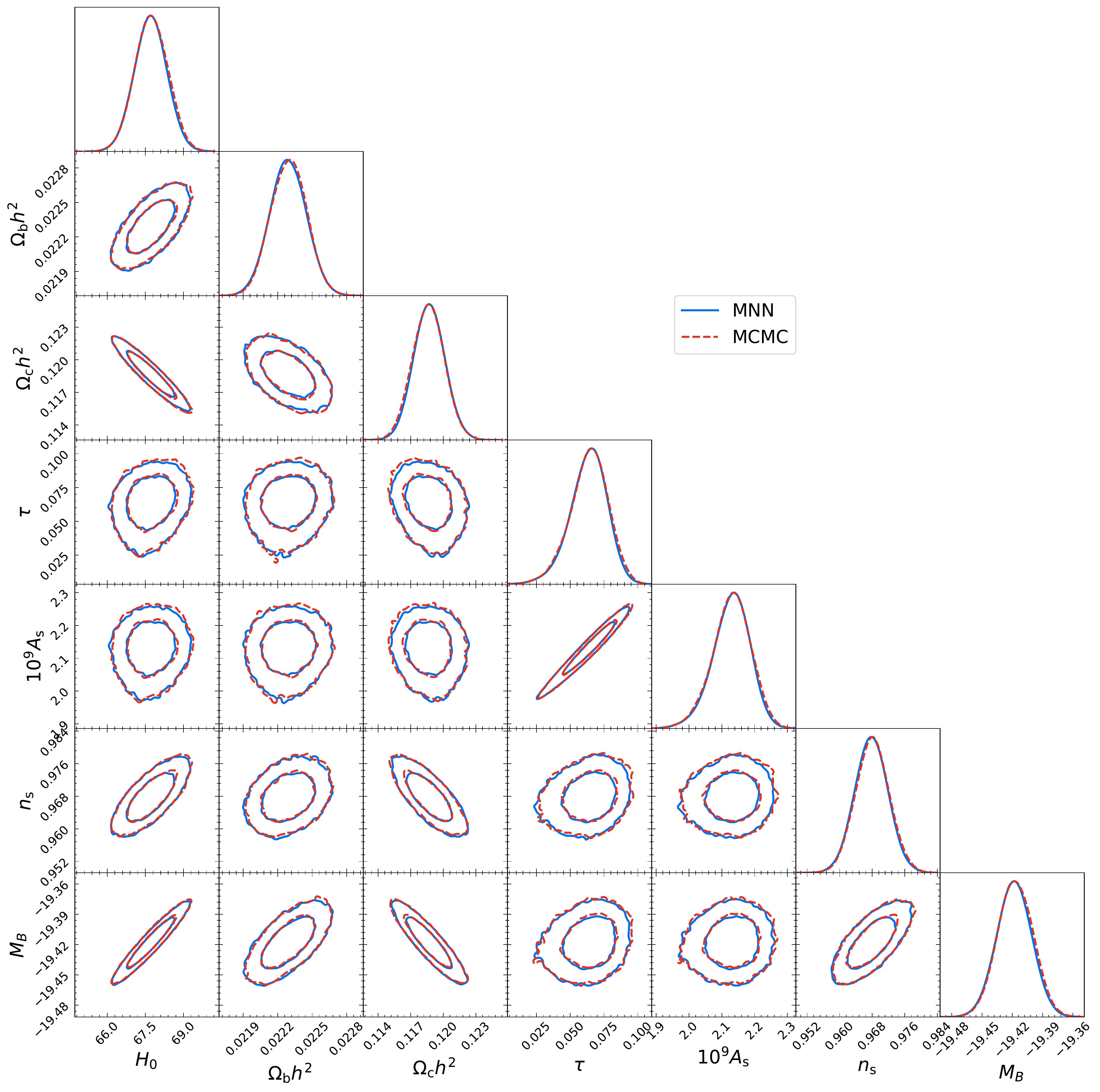}
	\caption{One-dimensional and two-dimensional marginalized distributions with 1$\sigma$ and 2$\sigma$ contours of $H_0$, $\Omega_{\rm b}h^2$, $\Omega_{\rm c}h^2$, $\tau$, $A_{\rm s}$, $n_{\rm s}$, and $M_B$ constrained from {\it Planck}-2015 CMB spectra ($C^{\rm TT}_{\ell}$, $C^{\rm TE}_{\ell}$ and $C^{\rm EE}_{\ell}$) and Pantheon SN Ia.}\label{fig:contour_planck_pantheon}
\end{figure*}

\begin{figure*}
	\centering
	\includegraphics[width=0.303\textwidth]{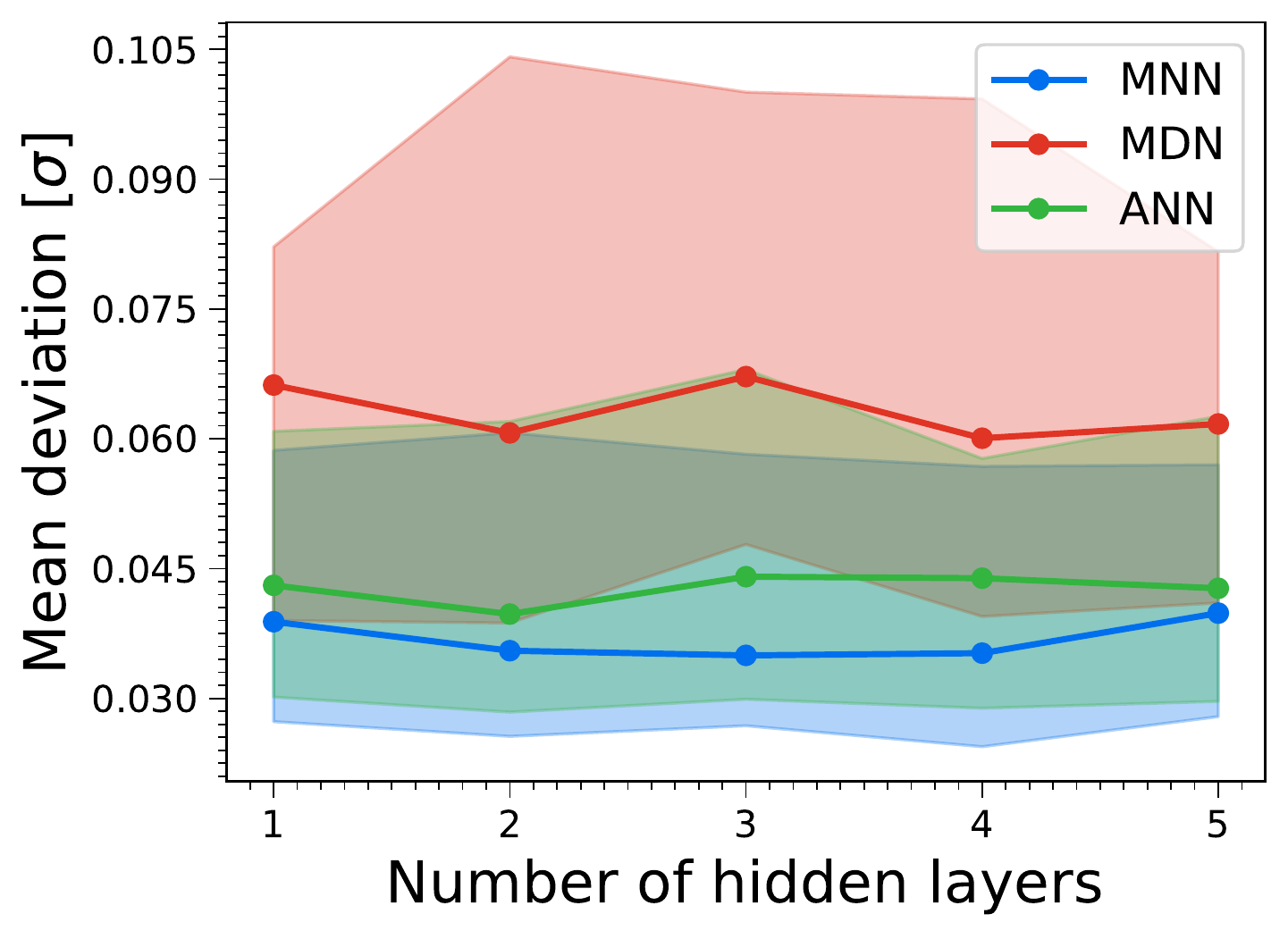}
	\includegraphics[width=0.3\textwidth]{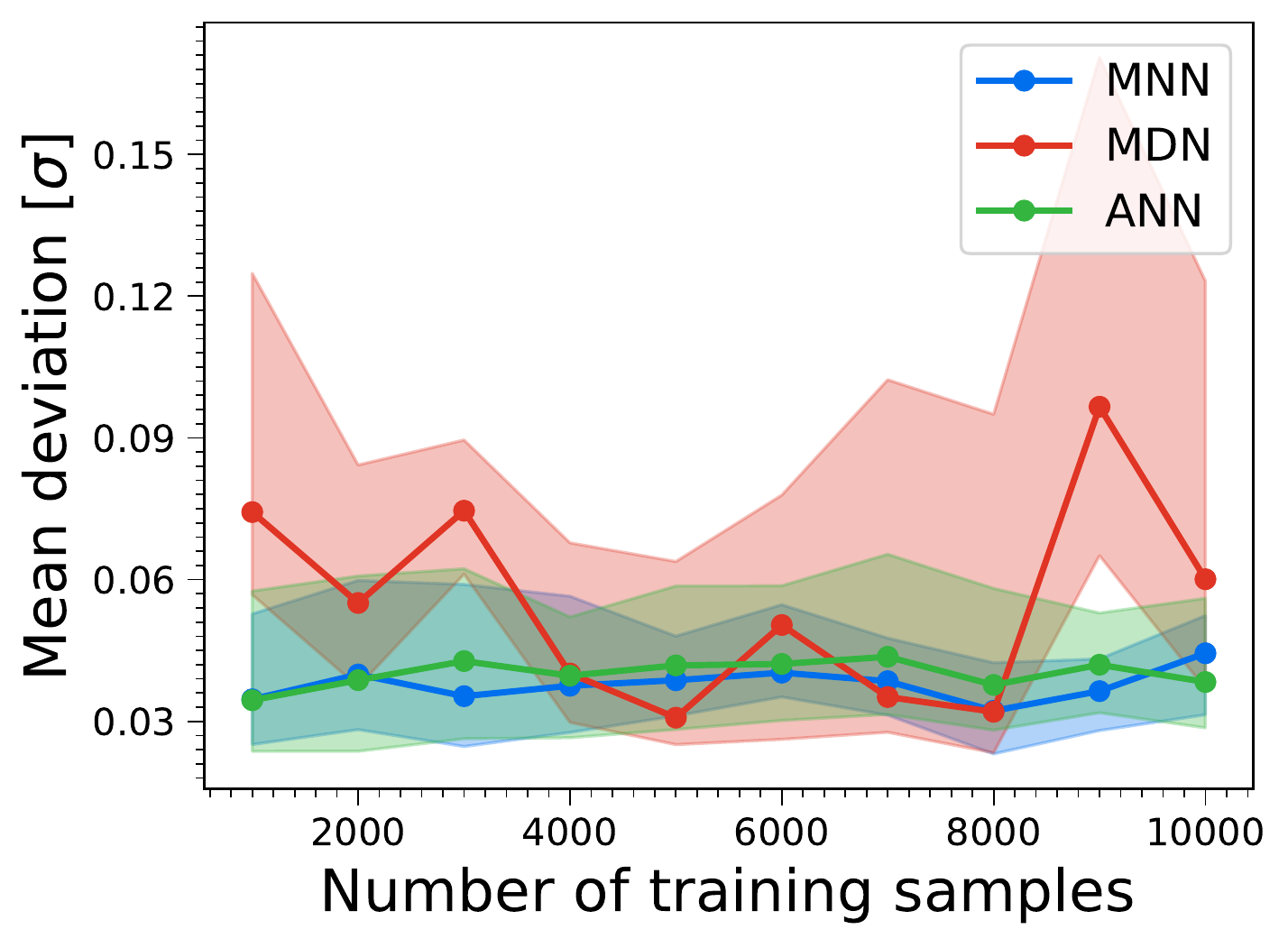}
	\includegraphics[width=0.3\textwidth]{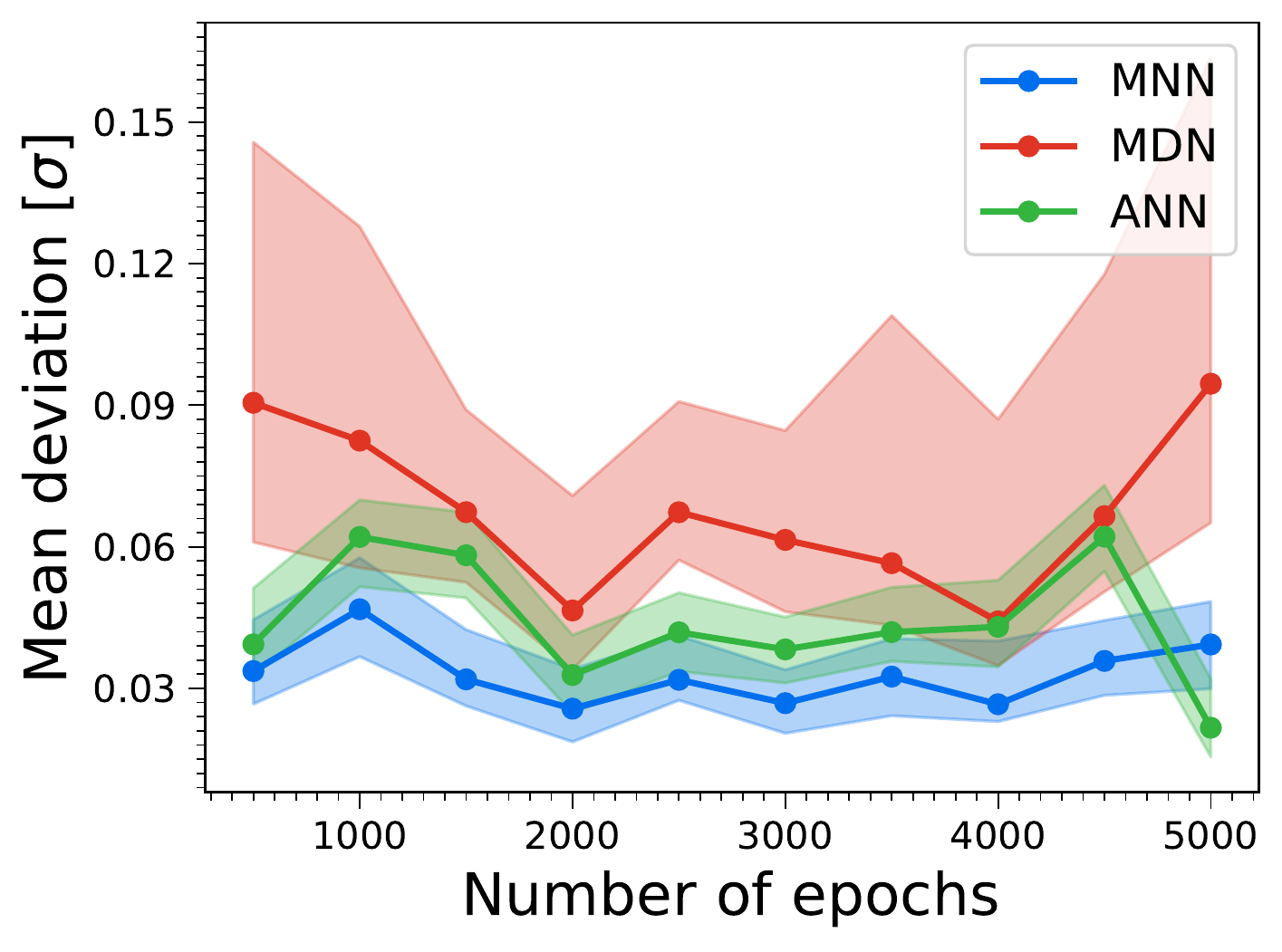}
	\includegraphics[width=0.91\textwidth]{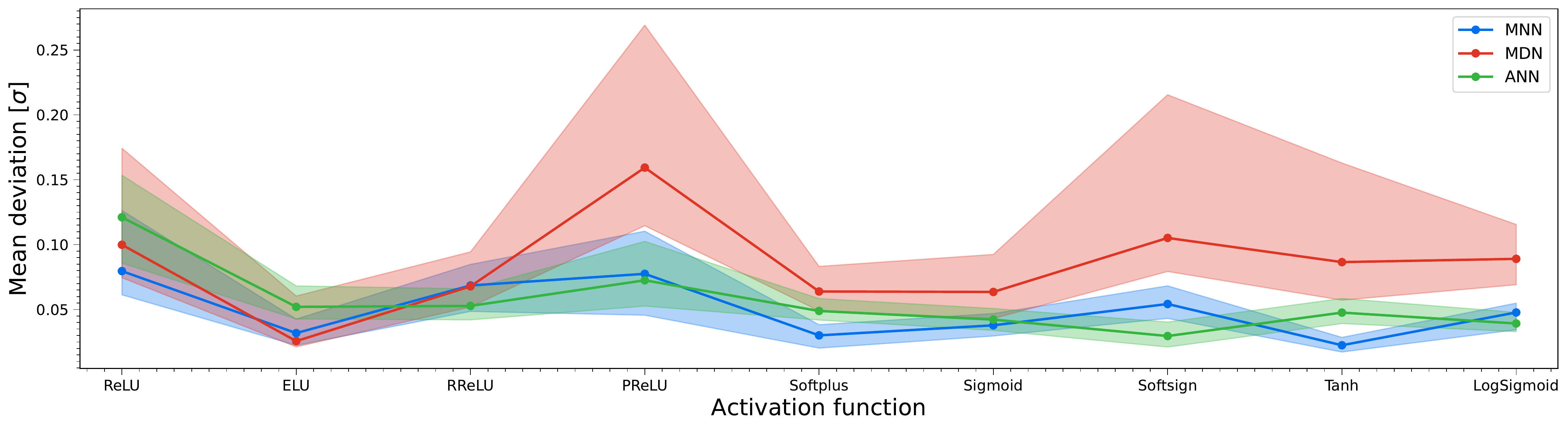}
	\caption{Mean deviations between the [MNN, MDN, and ANN] results and the fiducial values as a function of the number of hidden layers, the number of training samples, the number of epochs, and the activation function. The areas here indicate the $1\sigma$ error range.}\label{fig:effect_of_hyperparameters}
\end{figure*}

\section{Joint Constraint on Parameters}\label{sec:joint_constraint}

To test the capability of the multibranch MNN, we constrain the cosmological parameters of the $\Lambda$CDM model using the {\it Planck}-2015 temperature angular power spectrum (TT) + temperature‐polarization cross‐power spectrum (TE) + polarization E‐mode angular power spectrum (EE) {COM\_PowerSpect\_CMB\_R2.02.fits} and Pantheon SN Ia data. The multipole range is $\ell\in[2, 2500]$ for $C^{\rm TT}_{\ell}$, and  $\ell\in[2, 1996]$ for $C^{\rm TE}_{\ell}$ and $C^{\rm EE}_{\ell}$. Also, the systematic covariance matrix $\bm{C}_{\rm sys}$ of the Pantheon SN Ia data is used in our analysis. The parameters to be estimated are the six cosmological parameters ($H_0$, $\Omega_{\rm b}h^2$, $\Omega_{\rm c}h^2$, $\tau$, $A_{\rm s}$, and $n_{\rm s}$) and the $B$-band absolute magnitude $M_B$ (nuisance parameter in SN Ia data). Because there are four sets of observational data ($C^{\rm TT}_{\ell}$, $C^{\rm TE}_{\ell}$, $C^{\rm EE}_{\ell}$, and the corrected apparent magnitudes $m_{B,{\rm corr}}^*$ of Pantheon SN Ia), there are four branches in the multibranch MNN model. CMB $C_{\ell}$s are calculated with the public code {\sc camb}\footnote{\url{https://github.com/cmbant/CAMB}}~\citep{camb}.

With the same procedure, we first constrain these parameters using the MCMC method by generating an MCMC chain with 100,000 steps after burn-in. Then, the best-fit values with $1\sigma$ errors of the parameters are calculated using the MCMC chain, as shown in Table \ref{tab:params_planck_pantheon}. We plot the corresponding one-dimensional and two-dimensional marginalized distributions with red dashed lines in Figure~\ref{fig:contour_planck_pantheon}. 

Then, following the same procedure as Section \ref{sec:application_to_CMB}, we constrain the parameters using the MNN method. There are 5000 samples in the train set and 500 samples in the validation set. There are three hidden layers in the multibranch network, one in the branch part, and the other two in the rest of the network. The activation function used here is Softplus (Equation (\ref{equ:softplus})), and the network is trained after 2000 epochs. After the training process, we obtain three ANN chains after burn-in to estimate the parameters, where each ANN chain contains 10,000 samples. The best-fit values and $1\sigma$ errors calculated using the ANN chain are shown in Table \ref{tab:params_planck_pantheon}, and the corresponding one-dimensional and two-dimensional contours are shown in Figure~\ref{fig:contour_planck_pantheon}. We can see that both the best-fit values and the $1\sigma$ errors are almost the same as those of the MCMC method. The deviations of the MNN results and the MCMC results are $0.012\sigma$, $0.090\sigma$, $0.009\sigma$, $0.002\sigma$, $0.064\sigma$, $0.074\sigma$, and $0.003\sigma$, respectively, which are small in general. Therefore, the multibranch MNN model is capable of combining several sets of observational data to constrain multiple cosmological parameters.

\section{Effect of Hyperparameters}\label{sec:effect_of_hyperparameters}

There are many hyperparameters that can be selected manually in CoLFI, which may have an effect on the parameter estimation. Specifically, below, we discuss the effect of the number of hidden layers in Section \ref{sec:effect_of_hiddenLayer}, the number of training samples in Section \ref{sec:effect_of_trainingSample}, the number of epochs in Section \ref{sec:effect_of_epoch}, activation function in Section \ref{sec:effect_of_activationFunction}, the number of components in Section \ref{sec:effect_of_components}, and the parameter space sampling methods in Section \ref{sec:effect_of_spaceSamplingMethod}.

To test the impact of hyperparameters in Sections \ref{sec:effect_of_hiddenLayer}, \ref{sec:effect_of_trainingSample}, \ref{sec:effect_of_epoch}, and \ref{sec:effect_of_activationFunction}, we simulate CMB $C^{\rm TT}_{\ell}$ based on the Polarized Radiation Imaging and Spectroscopy Mission (PRISM; \citet{Andre:2014}), by using the Parameter Forecast for Future CMB Experiments code \citep{Perotto:2006}. The fiducial cosmological parameters of the $\Lambda$CDM model are set as follows:
\begin{align}\label{equ:fiducial}
\nonumber H_0&=67.31 \rm ~km ~s^{-1} ~Mpc^{-1}, & \Omega_{\rm b}h^2 &= 0.02222,\\
\Omega_{\rm c}h^2 &= 0.1197, & \tau &= 0.078,\\
\nonumber A_{\rm s} &= 2.19551\times 10^{-9}, & n_{\rm s} &= 0.9655.
\end{align}
CMB $C^{\rm TT}_{\ell}$ is simulated based on the experimental specifications of PRISM (Table \ref{tab:prism_specifications}) with the frequency lying within $90-220$\,GHz. In order to reduce training time, we set the range of parameters of the training data to $[P_{\rm fid}-5\sigma_p, P_{\rm fid}+5\sigma_p]$, where $P_{\rm fid}$ are the fiducial cosmological parameters in Equation (\ref{equ:fiducial}), and $\sigma_p$ is the $1\sigma$ error of the posterior distribution, which is estimated via the MCMC method.

\begin{table}
	\centering
	\caption{Experimental Specifications of the PRISM CMB Experiment: Frequency channels, Beam width, and Temperature Sensitivities for Each Channel. The sky fraction $f_{\rm sky}=0.8$ is for all frequency channels.}\label{tab:prism_specifications}
	\begin{tabular}{c|c|c}
		\hline\hline
		Channel & FWHM & $\triangle T$ \\
		(GHz) & (arcmin) & ($\mu$K $\cdot$ arcmin) \\
		\hline
		90  & 5.7 & 3.30 \\
		105 & 4.8 & 2.88 \\
		135 & 3.8 & 2.59 \\
		160 & 3.2 & 2.43 \\
		185 & 2.8 & 2.52 \\
		200 & 2.5 & 2.59 \\
		220 & 2.3 & 2.72 \\
		\hline\hline
	\end{tabular}
\end{table}

\subsection{The Number of Hidden Layers}\label{sec:effect_of_hiddenLayer}

To test the impact of the number of hidden layers, we consider five different MNN structures with the number of hidden layers varying from one to five. The Softplus (Equation (\ref{equ:softplus})) is taken as the activation function, and the number of samples in the training set is 3000. After training the MNN models with 2000 epochs, we obtain the corresponding ANN chains. Then, we calculate the mean deviation between the MNN results and the fiducial parameters:
\begin{equation}
{\rm Mean\ deviation}=\frac{1}{N}\left( \sum_{i=1}^N \frac{|\theta_{i, \rm pred}-\theta_{i, \rm fid}|}{\sigma_{i, \rm pred}} \right) ,
\end{equation}
where $N$ is the number of cosmological parameters, $\theta_{i, \rm fid}$ is the fiducial parameter, and $\theta_{i, \rm pred}$ and $\sigma_{i, \rm pred}$ are the predicted best-fit value and error of cosmological parameter, respectively. To test the effect of the initialization of the network on the final results, we train 100 MNNs with different initialization for each case. The mean deviations of the five cases are $0.043_{-0.013}^{+0.018}\sigma$, $0.040_{-0.011}^{+0.022}\sigma$, $0.044_{-0.014}^{+0.024}\sigma$, $0.044_{-0.015}^{+0.014}\sigma$, and $0.043_{-0.013}^{+0.020}\sigma$, respectively. We show them with the blue solid line with areas in the upper left panel of Figure~\ref{fig:effect_of_hyperparameters}. Obviously, all of the deviations are small and acceptable, but it should be noted that for structures with fewer hidden layers (e.g. with one hidden layer) it may be more difficult to learn good mappings for complex cosmological models, whereas, for structures with more hidden layers (e.g. with five hidden layers), it takes longer to train a network model. Therefore, a suitable structure should be chosen such as a structure with three hidden layers.

For comparison, we also estimate parameters using the MDN and the ANN methods illustrated in \citetalias{Wanggj:2020} and \citetalias{Wanggj:2022}, respectively. The activation function, the number of training samples, and the number of epochs are the same as those of the MNN method. Three Gaussian components are used for the MDN method. The mean deviations of the results are shown in the upper left panel of Figure~\ref{fig:effect_of_hyperparameters} with red and green solid lines, respectively. Obviously, the mean deviations of the five cases are consistent within $1\sigma$ error for the ANN and MNN methods. Both the ANN and MNN methods perform slightly better than the MDN method on both the best-fit values and $1\sigma$ errors.

\subsection{The Number of Training Samples}\label{sec:effect_of_trainingSample}

The MNN learns a mapping between the measurements and the cosmological parameters, in which case enough samples for the selected parameter space should be provided to ensure a good mapping. Therefore, the number of training samples also affects the parameter estimation. Here, we test the effect of the number of training samples on the parameter estimations by training MNNs with the number of samples varying from 1000 to 10,000. The MNN models contain three hidden layers. The Softplus (Equation (\ref{equ:softplus})) is taken as the activation function, and the MNNs are trained with 2000 epochs. With the same procedure, we train 100 MNNs with different initialization for each case. In the upper middle panel of Figure~\ref{fig:effect_of_hyperparameters}, we show, with the blue solid line with areas, the mean deviation of parameters between the MNN result and the fiducial parameters. Most of the deviations are smaller than $0.060\sigma$, which is quite small. At the same time, we estimate parameters using the MDN and the ANN methods. The corresponding deviations for these two methods are indicated by red and green solid lines, respectively. We can see that the deviations of the ANN method are similar to those of the MNN method, while the deviations of the MDN method are mostly larger than those of the MNN method for the cases of the number of training samples smaller than 4000 and larger than 8000.

\subsection{The Number of Epochs}\label{sec:effect_of_epoch}

The parameters $w$ and $b$ in Equation~(\ref{equ:neuron_function}) are optimized by minimizing the loss function $\mathcal{L}$ (Equations~(\ref{equ:loss_mnn_1}) and (\ref{equ:loss_mnn_multi})). Thus, the MNN model should be trained with a large number of epochs. To test the effect of the number of epochs on the parameter estimations, we train MNNs with epochs varying from 500 to 5000. There are three hidden layers in the MNN models, the Softplus (Equation (\ref{equ:softplus})) is taken as the activation function, and the number of samples in the training set is 3000. We train  100 MNNs with different initialization for each case. The mean deviations of parameters between the MNN result and the fiducial parameters are shown in the upper right panel of Figure~\ref{fig:effect_of_hyperparameters}, with the blue solid line. Most of the deviations are smaller than $0.060\sigma$. This means that the MNN can be well-trained with thousands (or even hundreds) of epochs. With the same procedure, we estimate parameters using the MDN and ANN methods, and the corresponding deviations are indicated by red and green solid lines, respectively. The results of the ANN method are similar to those of the MNN method, while the deviations of the MDN method are mostly larger than those of the MNN and ANN methods, especially for cases with small and large epochs.

\begin{table}
	\centering
	\caption{Activation functions used to test the performance of the MNN, MDN, and ANN methods.}\label{tab:activation_functions}
	\begin{tabular}{c|c|l}
		\hline\hline
		Name & Plot & Function, $f(x)$ \\ \hline
		
		ReLU 
		&\begin{minipage}{0.1\textwidth}
			\includegraphics[width=\textwidth]{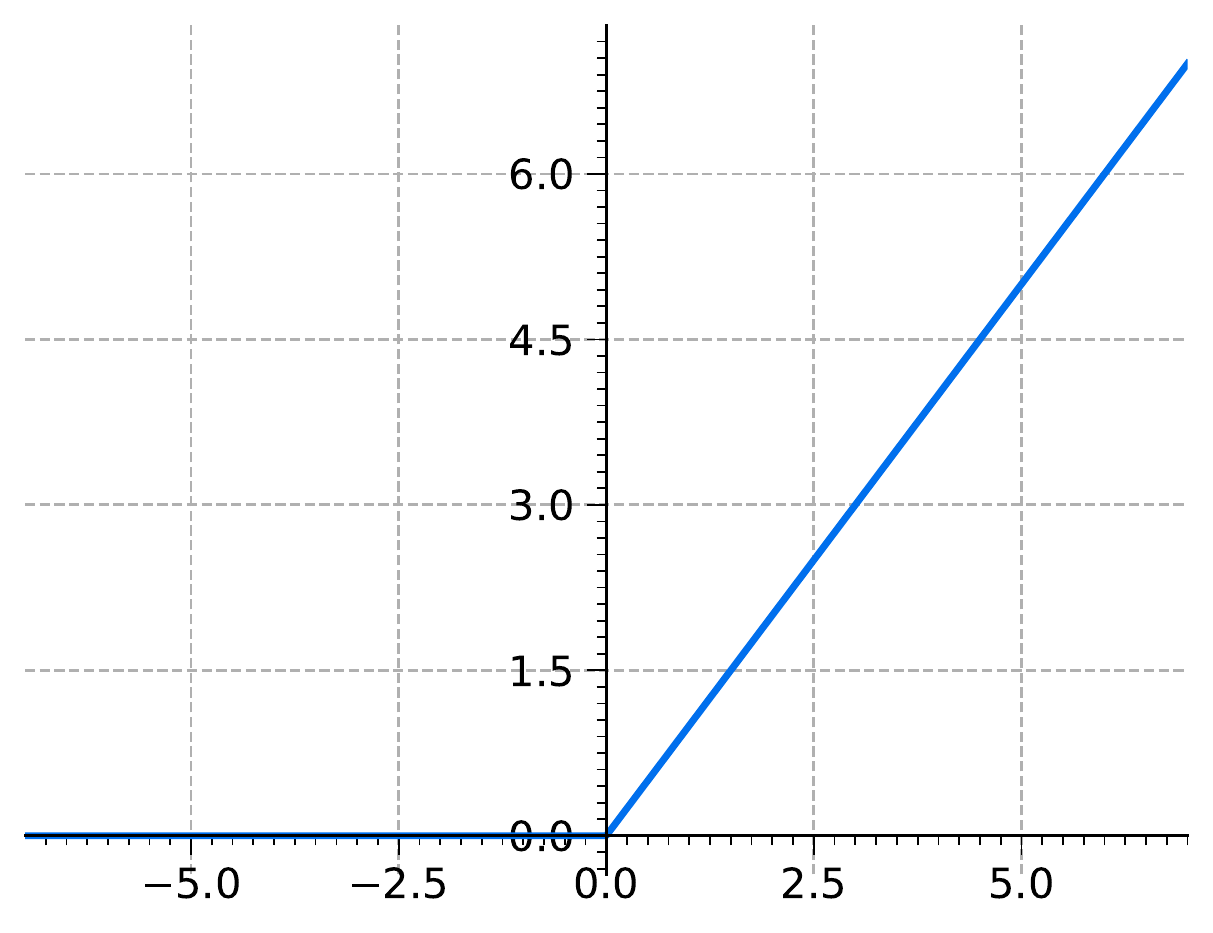}
		\end{minipage}
		&\begin{minipage}{0.1\textwidth}
			\begin{equation*}
			\left\{\begin{matrix}
			x & \text{if } x \geq 0 \\
			0 & \text{if } x < 0 
			\end{matrix}\right.
			\end{equation*}
		\end{minipage} \\ \hline
		
		RReLU 
		&\begin{minipage}{0.1\textwidth}
			\includegraphics[width=\textwidth]{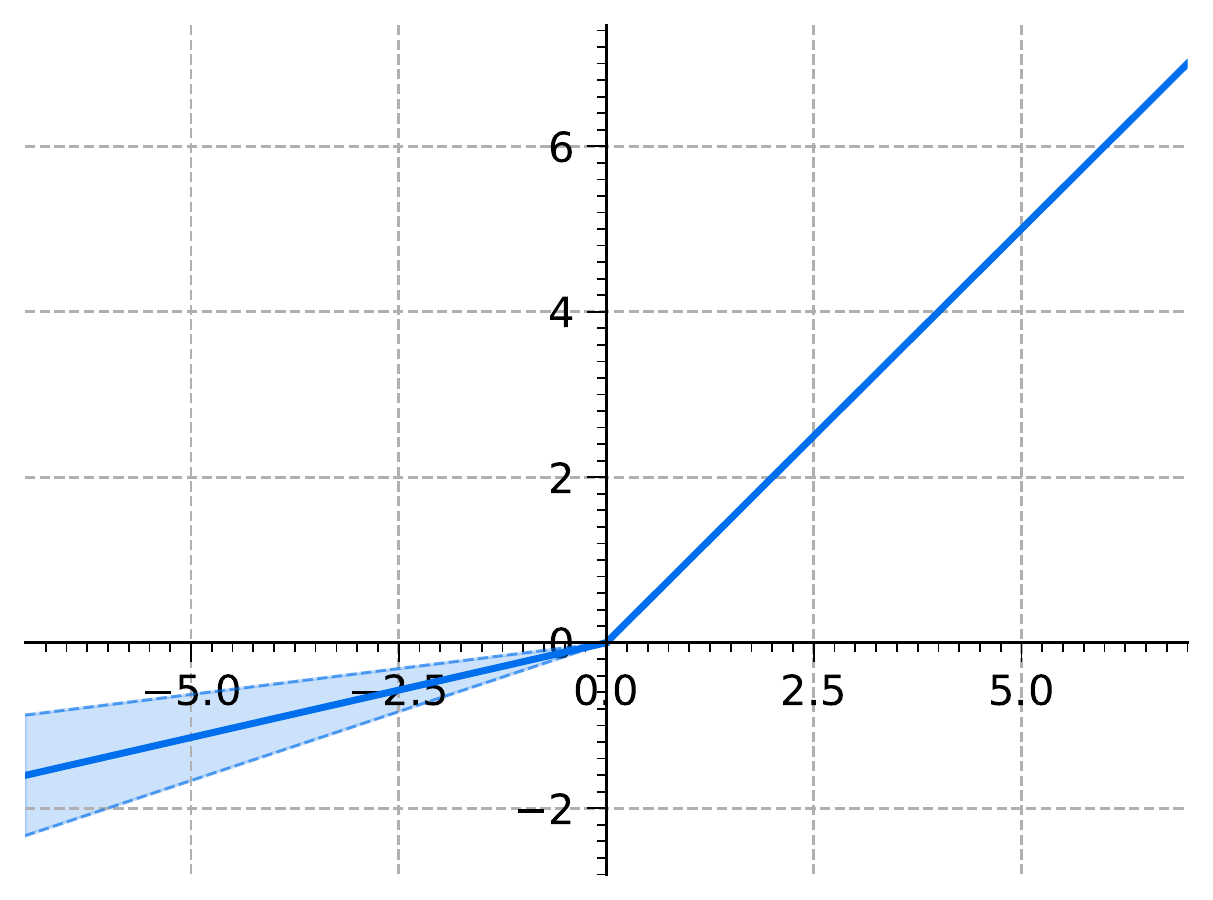}
		\end{minipage}
		&\begin{minipage}{0.1\textwidth}
			\begin{align*}
			&\left\{\begin{matrix}
			x & \text{if } x \geq 0 \\
			ax & \text{if } x < 0,
			\end{matrix}\right.\\
			&\text{with}\ a\sim U(1/8, 1/3)
			\end{align*}
		\end{minipage} \\ \hline
		
		PReLU 
		&\begin{minipage}{0.1\textwidth}
			\includegraphics[width=\textwidth]{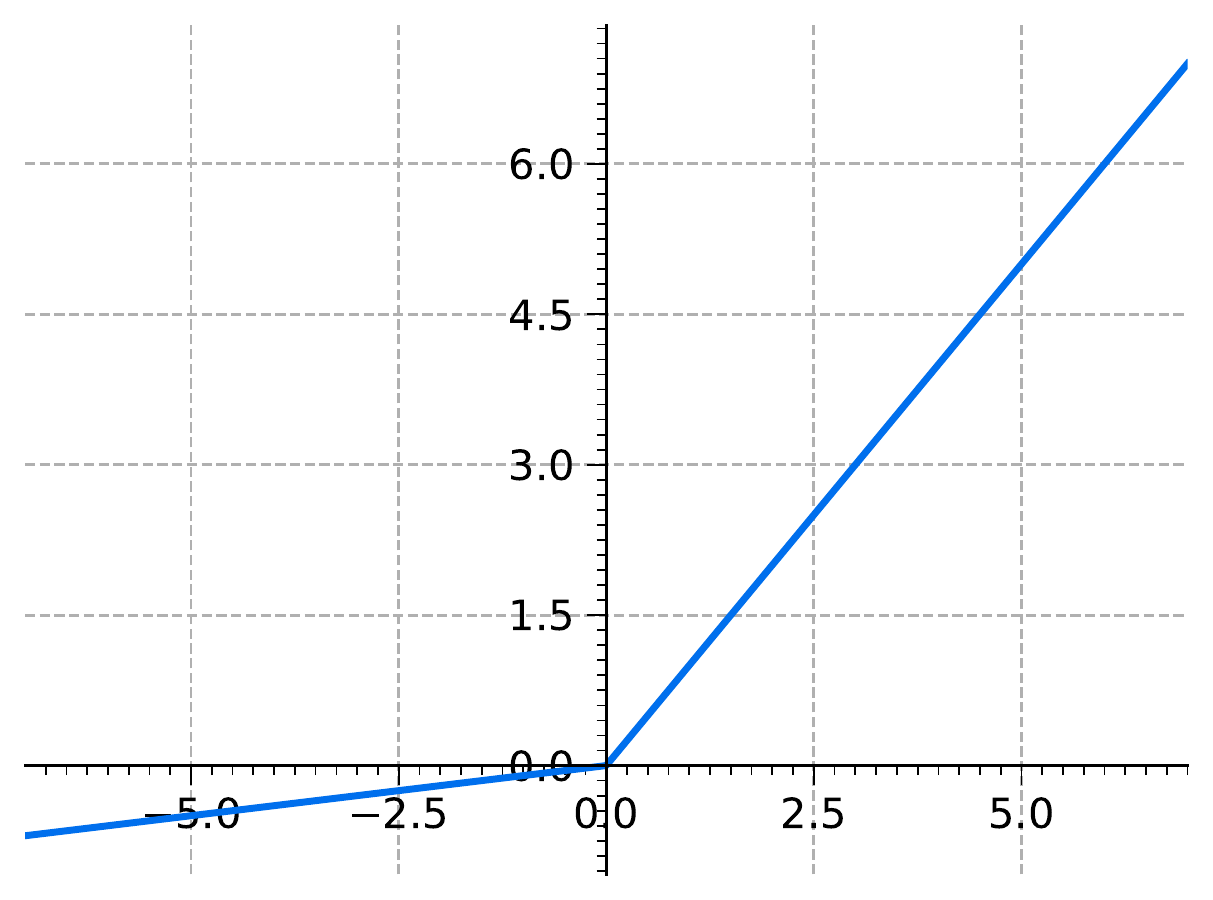}
		\end{minipage}
		&\begin{minipage}{0.1\textwidth}
			\begin{align*}
			&\left\{\begin{matrix}
			x & \text{if } x \geq 0 \\
			ax & \text{if } x < 0,
			\end{matrix}\right.\\
			&a\ \text{is a learnable parameter}
			\end{align*}
		\end{minipage} \\ \hline
		
		Sigmoid &\begin{minipage}{0.1\textwidth}
			\includegraphics[width=\textwidth]{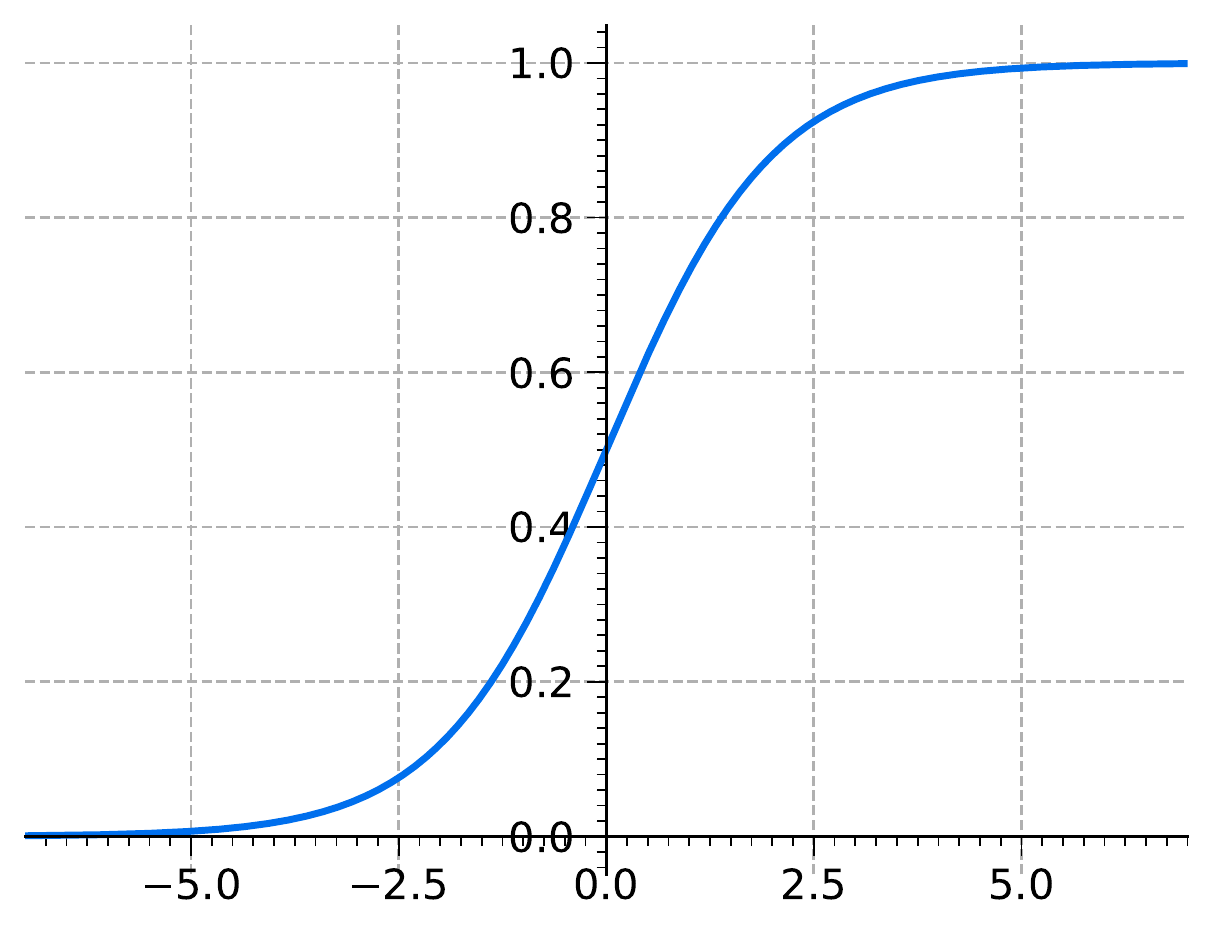}
		\end{minipage} &\begin{minipage}{0.1\textwidth}
			\begin{equation*}
			\frac{1}{1 + e^{-x}}
			\end{equation*}
		\end{minipage} \\ \hline
		
		Softsign &\begin{minipage}{0.1\textwidth}
			\includegraphics[width=\textwidth]{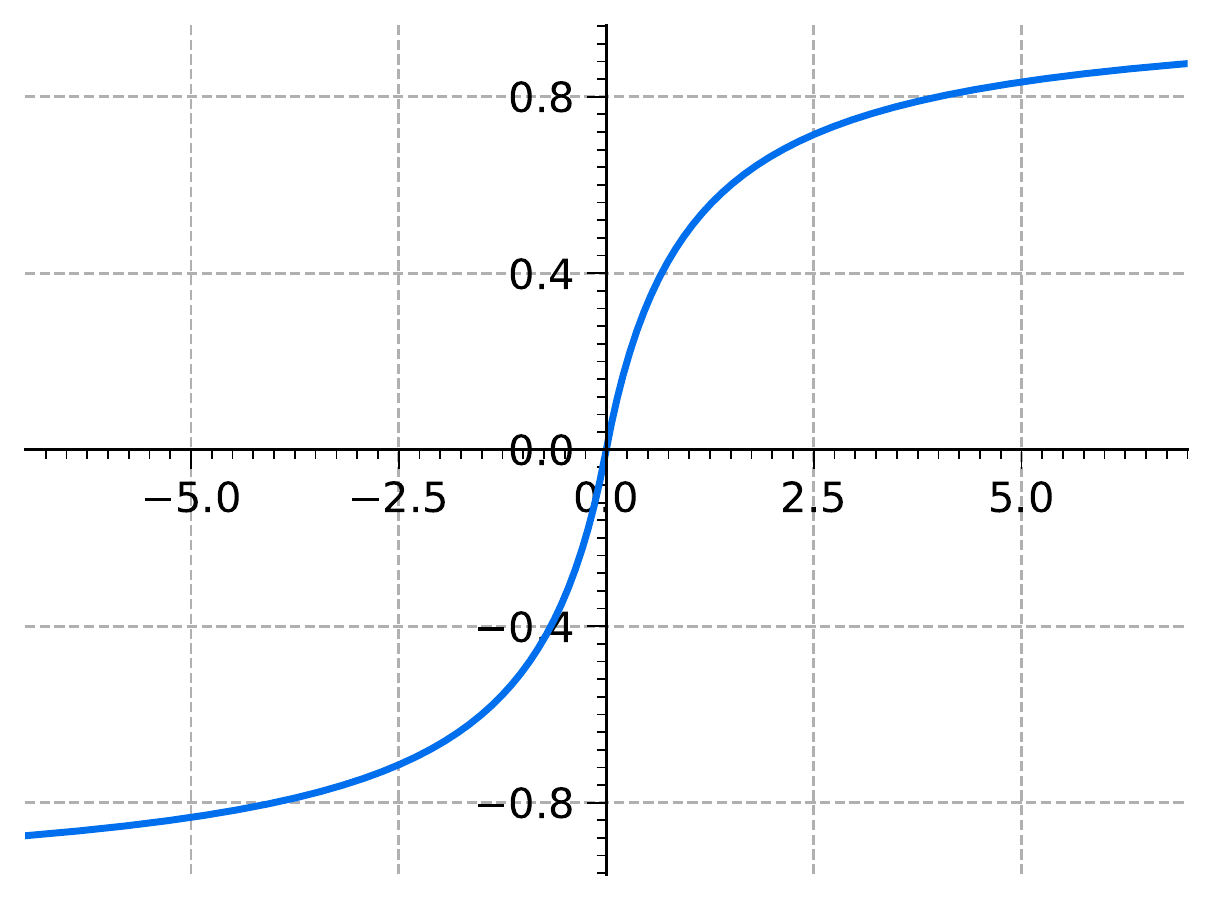}
		\end{minipage} &\begin{minipage}{0.1\textwidth}
			\begin{align*}
			\frac{x}{1+|x|}
			\end{align*}
		\end{minipage} \\ \hline
		
		Tanh 
		&\begin{minipage}{0.1\textwidth}
			\includegraphics[width=\textwidth]{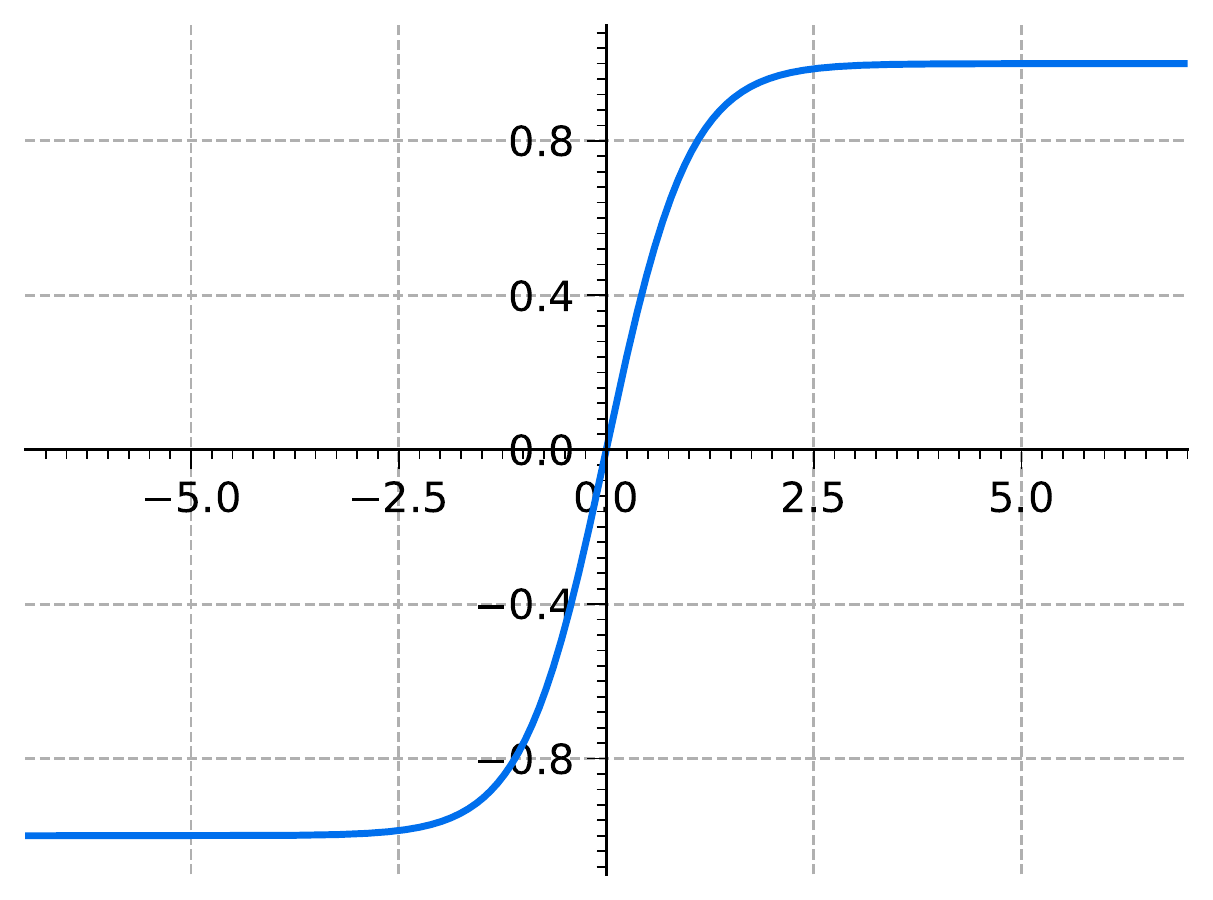}
		\end{minipage} &\begin{minipage}{0.1\textwidth}
			\begin{equation*}
			\frac{e^x - e^{-x}}{e^x + e^{-x}}
			\end{equation*}
		\end{minipage} \\ \hline
		
		ELU 
		&\begin{minipage}{0.1\textwidth}
			\includegraphics[width=\textwidth]{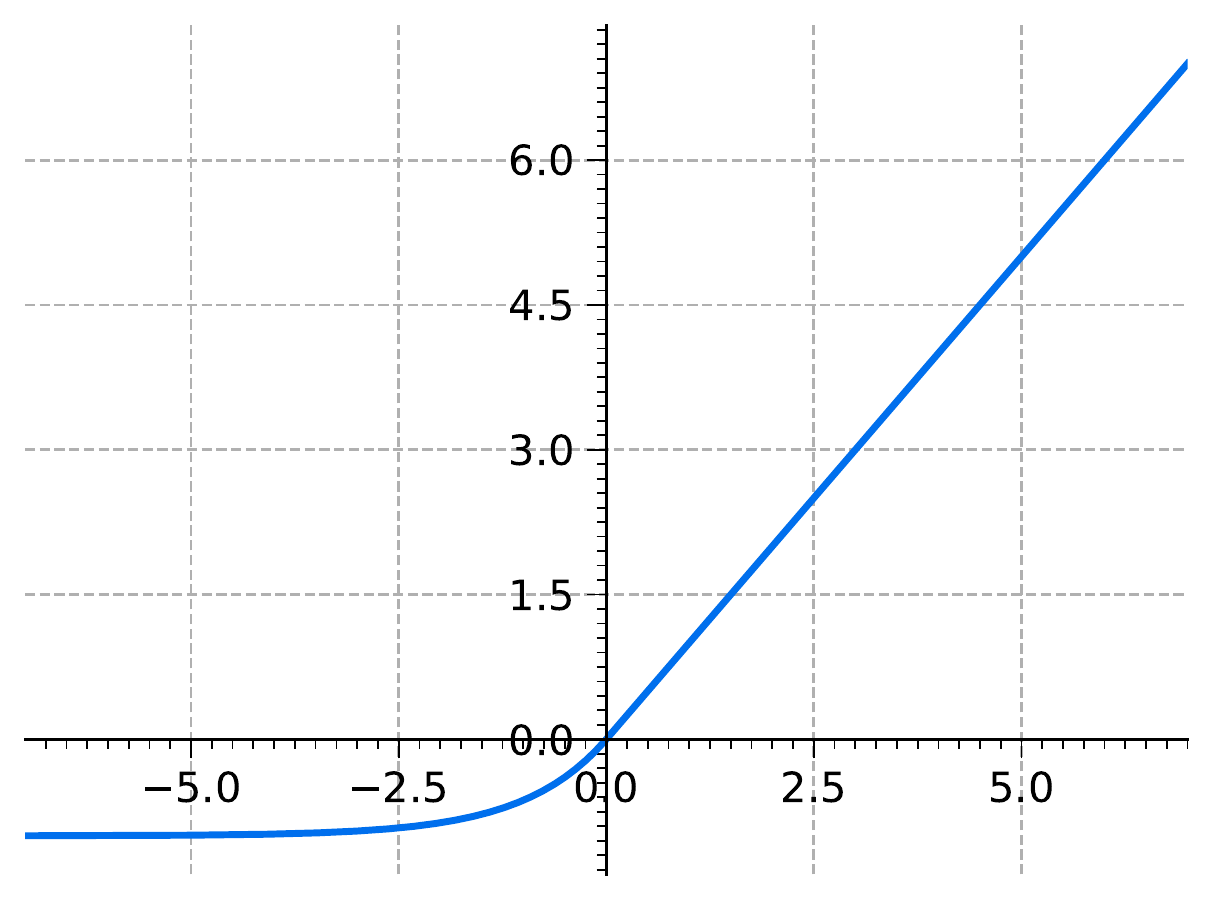}
		\end{minipage} &\begin{minipage}{0.1\textwidth}
			\begin{equation*}
			\left\{\begin{matrix}
			x & \text{if } x > 0 \\
			\alpha(e^x - 1) & \text{if } x \leq 0 
			\end{matrix}\right.
			\end{equation*}
		\end{minipage} \\ \hline
		
		Softplus
		&\begin{minipage}{0.1\textwidth}
			\includegraphics[width=\linewidth]{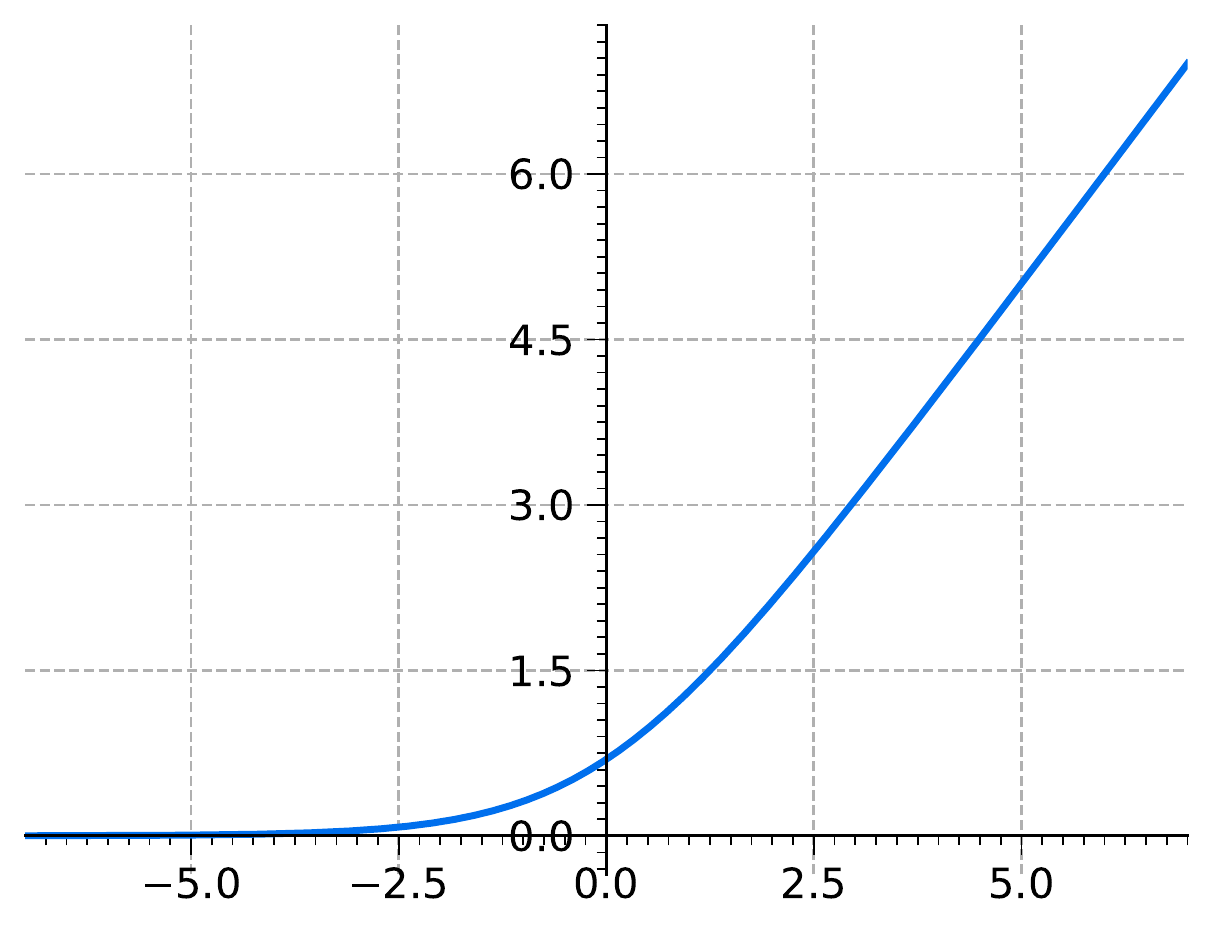}
		\end{minipage} &\begin{minipage}{0.1\textwidth}
			\begin{align*}
			&\frac{1}{\beta}\ln(1 + e^{\beta x}),\\
			&\text{with}\ \beta=1
			\end{align*}
		\end{minipage} \\ \hline
		
		LogSigmoid 
		&\begin{minipage}{0.1\textwidth}
			\includegraphics[width=\textwidth]{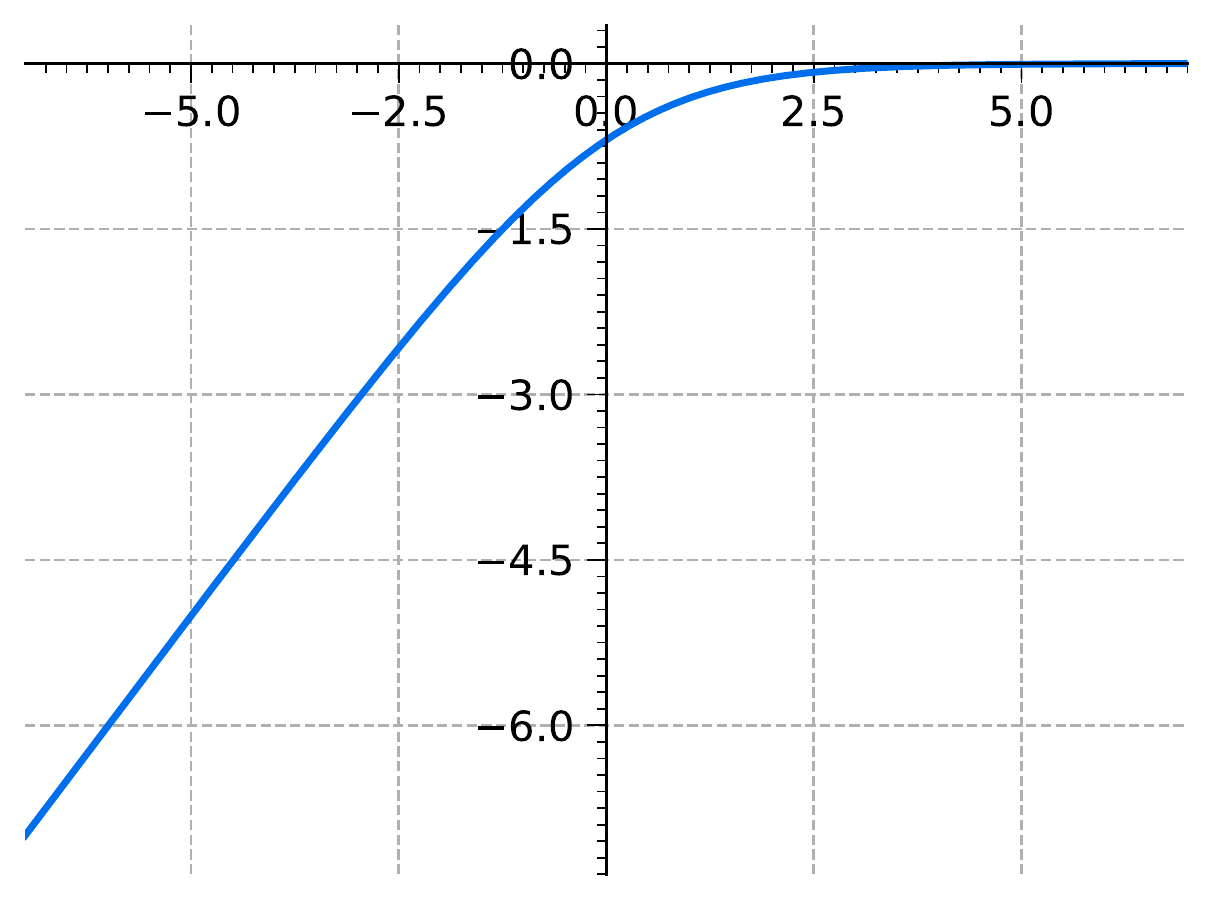}
		\end{minipage} &\begin{minipage}{0.1\textwidth}
			\begin{align*}
			\ln\left(\frac{1}{1+e^{-x}}\right)
			\end{align*}
		\end{minipage} \\	
		\hline\hline
	\end{tabular}
\end{table}

\begin{figure*}
	\centering
	\includegraphics[width=0.45\textwidth]{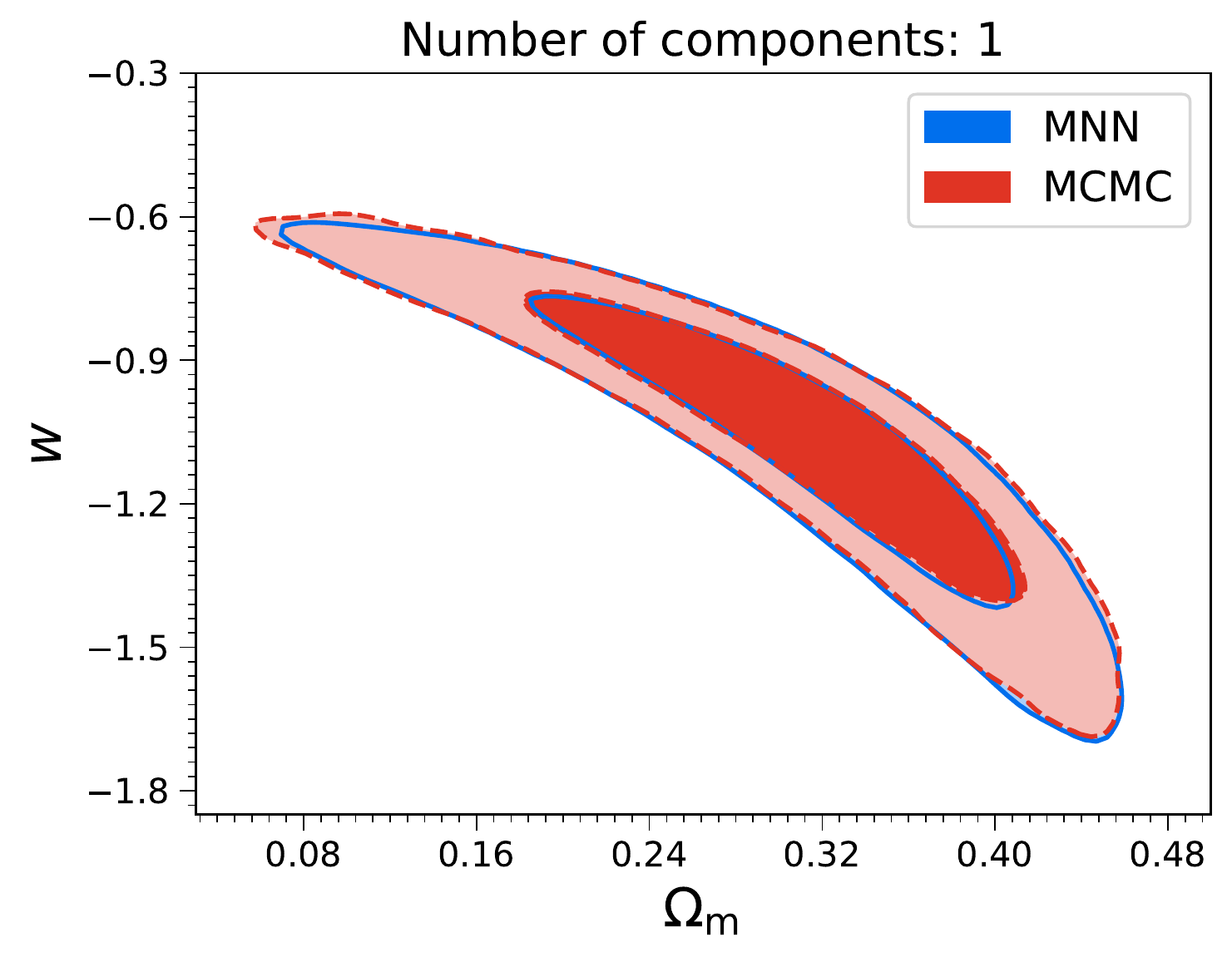}
	\includegraphics[width=0.45\textwidth]{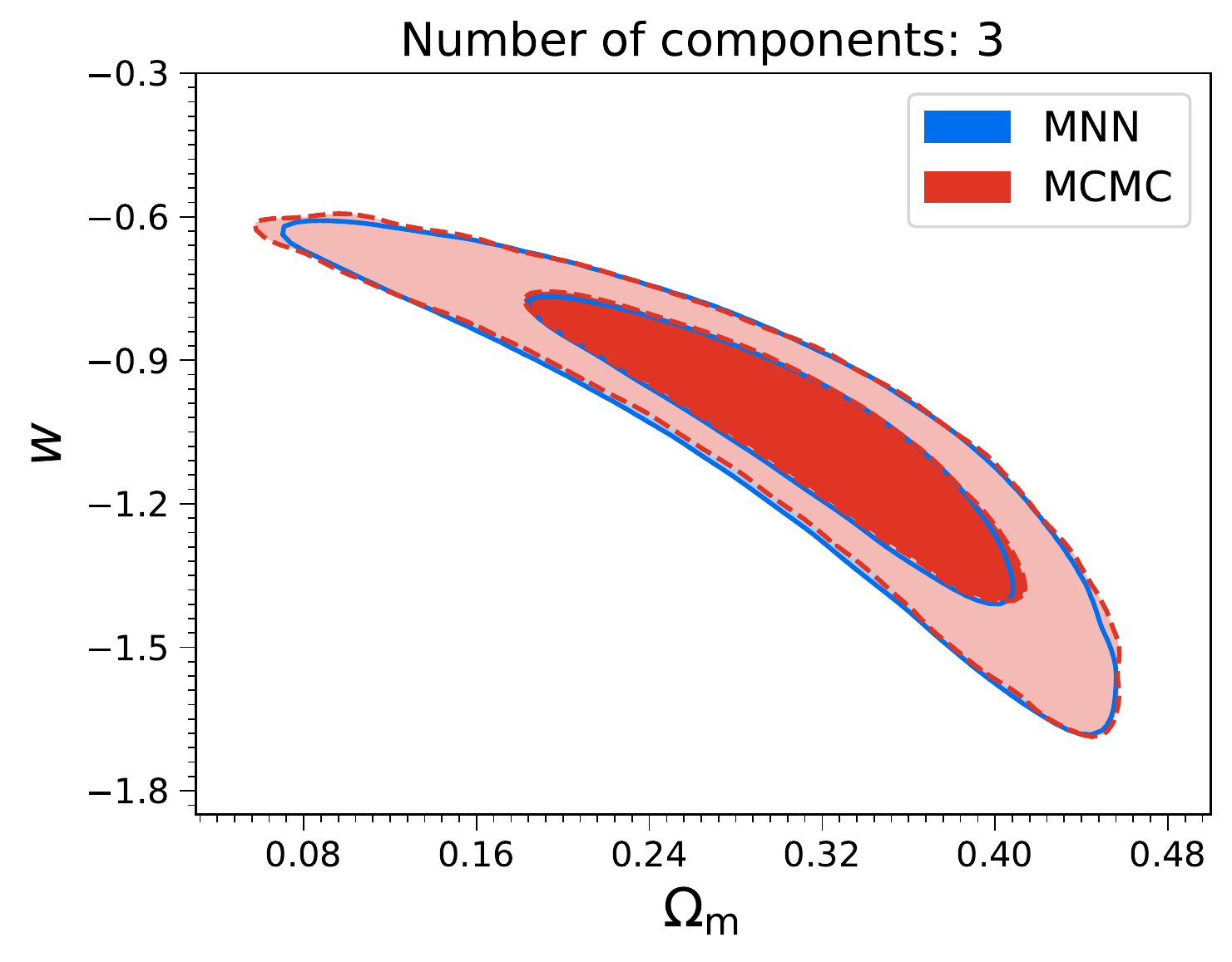}
	\includegraphics[width=0.45\textwidth]{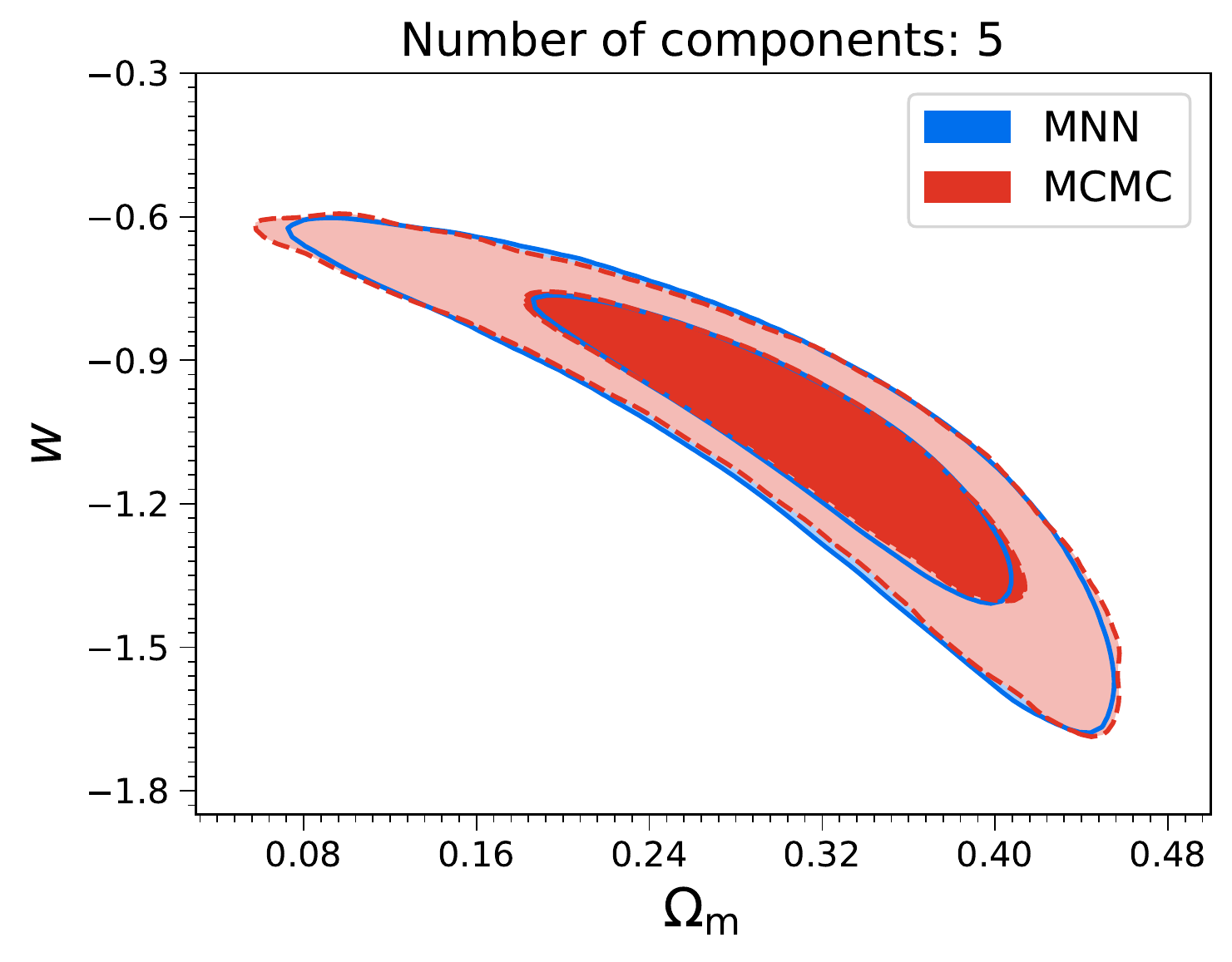}
	\includegraphics[width=0.45\textwidth]{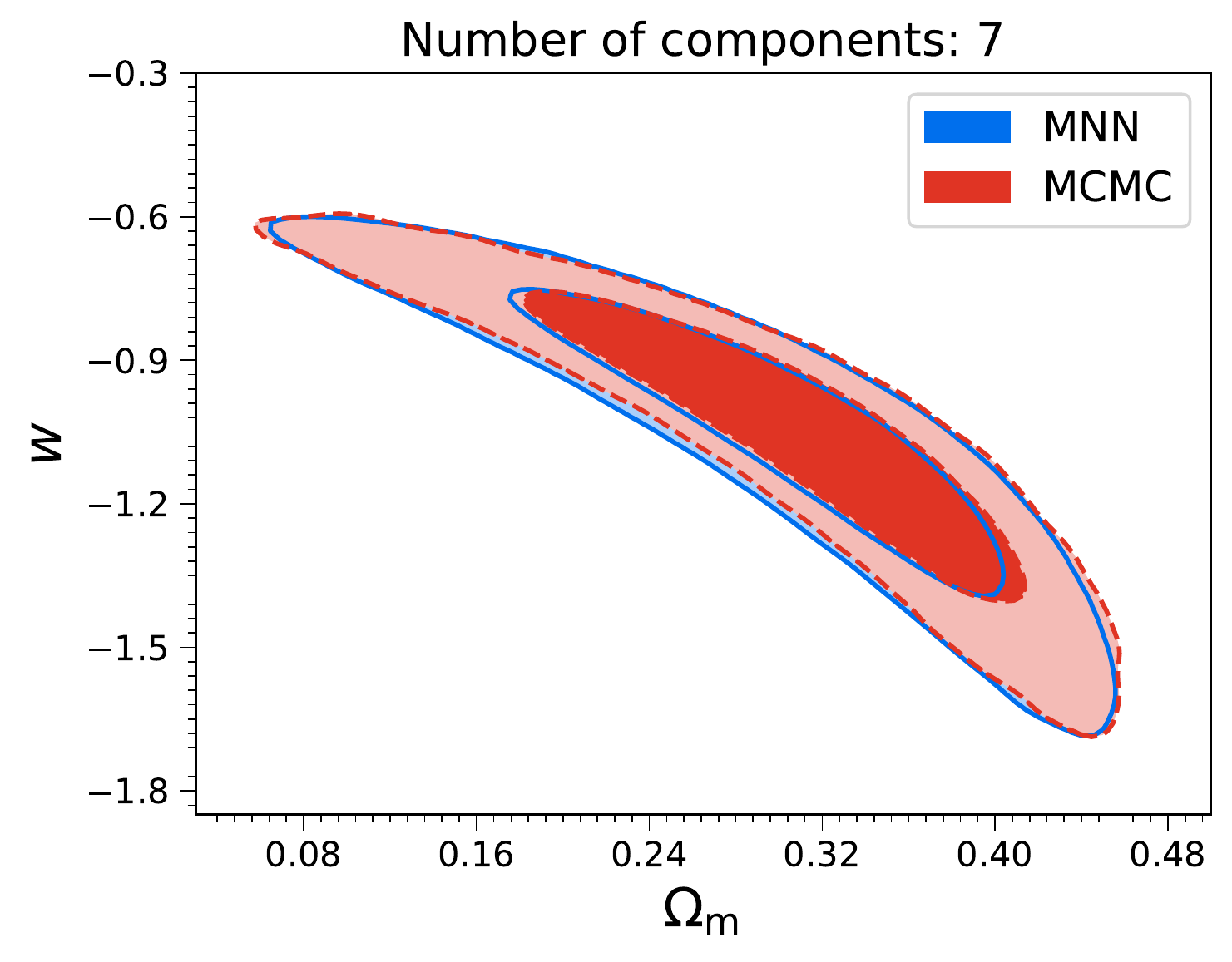}
	\caption{Two-dimensional marginalized distributions with 1$\sigma$ and 2$\sigma$ contours of $w$ and $\Omega_{\rm m}$ constrained from Pantheon SN Ia.}\label{fig:effect_of_component}
\end{figure*}

\subsection{Activation Function}\label{sec:effect_of_activationFunction}

The activation function used in the ANN may also affect the parameter estimation. To test this, we train network models with different activation functions. We select nine activation functions that are commonly used in machine-learning tasks: rectified linear unit (ReLU; \citealt{Nair:2010}), randomized leaky rectified linear unit (RReLU; ~\citealt{RReLU}), parametric rectified linear unit (PReLU; \citealt{prelu}), Sigmoid \citep{Han:1995}, Softsign \citep{softsign}, hyperbolic tangent (Tanh; \citealt{Malfliet:1992}), exponential linear unit (ELU; \citealt{Clevert:2016}), Softplus, and LogSigmoid, respectively. Formulas and shapes of these activation functions are shown in Table \ref{tab:activation_functions}. We estimate parameters for these nine cases using the MNN, MDN, and ANN methods, respectively. There are three hidden layers in the networks, the number of samples in the training set is 3000, and the networks are trained with 2000 epochs. The deviations are shown in the lower panel of Figure~\ref{fig:effect_of_hyperparameters}. It shows that the results of the MNN method are similar to those of the ANN method, and both perform better than the MDN method in most cases.

\subsection{The Number of Components}\label{sec:effect_of_components}

In the analysis above, the number of components in Equations~(\ref{equ:loss_mnn_1}) and (\ref{equ:loss_mnn_multi}) are set to unity. Here, we test whether the number of components has a large effect on the parameter estimation. With the same procedure as Section \ref{sec:application_to_SN}, we constrain $w$ and $\Omega_{\rm m}$ using the Pantheon SN Ia. The reason that we constrain $w$ and $\Omega_{\rm m}$ is that the distribution of these two parameters deviates significantly from the Gaussian distribution. Here, we consider four cases with different numbers of components: one, three, five, and seven components, respectively. In Figure~\ref{fig:effect_of_component}, we show the constraints on $w$ and $\Omega_{\rm m}$ using the MNN and MCMC methods. Obviously, we can see that the MNN results are almost the same as the MCMC results for all the cases. Furthermore, we calculate the mean deviations of parameters between the MNN results and the MCMC results for the four cases: $0.051\sigma$, $0.046\sigma$, $0.067\sigma$, and $0.044\sigma$, respectively, which are similar to each other. Therefore, this indicates that the number of components has no effect on the parameter estimation, which is very different from the MDN method (see \citetalias{Wanggj:2022}).

We find the reason is, for the MNN method, the posterior distribution of the cosmological parameters is output directly by the network. Since the strong nonlinearity of ANN, it can approximate a variety of functions \citep{Cybenko:1989,Hornik:1991}. Therefore, this allows MNNs to obtain accurate estimates for parameters with non-Gaussian distributions using only one component. However, for the MDN method, the parameters of the mixture model need to be known before obtaining the posterior distribution. This situation indicates that we need to use a lot of components for the parameters with non-Gaussian distributions, which makes it difficult to train MDN models.

\begin{figure}
	\centering
	\includegraphics[width=0.45\textwidth]{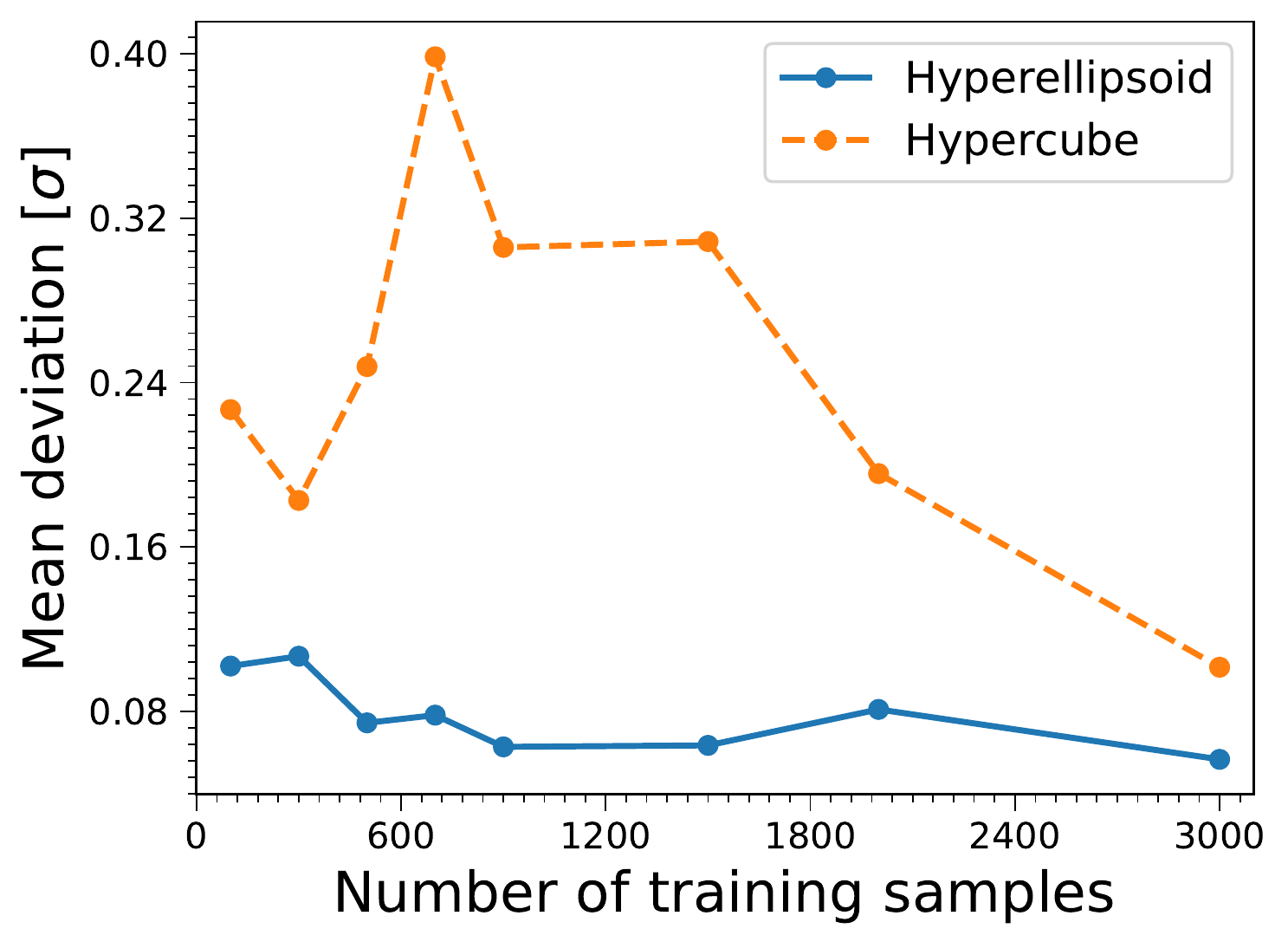}
	\caption{Mean deviations between the MNN results and the MCMC results as a function of the number of training samples for the two sampling methods mentioned in Section \ref{sec:training_set}.}\label{fig:effect_of_samplingMethod}
\end{figure}

\begin{table*}
	\centering
	\caption{The same as Table~\ref{tab:params_planck_TT_6params} but with an additional neutrino mass parameter ($\sum m_{\nu}$).}\label{tab:params_planck_TT_7params}
	\begin{tabular}{c|c|c|c|c}
		\hline\hline
		& \multicolumn{4}{c}{Methods} \\
		\cline{2-5}
		Parameters & MCMC & MNN & MDN & ANN \\
		\hline
		$H_0$               & $65.395\pm2.544$    & $65.653\pm2.647$    & $63.735\pm2.383$    & $65.919\pm2.672$ \\
		$\Omega_{\rm b}h^2$ & $0.02231\pm0.00023$ & $0.02233\pm0.00024$ & $0.02231\pm0.00025$ & $0.02234\pm0.00025$ \\
		$\Omega_{\rm c}h^2$ & $0.11826\pm0.00273$ & $0.11804\pm0.00285$ & $0.11871\pm0.00290$ & $0.1177\pm0.00282$ \\
		$\tau$		        & $0.15172\pm0.03533$ & $0.15317\pm0.03613$ & $0.15605\pm0.03552$ & $0.1471\pm0.03477$ \\
		$10^9A_{\rm s}$     & $2.51691\pm0.16700$ & $2.53421\pm0.17225$ & $2.54984\pm0.16977$ & $2.50745\pm0.15981$ \\
		$n_{\rm s}$         & $0.96691\pm0.00752$ & $0.96603\pm0.00793$ & $0.96476\pm0.00802$ & $0.96549\pm0.00765$ \\
		$\sum m_\nu$        & $<0.544$ & $<0.580$ & $<0.801$ & $<0.726$ \\
		\hline\hline
	\end{tabular}
\end{table*}
\begin{figure*}
	\centering
	\includegraphics[width=0.9\textwidth]{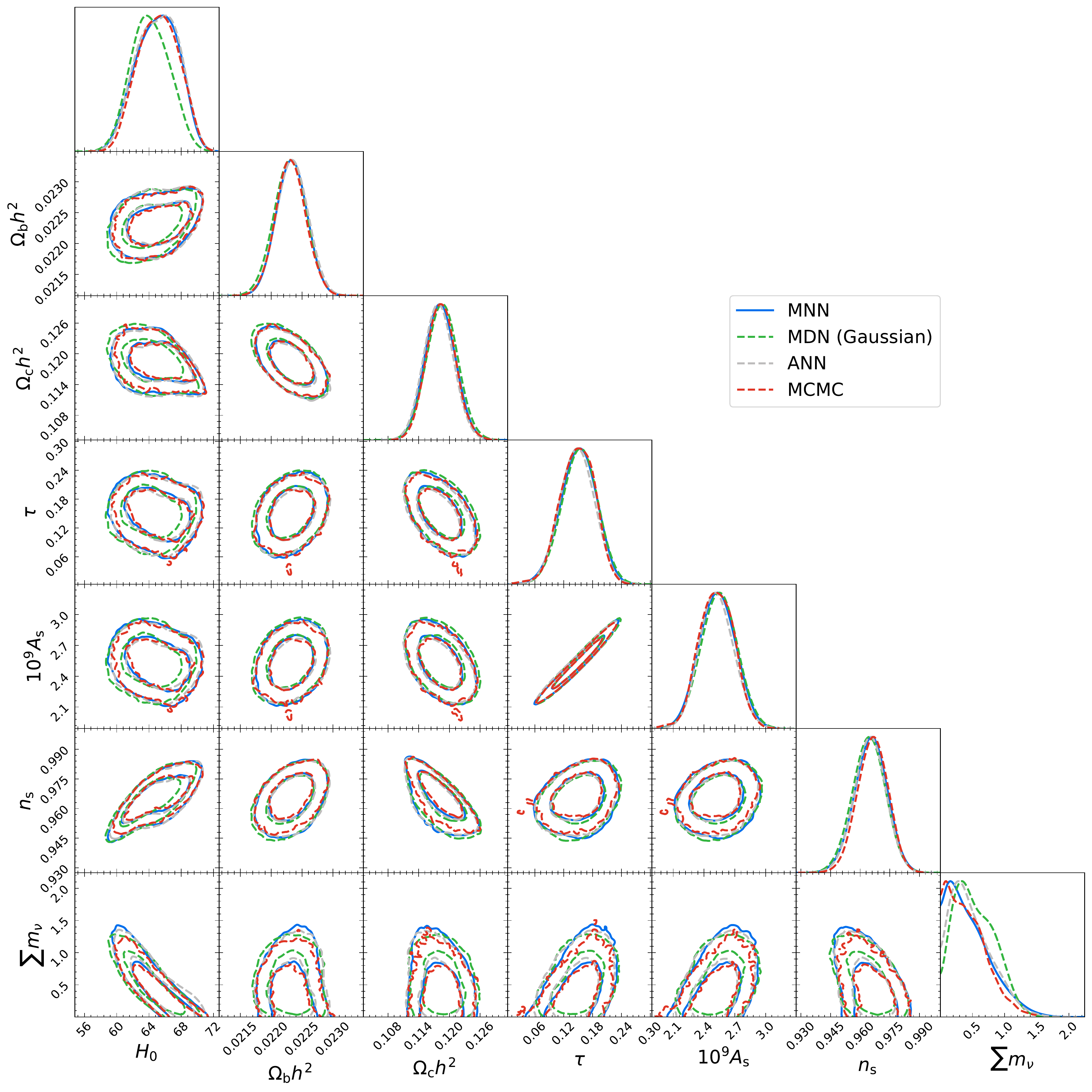}
	\caption{Posterior distributions of $H_0$, $\Omega_{\rm b}h^2$, $\Omega_{\rm c}h^2$, $\tau$, $A_{\rm s}$, $n_{\rm s}$, and $\sum m_\nu$ constrained from {\it Planck}-2015 $C^{\rm TT}_{\ell}$ with $1\sigma$ and $2\sigma$ contours. Three Gaussian components are used for the MDN method.}\label{fig:contour_planck_TT_7params}
\end{figure*}
\begin{figure}
	\centering
	\includegraphics[width=0.45\textwidth]{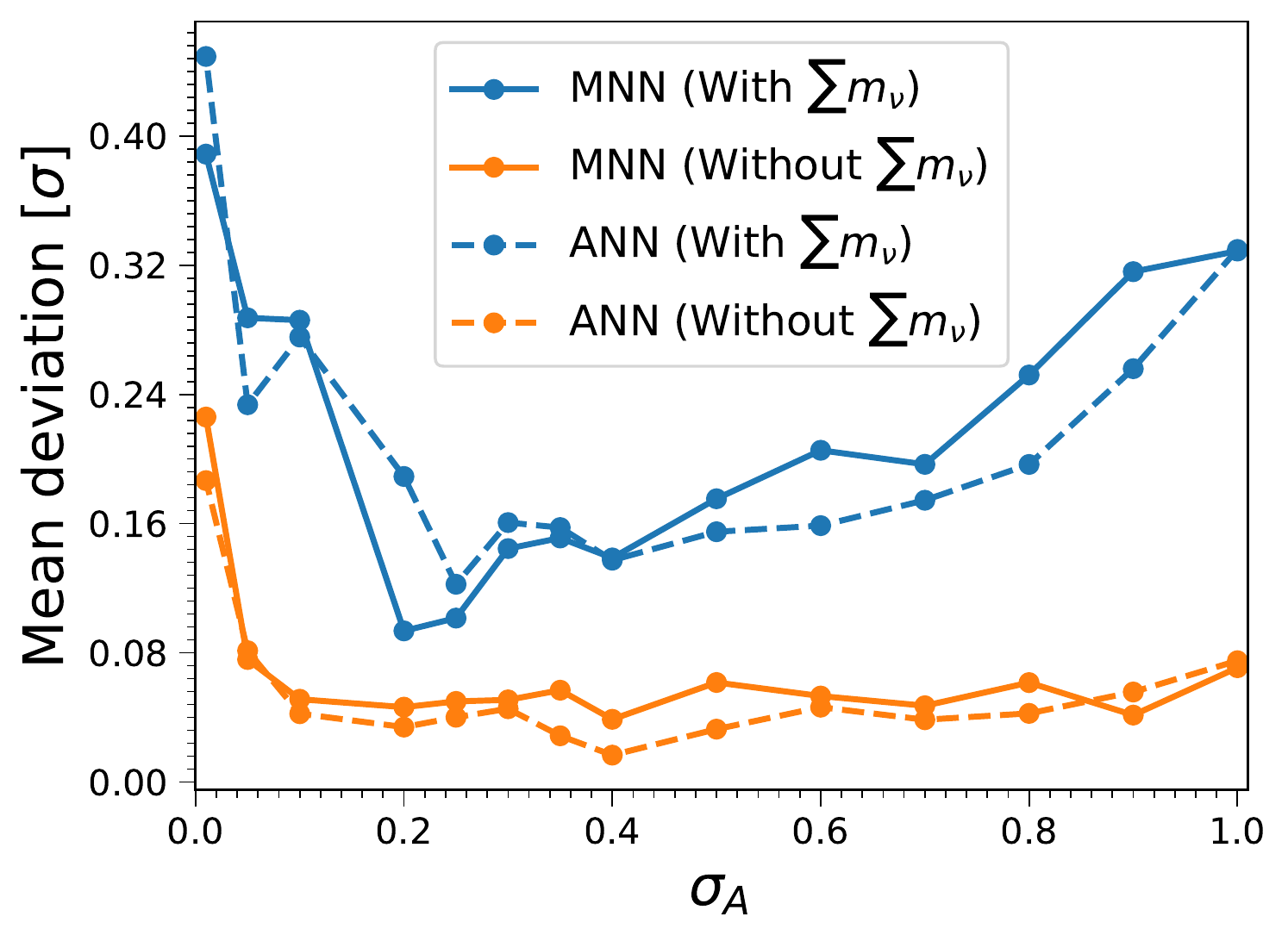}
	\caption{Mean deviations between the MNN/ANN results and the MCMC results as a function of the standard deviation of the coefficient $A$ (see Section \ref{sec:add_noise}). Parameters are constrained using the {\it Planck}-2015 $C^{\rm TT}_{\ell}$.}\label{fig:effect_of_coefficient_planck}
\end{figure}

\subsection{Parameter Space Sampling Methods}\label{sec:effect_of_spaceSamplingMethod}

As we illustrated in Section \ref{sec:training_set} there are two parameter space sampling methods that can be used when generating a training set: sampling uniformly in a hypercube and in a hyper-ellipsoid. All analyses above use the latter one because of the high efficiency. We further assess the impact of sampling methods by including the hypercube function. Specifically, for each sampling method, we train eight MNN models, with the number of training samples varying from 100 to 3000, to constrain $H_0$, $\Omega_{\rm b}h^2$, $\Omega_{\rm c}h^2$, $\tau$, $A_{\rm s}$, and $n_{\rm s}$ using the {\it Planck}-2015 $C^{\rm TT}_{\ell}$. Other settings of the MNN model are the same as those of the MNN model in Section \ref{sec:application_to_CMB}. In Figure~\ref{fig:effect_of_samplingMethod}, we show the mean deviations of cosmological parameters of the MNN results and the MCMC results. We can see that, for the method sampling in the hyper-ellipsoid, the deviation is $\sim 0.100\sigma$ for the cases using less than 300 samples, while $<0.081\sigma$ for the cases using more than 300 samples. However, for the method sampling in the hypercube, the deviations are larger than $0.100\sigma$ for all cases, which are much larger than those of the method sampling in the hyper-ellipsoid. Therefore, the method sampling in the hyper-ellipsoid is more efficient for parameter estimation tasks, so we highly recommend using it. It should be noted that the method sampling in the hyper-ellipsoid makes it possible to obtain accurate estimations of parameters using $\mathcal{O}(10^2)$ samples. This is very helpful for complex and resource-consuming cosmological models.

\section{Discussions}\label{sec:discussions}

\subsection{Parameters with Physical Limits}\label{sec:parameters_with_physical_limits}

As we illustrated in Section \ref{sec:training_set}, the parameter space to be learned is a $\pm5\sigma_p$ range of the posterior distribution. However, it should be noted that this only works for the distributions of parameters that are not truncated. For parameters that have truncated distributions due to the physical limits, the range of parameters to be learned will be smaller than $\pm5\sigma_p$, which will be harder for ANN to obtain accurate estimates and ultimately affect the parameter estimation. Therefore, in this section, we test this by using two cases with truncated distributions: the case of multiple parameters and the case of only one parameter.

\subsubsection{Multiple Parameters}\label{sec:multiple_parameters}

With the same procedure and MNN settings as those in Section \ref{sec:application_to_CMB}, we constrain $H_0$, $\Omega_{\rm b}h^2$, $\Omega_{\rm c}h^2$, $\tau$, $A_{\rm s}$, $n_{\rm s}$, and $\sum m_\nu$ with the MCMC and MNN methods, by using the {\it Planck}-2015 $C^{\rm TT}_{\ell}$. The $1\sigma$ constraints on parameters are shown in Table \ref{tab:params_planck_TT_7params}, and the corresponding one-dimensional and two-dimensional marginalized distributions with $1\sigma$ and $2\sigma$ contours are shown in Figure~\ref{fig:contour_planck_TT_7params}. We can see that the contours of the MNN method are almost the same as those of the MCMC method. Quantitatively, the deviations of parameters between the MNN results and the MCMC results are $0.070\sigma$, $0.069\sigma$, $0.055\sigma$, $0.001\sigma$, $0.040\sigma$, $0.081\sigma$, and $0.010\sigma$, respectively, which are quite small. Therefore, the MNN method is capable
of dealing with cosmological parameters with truncated distributions.

As a comparison, we further constrain parameters using the MDN and ANN methods. The MDN and ANN have the same settings (i.e., the number of hidden layers, training samples, epochs, and the activation function) as the MNN model. Three Gaussian components are used for the MDN method. The best-fit values with $1\sigma$ errors are shown in Table \ref{tab:params_planck_TT_7params}, and the corresponding distributions are shown in Figure~\ref{fig:contour_planck_TT_7params}, with the green solid and gray solid lines, respectively. Obviously, we can see that both of them have larger deviations with the MCMC method, especially for $\sum m_\nu$. This indicates that both MDNs and ANNs have limitations in estimating the parameter with truncated distributions. Therefore, for the case of multiple parameters (especially for parameters with truncated distribution), we recommend using the MNN method instead of the MDN and ANN methods.

For the MNN and ANN methods, the standard deviation of the coefficient $A$ ($\sigma_A$, see Section \ref{sec:add_noise}) is set to 0.2. However, it should be noted that $\sigma_A$ has a large effect on the parameter estimations for parameters with truncated distributions. To illustrate this, we train MNN models with $\sigma_A$ varying from 0.01 to 1 to constrain the seven cosmological parameters. Then, we calculate the mean deviations between the MNN results and the MCMC results, as shown in Figure~\ref{fig:effect_of_coefficient_planck} (the blue solid line). We can see that there is a minimum deviation at $\sigma_A=0.2$. The reason for the large deviation when $\sigma_A<0.2$ is that $|A|_{\rm max}$ will be less than 1, which makes the network unable to learn the observation error level. 

For the cases of $\sigma_A>0.2$, the range of $A$ will cover 1; thus, theoretically, the deviation should not increase with the increase of $\sigma_A$, which is contrary to the result. This result should be caused by $\sum m_\nu$ with a truncated distribution. To test this, we only constrain $H_0$, $\Omega_{\rm b}h^2$, $\Omega_{\rm c}h^2$, $\tau$, $A_{\rm s}$, and $n_{\rm s}$ by fixing $\sum m_\nu$. The corresponding deviations of the MNN results and the MCMC results are indicated by the orange solid line. We can see that the deviations are similar for cases of $\sigma_{A}>0.2$. This means that MNN can not accurately estimate the parameters with truncated distributions when using a coefficient with $\sigma_A>0.2$. Therefore, $\sigma_A$ is set to 0.2 by default in CoLFI.

With the same procedure, we train ANN models with $\sigma_A$ varying from 0.01 to 1 to constrain cosmological parameters (with and without $\sum m_\nu$). The mean deviations between the ANN results and the MCMC results are shown in Figure~\ref{fig:effect_of_coefficient_planck} (the blue and orange dashed lines). We can see that, for parameters with truncated distributions (with $\sum m_\nu$), the MNN method outperforms the ANN method at $\sigma_A=0.2$, while, for parameters without truncated distributions (without $\sum m_\nu$), the ANN method outperforms the MNN method for many cases. Therefore, for the parameters without physical limits (or without truncated distributions), the ANN method is recommended for parameter estimation. We will investigate this further to optimize CoLFI.

\begin{table}
	\centering
	\caption{Constraints on $\Omega_{\rm m}$ of the $w$CDM model using the Pantheon SN Ia data while fixing $w=-0.5$, quoted with best-fit values and $1\sigma$~C.L.}\label{tab:params_pantheon_modified}
	\begin{tabular}{c|c}
		\hline\hline
		Methods & $\Omega_{\rm m}$ \\
		\hline
		MCMC & $0.0003_{-0.0002}^{+0.0151}$ \\
		\hline
		MNN  & $0.0005_{-0.0003}^{+0.0148}$ \\
		\hline
		MDN  & $0.0003_{-0.0002}^{+0.0147}$ \\
		\hline
		ANN  & $0.0003_{-0.0002}^{+0.0150}$ \\
		\hline\hline
	\end{tabular}
\end{table}

\subsubsection{One Parameter}\label{sec:one_parameter}

\begin{figure}
	\centering
	\includegraphics[width=0.45\textwidth]{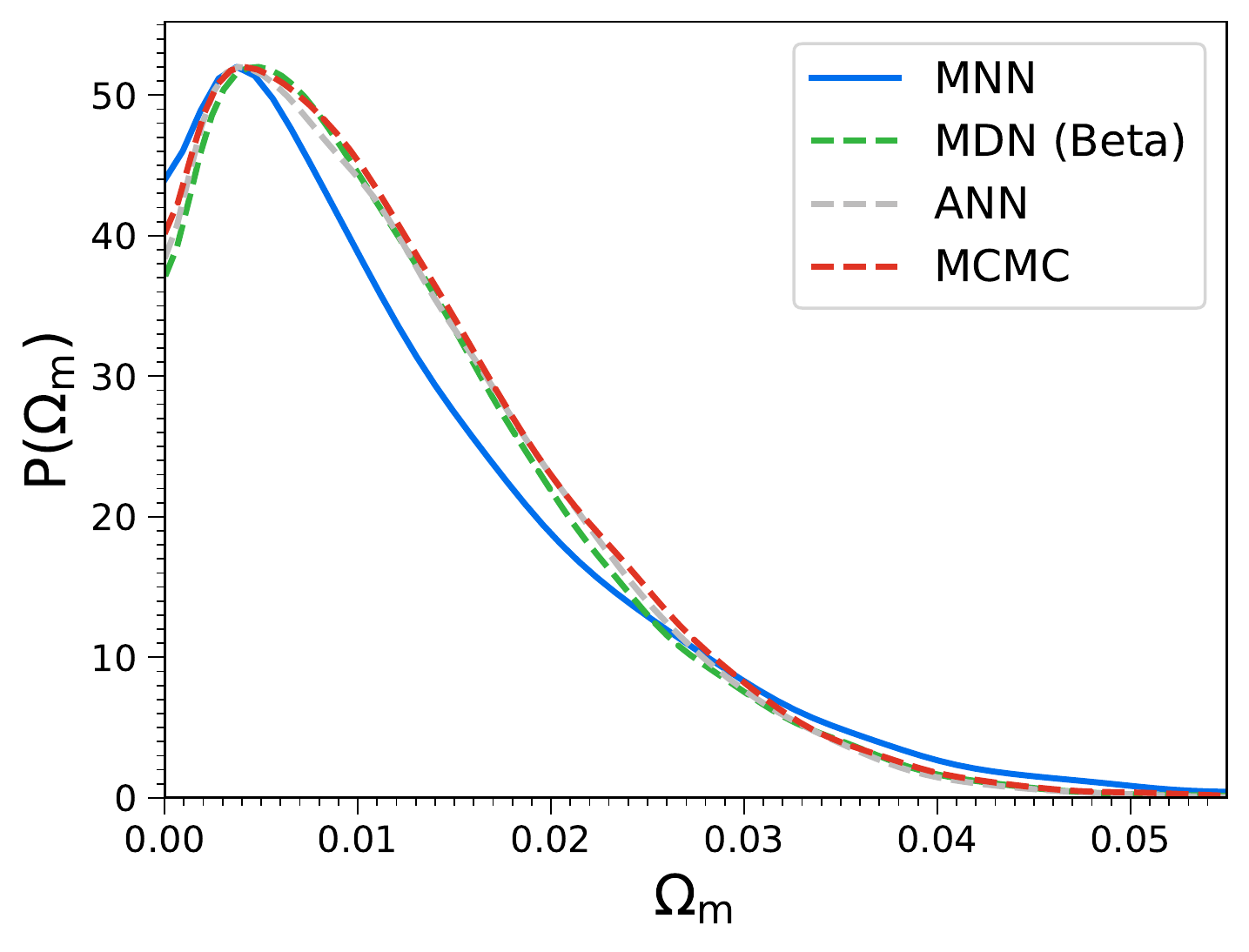}
	\caption{One-dimensional distributions of $\Omega_{\rm m}$ constrained from Pantheon SN Ia. Here, $w$ is manually set to $-0.5$, and one Beta component is used for the MDN method.}\label{fig:contour1D_pantheon_w-0.5}
\end{figure}

For the case of one parameter, we constrain $\Omega_{\rm m}$ of the $w$CDM model using the Pantheon SN Ia data. The analysis here is only to test the performance of CoLFI on truncated parameters. Therefore, the Hubble constant, the absolute magnitude, and the equation of state of dark energy are manually set to $H_0=70 ~\rm km\ s^{-1}\ Mpc^{-1}$, $M_B=-19.3$, and $w=-0.5$, respectively. We constrain $\Omega_{\rm m}$ using the MCMC, MNN, MDN, and ANN methods, respectively. The MNN, MDN, and ANN have the same settings (i.e., the number of hidden layers, training samples, epochs, and the activation function) as the MNN model in Section \ref{sec:application_to_SN}. One Beta component is used for the MDN method. The best-fit values and $1\sigma$ errors are shown in Table \ref{tab:params_pantheon_modified}, and the corresponding distributions are shown in Figure~\ref{fig:contour1D_pantheon_w-0.5}. We can see that the results of the MNN, MDN, and ANN methods all converge to the MCMC result. For the MNN method, the shape of the distribution of $\Omega_{\rm m}$ is slightly different from that of the MCMC method, while, for the MDN and ANN methods, the shape of distributions is almost the same as that of the MCMC method. The reason why the MNN result is slightly different from the MCMC result is that MNNs are learning errors of parameters, which is redundant information for the case of only one cosmological parameter.

\begin{figure}
	\centering
	\includegraphics[width=0.45\textwidth]{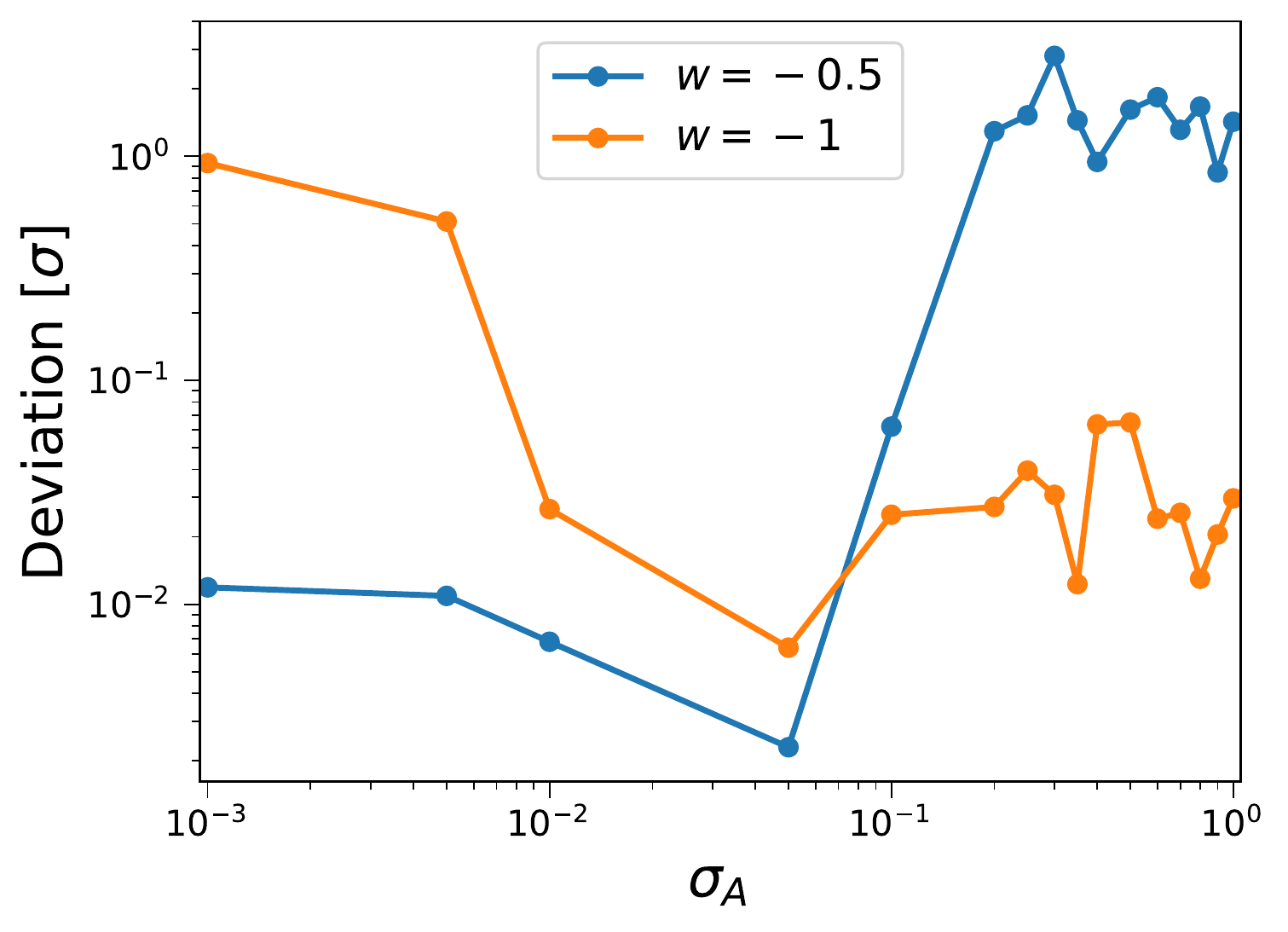}
	\caption{Deviations between the ANN results and the MCMC results as a function of the standard deviation of the coefficient $A$ (see Section \ref{sec:add_noise}). The cosmological parameter here is constrained using the Pantheon SN Ia data.}\label{fig:effect_of_coefficient_pantheon}
\end{figure}

It should be noted that, for the MNN and ANN methods, $\sigma_A$ is set to 0.05 for the case of only one cosmological parameter. To illustrate how $\sigma_A$ affects the parameter estimation, we train ANN models with $\sigma_A$ varying from 0.001 to 1 to constrain $\Omega_{\rm m}$. The deviations between the ANN results and the MCMC results as a function of $\sigma_A$ are shown in Figure~\ref{fig:effect_of_coefficient_pantheon} with the blue solid line. We can see that there is a minimum deviation at $\sigma_A=0.05$. In addition, we also test the effect of $\sigma_A$ on the parameter without truncated distribution. Specifically, with the same procedure, we constrain $\Omega_{\rm m}$ using the ANN method by setting $w=-1$ manually. The deviations between the ANN results and the MCMC results are shown in Figure~\ref{fig:effect_of_coefficient_pantheon} with the orange solid line. We can see that the minimum deviation is located at $\sigma_A=0.05$, and there is a low-level deviation at $\sigma_A>0.05$, which is very different from that of the truncated parameter. Therefore, when using ANN (or MNN), we recommend using $\sigma_A=0.05$ for the case of only one cosmological parameter.


Comparing with the analysis in Section \ref{sec:multiple_parameters}, one can see that the setting of $\sigma_A$ is different between the case of only one parameter and the case of   multiple parameters. According to our understanding, the setting of $\sigma_A$ should ensure that unity is covered by the range of $|A|$. For the case of only one parameter, the setting of $\sigma_A=0.05$ will results in $|A|_{\rm max}<1$. Thus, further research should be done to understand this better. Therefore, for cases of only one cosmological parameter, we recommend using the MDN method (with Beta components) instead of the MNN (or ANN) method.

\begin{figure}
	\centering
	\includegraphics[width=0.45\textwidth]{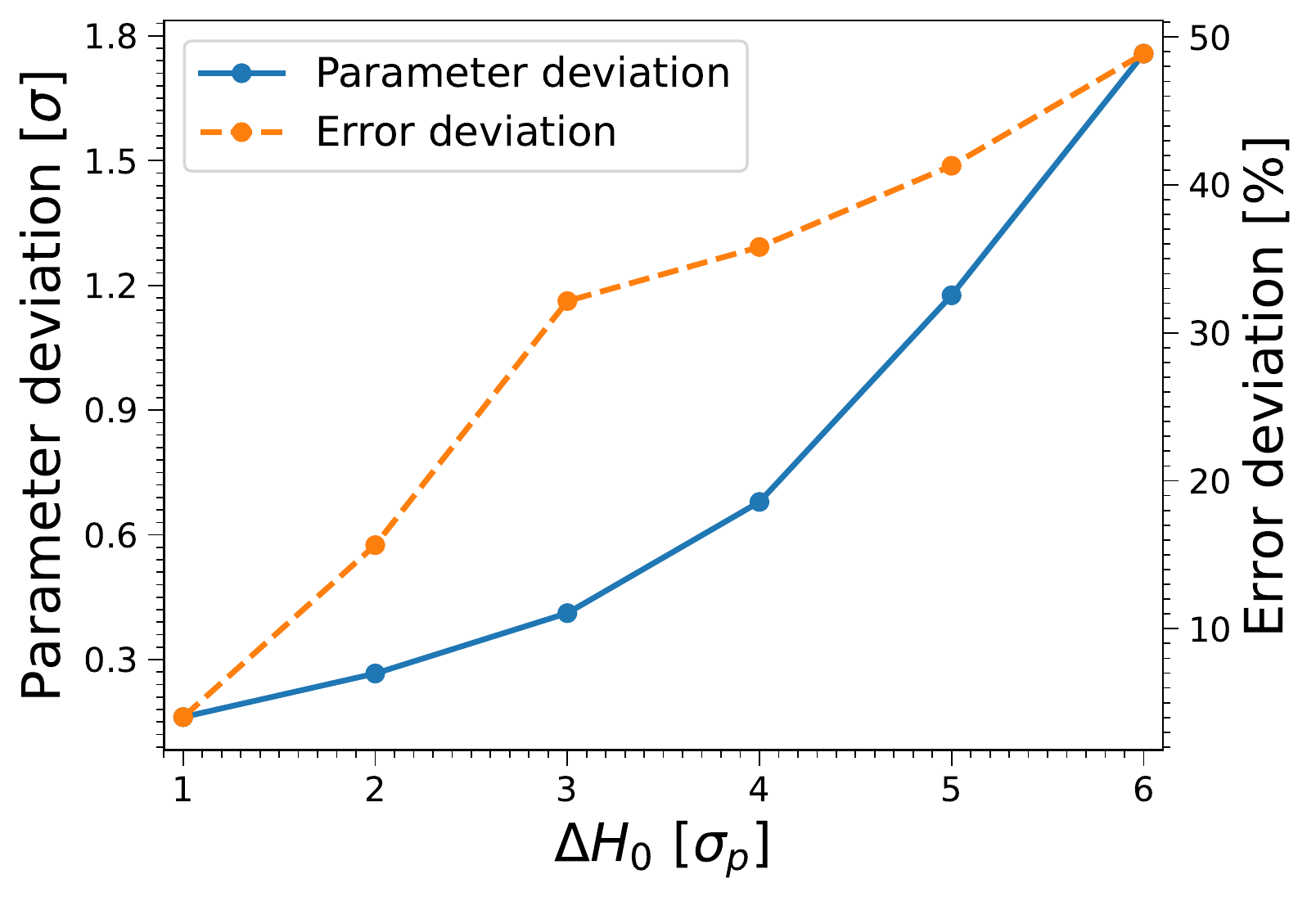}
	\caption{Mean deviation of parameters and relative deviation of errors between the $(i+1)$th posterior and the $i$th posterior as a function of $\Delta H_0$ (Equation~(\ref{equ:deviation_H0})).}\label{fig:biased_H0}
\end{figure}

\subsection{Update Parameter Space}\label{sec:update_parameter_space}

\begin{figure}
	\centering
	\includegraphics[width=0.45\textwidth]{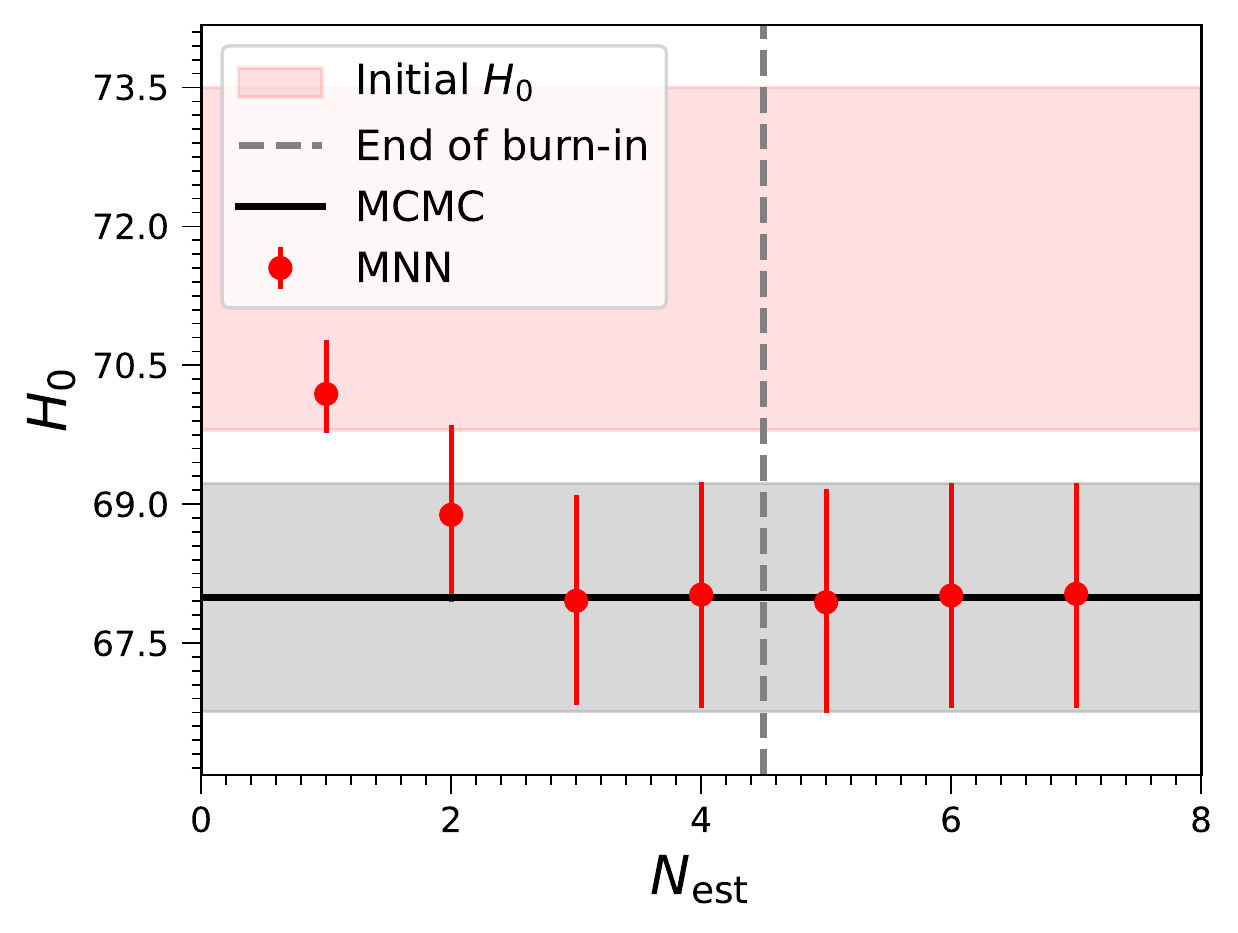}
	\caption{Best-fit values and $1\sigma$ errors of $H_0$ as a function of the number of estimations. Here, the $H_0$ is constraint with {\it Planck}-2015 $C^{\rm TT}_{\ell}$. The initial $H_0$ here is the initial setting of parameter space when using CoLFI (see Figure \ref{fig:colfi_schematic}). The black solid line with the gray area is the result of the MCMC method. Each estimation corresponds to an estimate in Figure \ref{fig:colfi_schematic}.}\label{fig:biased_initial_H0}
\end{figure}

The reason that the parameter space can be updated in the training process (see Figure~\ref{fig:colfi_schematic}) is that if the best-fit values are not located near the center of the parameter space (or not covered by the parameter space) of the training set; the network will make a biased estimation. To illustrate this, with the same procedure and MNN settings as Section \ref{sec:application_to_CMB}, we constrain the six cosmological parameters ($H_0$, $\Omega_{\rm b}h^2$, $\Omega_{\rm c}h^2$, $\tau$, $A_{\rm s}$, and $n_{\rm s}$) using {\it Planck} $C^{\rm TT}_{\ell}$. Specifically, we assume the best-fit value of $H_0$ in the $i$th estimated posterior (see Figure~\ref{fig:colfi_schematic}) deviates from $67.701\pm0.633$ (the MCMC result in Table \ref{tab:params_planck_TT_6params}) with $1\sigma_p$, $2\sigma_p$, $3\sigma_p$, $4\sigma_p$, $5\sigma_p$, and $6\sigma_p$, respectively, and other parameters are fixed to the MCMC results. The deviation here is deﬁned as
\begin{equation}\label{equ:deviation_H0}
\Delta H_0 = \frac{H_0 - H_{0,p}}{\sigma_p},
\end{equation}
where $H_{0,p}=67.701$, and $\sigma_p=0.633$. Therefore, the $H_0$ values for the six cases are $68.334$, $68.967$, $69.600$, $70.233$, $70.866$, and $71.499$, respectively. Then, we train six MNN models for the $(i+1)$th estimation, with samples generated in six different parameter spaces. Finally, we calculate mean deviations of parameters between the $(i+1)$th posterior and the $i$th posterior, as shown in Figure~\ref{fig:biased_H0} with the blue solid line. We can see that the deviation between two posteriors increases with $\Delta H_0$. At the same time, we also calculate the mean relative errors of parameters between these two posteriors, which is indicated by the orange dashed line. The relative errors also increase with $\Delta H_0$. This means that, if the true parameters are not located near the center of the parameter space (or not covered by the parameter space), then both the best-fit values and errors of parameters in the $(i+1)$th estimation will be very different from those of the $i$th estimation. Therefore, the parameter space should be updated for the next training and estimation. Thus, the MNN method is not sensitive to the initial parameters, which indicates that the initial parameters can be set free with an arbitrary range of values. This fact outperforms the MCMC method and favors models that lack prior knowledge. To prove this, in Figure \ref{fig:biased_initial_H0}, we show an example with biased initial $H_0$ when constraining parameters using the {\it Planck}-2015 $C^{\rm TT}_{\ell}$. We can see that, even if the biased initial $H_0$ does not cover the best-fit Hubble constant, the MNN gradually finds out the true posterior, and the final result is almost the same as that of the MCMC method. We note that the setting of initial parameters will affect the number of estimations in the burn-in phase, which will further affect the training time. Therefore, it is recommended to use appropriate initial parameters covering the posterior distribution to reduce the burn-in phase.

As we illustrated in Section \ref{sec:training_and_parameter_estimation} ANN chains after burn-in can be
used for parameter estimations. The end of burn-in here is judged based on deviations of parameters and errors. Specifically, in the setting of CoLFI, the end of burn-in is defined as the maximum deviation of parameters between the $(i+1)$th posterior and the $i$th posterior is less than $0.25\sigma$ and the maximum relative deviation of errors less than $25\%$. Of course, these settings can be optimized in further research.

\begin{table*}
	\centering
	\caption{Comparison of the ANN, MDN, and MNN methods. $N$ is the number of cosmological parameters. $\hat{\bm\theta}$ is the ground truth (i.e., the target) in the training set. $\mathcal{L}_{\rm G}$ is the loss function for the MDN with Gaussian components, and $\mathcal{L}_{\rm B}$ is the loss function for the MDN with Beta components. Inference input here refers to the input of the network when estimating parameters. $\bm{d}_0$ is the observational data.}\label{tab:compare_with_MDN_ANN}
	\resizebox{\linewidth}{!}{%
		\begin{tabular}{c|c|c|c|c}
			\hline\hline
			Methods & Principle & Loss Function & Noise Type & $\begin{aligned}
			&\text{Inference} \\ 
			&\text{Input}
			\end{aligned}$ \\
			\hline
			ANN 
			& $p(\bm\theta|\bm{d})$
			& $\begin{aligned}
			\mathcal{L} &= \mathbb{E}\left( \frac{1}{N}\sum_{i=1}^{N}|\bm\theta_i - \hat{\bm\theta}_i| \right)
			\end{aligned}$ 
			& $\mathcal{N}(0, A^2\bm\Sigma)$ & $\mathcal{N}(\bar{\bm{d}}_0, \bm\Sigma)$ \\
			\hline
			MDN & $\begin{aligned}
			\nonumber p(\bm\theta|\bm{d}) &= \sum_{i=1}^K \omega_i\mathcal{N}(\bm\theta; \bm\mu_i, \bm\Sigma_i), \\
			&\text{or} \\
			p(\theta|\bm{d}) &= \sum_{i=1}^K \omega_i {\rm Beta}(\theta; \alpha_i, \beta_i)
			\end{aligned}$
			& $\begin{aligned}
			\mathcal{L}_{\rm G} &= \mathbb{E}\left[ -\ln\left( \sum_{i=1}^K \omega_i\cdot\frac{\exp{\left( -\tfrac{1}{2} (\hat{\bm\theta} - \bm\mu_i)^\top \bm\Sigma_i^{-1} (\hat{\bm\theta} - \bm\mu_i) \right)}}{\sqrt{\left( 2\pi \right)^N |\bm\Sigma_i|}} \right) \right], \\
			\mathcal{L}_{\rm B} &= \mathbb{E}\left[ -\ln\left(\sum_{i=1}^K \omega_i \cdot \frac{\Gamma(\alpha_i+\beta_i)}{\Gamma(\alpha_i)\Gamma(\beta_i)} \theta^{\alpha_i-1} (1-\theta)^{\beta_i-1}\right) \right] \end{aligned}$ 
			& $\mathcal{N}(0, \bm\Sigma)$ & $\bar{\bm{d}}_0$ \\
			\hline
			MNN & $\begin{aligned}
			p(\bm\theta|\bm{d}) = \sum_{i=1}^K \omega_i p_i(\bm\theta|\bm{d})
			\end{aligned}$
			& $\begin{aligned}
			\mathcal{L} &= \mathbb{E}\left[ -\ln\left( \sum_{i=1}^K \omega_i\cdot\frac{\exp{\left( -\tfrac{1}{2} (\bm\theta_i - \hat{\bm\theta})^\top \bm\Sigma_i^{-1} (\bm\theta_i - \hat{\bm\theta}) \right)}}{\sqrt{\left( 2\pi \right)^N |\bm\Sigma_i|}} \right) \right]
			\end{aligned}$
			& $\mathcal{N}(0, A^2\bm\Sigma)$ & $\mathcal{N}(\bar{\bm{d}}_0, \bm\Sigma)$ \\
			\hline\hline
		\end{tabular}%
	}
\end{table*}

\subsection{Comparing with MDN and ANN}\label{sec:comparing_with_MDN_and_ANN}

Here, we compare the MNN method with the MDN and ANN methods to have a deeper understanding to use them for parameter estimations. In Table \ref{tab:compare_with_MDN_ANN}, we show the difference between the three methods. We can see that all methods aim to obtain the conditional probability density $p(\bm\theta|\bm{d})$. For the ANN method, the network will output cosmological parameters directly, and the least absolute deviation is used as the loss function to quantify the difference between the predicted result and the ground truth. The $\bm\theta$ in the loss function is the estimated cosmological parameters; therefore, it should be interpreted as point estimates of the cosmological parameters. The basic idea of the MDN method is to model the posterior distribution of cosmological parameters using the Gaussian (or Beta) mixture model, which is very different from that of the ANN method. The MDN method will first obtain the parameters of the mixture model, and then get the posterior distribution by sampling with the mixture model. Therefore, the loss function here is the negative logarithm of the probability density of the mixture model. For the MNN method, the network also outputs cosmological parameters directly, which is similar to the ANN method. Therefore, the $\bm\theta_i$ in the loss function should be interpreted as the point estimates of the cosmological parameters, which is quite different from the MDN with Gaussian components, although they have a similar loss function. Unlike the ANN method, the covariance information is output by the network in MNN to ensure that the estimated cosmological parameters have the correct correlations. In addition, the MNN method outputs multiple sets of cosmological parameters, and the posterior distribution is a combination of these sets of cosmological parameters, which is similar to the MDN method.

Furthermore, the training and parameter estimation of these three methods are different. Since the ANN and the MNN methods learn the mapping between the data space of measurements and the parameter space of cosmological parameters, different levels of noise $\mathcal{N}(0, A^2\bm\Sigma)$ can be added to the training set to make the network learn a good mapping between the measurements and the cosmological parameters. The posterior distribution can then be obtained by feeding samples subject to $\mathcal{N}(\bar{\bm{d}}_0, \bm\Sigma)$. In contrast, the MDN method learns a mapping between measurements and parameters of the mixture model. Therefore, the noise added to the training set should be $\mathcal{N}(0, \bm\Sigma)$, and the parameters of the mixture model can be obtained by feeding the mean (or best values) $\bar{\bm{d}}_0$ of the measurement.

As we illustrated in Sections \ref{sec:multiple_parameters} and \ref{sec:one_parameter}, the MNN method outperforms the ANN and MDN methods for multiple parameters with truncated distributions, while, for the case of only one parameter, the MDN (or ANN) method outperforms the MNN method. Therefore, CoLFI is designed to contain these three methods, which makes it possible to estimate the parameters for any kind of cosmological model.

In addition, there are many hyperparameters that can be selected manually when using CoLFI, such as the number of hidden layers, the number of training samples, the number of epochs, and activation functions. The analysis in Section \ref{sec:effect_of_hyperparameters} shows that the selection of these hyperparameters makes the final estimated parameters have certain instability. But this shortcoming does not affect the fact that ANN, MDN, and MNN can all get accurate parameter estimations with a mean deviation level of $\mathcal{O}(10^{-2}\sigma)$, after selecting appropriate hyperparameters (e.g.,~using a few thousands of samples or epochs).

\subsection{Extensibility of CoLFI}\label{sec:extensibility_of_colfi}

The measurements $\bm{d}_0$ here are assumed subject to multivariate Gaussian distribution. Therefore, Gaussian noise is added to the training set in the training process. This is why the MNN (or MDN, ANN) results are the same as the MCMC results. However, if the measurements deviate from the Gaussian distribution, CoLFI can be extended to add non-Gaussian noise to conduct parameter estimates.

In addition, it should be noted that the network used in CoLFI is a fully connected network that can only deal with one-dimensional data. But it is possible to extend CoLFI to deal with higher-dimensional data, such as two-dimensional maps or three-dimensional cubes of many sky survey experiments, to extract more useful information. This will improve our understanding of what information is encoded in two-dimensional and three-dimensional data and how we can extract all the information from the data, improving our understanding of the physics of the universe. We will investigate these interesting issues in future works.

\section{Conclusions}\label{sec:conclusions}

In this work, we propose a new method named MNN to estimate cosmological parameters by learning the conditional probability density $p(\bm\theta|\bm{d})$. We test the MNN by constraining parameters of the $\Lambda$CDM and $w$CDM models using the {\it Planck}-2015 CMB TT+TE+EE, and Pantheon SN Ia data, and find that it can obtain almost the same result as the traditional MCMC method with a slight difference of $\mathcal{O}(10^{-2}\sigma)$. In addition, we propose sampling parameters in a hyper-ellipsoid for the generation of the training set. This sampling method provides an efficient way to obtain accurate parameter estimations using $\mathcal{O}(10^2)$ forward simulation samples, which makes parameter estimation faster, especially for complex and resource-consuming cosmological models.

Besides, a code called CoLFI is developed for parameter estimations, which is suitable for any parameter estimation of complicated models in a wide range of scientific fields. The MNN method outperforms the ANN and MDN methods for multiple parameters with truncated distributions, while the MDN and ANN methods have advantages in the case of only one parameter with truncated distributions. Therefore, CoLFI is designed to incorporate the advantages of the MNN, ANN, and MDN methods.

The current stage of CoLFI can only deal with one-dimensional data. However, it can also be extended to two-dimensional and three-dimensional data such as galaxy surveys~\citep{LSST_book} and 21 cm intensity mapping~\citep{Cunntington2022}, which may extract useful information to measure modes of perturbations in the universe.

\section{Acknowledgement}

Y.-Z.M. acknowledges the support of the National Research Foundation with grants No.~150580, No.~120385, and No.~120378. J.-Q.X. is supported by the National Science Foundation of China under grant No.~U1931202. All computations were carried out using the computational cluster resources at the Centre for High-Performance Computing, Cape Town, South Africa. This work was part of the research program ``New Insights into Astrophysics and Cosmology with Theoretical Models Confronting Observational Data'' of the National Institute for Theoretical and Computational Sciences of South Africa.

\end{document}